\documentclass[iop]{emulateapj}

\usepackage{natbib,aas_macros,amsmath}
\usepackage{multirow,color}

\begin{document}
\shorttitle{Sizes of Galaxies at $\lowercase{z} \sim 7-12$ 
}
\shortauthors{Ono et al.}
%\slugcomment{Ver. December 15, 2012}
\slugcomment{submitted to ApJ}

\title{%
Evolution of the Sizes of Galaxies over $7<\lowercase{z} <12$ \\
Revealed by the 2012 Hubble Ultra Deep Field Campaign
}

\author{%
Yoshiaki Ono\altaffilmark{1}, 
Masami Ouchi\altaffilmark{1,2},
Emma Curtis-Lake\altaffilmark{3},
Matthew A. Schenker\altaffilmark{4}, 
Richard S. Ellis\altaffilmark{4}, \\
Ross J. McLure\altaffilmark{3},
James S. Dunlop\altaffilmark{3},
Brant E. Robertson\altaffilmark{5},
Anton M. Koekemoer\altaffilmark{6}, \\
Rebecca A.A. Bowler\altaffilmark{3},
Alexander B. Rogers\altaffilmark{3},
Evan Schneider\altaffilmark{5},
Stephane Charlot\altaffilmark{7}, \\
Daniel P. Stark\altaffilmark{5}, 
Kazuhiro Shimasaku\altaffilmark{8}, 
Steven R. Furlanetto\altaffilmark{9},
Michele Cirasuolo\altaffilmark{3,10} 
}

\email{ono@icrr.u-tokyo.ac.jp}

\altaffiltext{1}{%
Institute for Cosmic Ray Research, The University of Tokyo,
Kashiwa 277-8582, Japan
}
\altaffiltext{2}{%
Kavli Institute for the Physics andMathematics of the Universe 
(Kavli IPMU), WPI, The University of Tokyo, Chiba 277-8583, Japan
}
\altaffiltext{3}{%
Institute for Astronomy, University of Edinburgh, Royal Observatory,
Edinburgh EH9 3HJ, UK
}
\altaffiltext{4}{%
Department of Astrophysics, California Institute of Technology,
MS 249-17, Pasadena, CA 91125
}
\altaffiltext{5}{%
Department of Astronomy and Steward Observatory, University of Arizona, 
Tucson AZ 85721
}
\altaffiltext{6}{%
Space Telescope Science Institute, Baltimore, MD 21218
}
\altaffiltext{7}{%
UPMC-CNRS, UMR7095, Institut dfAstrophysique, F-75014 Paris, France
}
\altaffiltext{8}{%
Department of Astronomy, Graduate School of Science, 
The University of Tokyo, Tokyo 113-0033, Japan
}
\altaffiltext{9}{%
Department of Physics {\&} Astronomy, University of California, 
Los Angeles CA 90095
}
\altaffiltext{10}{%
UK Astronomy Technology Centre, Royal Observatory, 
Edinburgh EH9 3HJ, UK
}

%---------------------------------------------------------------------
\begin{abstract}
We analyze the redshift- and luminosity-dependent sizes of dropout galaxy candidates in
the redshift range $z \sim 7-12$ using deep images from the 2012 Hubble Ultra Deep 
Field (UDF12) campaign, data which offers two distinct advantages over that used in earlier work.
Firstly, we utilize the increased signal-to-noise ratio offered by the 
UDF12 imaging to provide improved size measurements 
for known galaxies at $z \simeq 6.5 - 8$ in the HUDF. 
Specifically, 
we stack the new deep F140W image with the existing F125W data 
in order to provide improved measurements of the half-light radii of $z'$-band dropouts
(at $z \simeq 7$).  
Similarly we stack this image with the new deep UDF12 F160W image to obtain 
new size measurements for a sample of $Y$-band dropouts (at $z \simeq 8$). 
Secondly, because the UDF12 data 
have allowed the construction of the first robust galaxy sample in the HUDF at $z > 8$,
we have been able to extend the measurement of average galaxy size out to significantly 
higher redshifts.
Restricting our size measurements to sources which are now detected at  $> 15 \sigma$, 
we confirm earlier indications that the average half-light radii of $z\sim 7-12$ galaxies 
are extremely small, $0.3-0.4$ kpc, comparable to the sizes of giant molecular associations in 
local star-forming galaxies. 
We also confirm that there is a clear trend of 
decreasing half-light radius with increasing redshift, and provide 
the first evidence that this trend continues beyond $z \simeq 8$.
Modeling the evolution of the average half-light radius as a power-law, $\propto\,(1+z)^{s}$, 
we obtain a best-fit index of $s=-1.28\pm 0.13$ over the redshift range $z\sim 4-12$, 
mid-way between the physically expected evolution for baryons embedded in dark halos of 
constant mass ($s=-1$) and constant velocity ($s=-1.5$).
A clear size-luminosity relation, such as that found at lower redshift, is also evident in 
both our $z$- and $Y$-dropout sample. 
This relation can be interpreted in terms of a 
constant surface density of star formation over a range in luminosity of $0.05-1.0 L^\ast_{z=3}$. 
Our results also strengthen previous claims that the star-formation surface density in 
dropout galaxies is broadly unchanged from $z \simeq 4$ to $z \simeq 8$ at 
$\Sigma_{\rm SFR} \simeq 2 \, {\rm M}_{\odot} \, {\rm yr}^{-1} \, {\rm kpc}^{-2}$. 
This value is $2-3$ orders of magnitude 
lower than that found in extreme starburst galaxies, but is very 
comparable to that seen today in the centers of normal disk galaxies. 
This provides 
further support for a steady smooth build-up of the stellar populations in 
galaxies in the young universe.
\end{abstract}

%---------------------------------------------------------------------
\keywords{%
galaxies: formation ---
galaxies: evolution ---
galaxies: high-redshift ---
galaxies: structure
}
%---------------------------------------------------------------------
%---------------------------------------------------------------------

%%%%%%%%%%%%%%%%%%%%%%%%%%%%%%%%%%%%%%%%%%%%%%%%%%%%%%%%%%%%%%%%%
%%%%%%%%%%%%%%%%%%%%%%%%%%%%%%%%%%%%%%%%%%%%%%%%%%%%%%%%%%%%%%%%%
\section{Introduction} \label{sec:introduction}
%%%%%%%%%%%%%%%%%%%%%%%%%%%%%%%%%%%%%%%%%%%%%%%%%%%%%%%%%%%%%%%%%
%%%%%%%%%%%%%%%%%%%%%%%%%%%%%%%%%%%%%%%%%%%%%%%%%%%%%%%%%%%%%%%%%

Considerable progress has been made 
in charting the abundance of galaxies at $z \sim 7-10$ 
from deep imaging with ground-based observations 
and various campaigns undertaken with 
%the WFC3/IR camera 
the Wide Field Camera 3 infrared channel (WFC3/IR) 
on \textit{Hubble Space Telescope} (HST). 
Sample selection makes use of the well-established dropout technique, 
which takes advantage of the unique spectral characteristics of 
high-redshift star-forming galaxies, 
i.e., a blue UV spectrum and a sharp drop in flux at wavelengths 
shorter than Ly$\alpha$. 
These complementary studies have 
identified a large number of dropout galaxies at $z \sim 7$ and beyond. 
Investigating the abundance of dropout galaxies over $7 < z < 10$  
has revealed a clear decrease in the number density of luminous galaxies 
with increasing redshift \citep[e.g.,][]{ouchi2009b,mclure2009,castellano2010,oesch2010,bouwens2011}.

Characterizing the evolution of galaxy morphologies and sizes 
is useful for understanding galaxy formation history. 
Analytical studies have calculated the size-redshift relation of disk galaxies, 
suggesting the typical size of galaxies of a given luminosity is expected 
to decrease with increasing redshift \citep{mo1998,mo1999}.
The virial radius of a dark matter halo scales with redshift and virial velocity or virial mass.  
Assuming that the exponential scale length of the baryonic disc scales with the virial radius, 
the sizes of disks are expected to scale with redshift, 
proportional to $H(z)^{-1}$ at a fixed mass or 
$H(z)^{-2/3}$ at a fixed circular velocity \citep[e.g.,][]{ferguson2004},
where $H(z)$ is the Hubble parameter 
which scales as $\sim (1+z)^{3/2}$ at high redshifts.

Earlier observations have reported that 
the sizes (half-light radii) of dropout galaxies decrease according to about $(1+z)^{-1}$ 
up to $z \sim 7$ \citep{oesch2010b},  
which is expected at fixed halo masses, 
and in good agreement with previous estimates at lower redshifts \citep{bouwens2004}. 
However, 
because they used only the first epoch of their survey data with the WFC3/IR, 
their analysis still shows large uncertainties, especially in their fainter sample. 
Therefore, it is also consistent with $(1+z)^{-1.5}$ \citep{ferguson2004,hathi2008}, 
as expected for sizes that scale with halo circular velocity.

\cite{oesch2010b} have also reported that 
the star-formation rate (SFR) surface densities of dropout galaxies 
remains constant from $z \sim 7$ to $z \sim 4$. 
They suggest a possible explanation is that 
the average star formation efficiency is very similar in all these galaxies, 
and that 
feedback effects change the mode of star formation by only a small amount.
It would be interesting to see if this possible trend continues toward higher redshifts, 
to infer star-formation activities in galaxies at earlier epochs of galaxy formation.

Recently, a new campaign was carried out with 
the WFC3/IR 
to significantly deepen 
the Hubble Ultra Deep Field in 2012 
(GO 12498; PI: R. Ellis, hereafter UDF12: \citealt{ellis2012b} and 
\citealt{koekemoer2012} for the project description); 
this yields the deepest near-infrared images ever obtained. 
Additional scientific results from this project are presented 
in \cite{dunlop2012}, \cite{schenker2012}, 
\cite{mclure2012b}, and \cite{robertson2012}. 
In this paper, we study morphologies of $z \sim 7-12$ galaxies 
based on the complete WFC3/IR UDF12 data set. 
The advantages of the new images are 
(i) a new F140W image and a deeper F160W data
from which we obtain robust estimates 
on the rest-frame UV morphologies of galaxies not only at $z \sim 7$, 
but also $z \sim 8-12$ for the first time,  
(ii) a deeper F105W image which enables us to safely exclude 
contaminations by foreground sources from our galaxy samples at $z \sim 8$ and beyond\footnote{We 
do not use our deep F105W image for the morphology analysis of $z \sim 7$ galaxies, since 
a redshifted Ly$\alpha$ and the continuum break of an object at $z \gtrsim 6.4$ enters the F105W band.}. 
The purpose of this paper is to investigate the galaxy size and 
SFR surface density evolution beyond $z \sim 7$, 
and the correlation of size with UV luminosity.

\tabletypesize{\tiny}
%ttttttttttttttttttttttttttttttttttttttttttttttttttttttttttttttttttttttttt%
\begin{deluxetable*}{ccccccccccc} 
\tablecolumns{11} 
\tablewidth{0pt} 
\tablecaption{Bright $z_{850}$- and $Y_{105}$-Dropout Galaxies in the HUDF Reported in the Literature\label{tab:sum_drop_hudf}}
\tablehead{
    \colhead{ID}     
    &  \colhead{(1)}    
    &  \colhead{(2)}    
    & \colhead{(3)}  
    & \colhead{(4)} 
    &  \colhead{(5)}    
    &  \colhead{(6)}    
    & \colhead{(7)}  
    & \colhead{(8)} 
    &  \colhead{(9)}    
    &  \colhead{(10)}    
}
\startdata 
 \multicolumn{11}{c}{\bf Bright $z_{850}$-dropouts}  \smallskip \\
UDF12-4258-6567 & UDF-640-1417 & --- & UDFz-42566566 & 688 & 1441 & A032 & zD1 & HUDF.z.4444 & UDFz-42566566 & HUDF-658 \\ 
UDF12-3746-6327 & --- & --- & --- & 837 & 769 & --- & --- & --- & --- & HUDF-796 \\ 
UDF12-4256-7314 & UDF-387-1125 & --- & UDFz-42577314 & 1144 & 2432 & A008 & zD3 & HUDF.z.6433 & UDFz-42567314 & --- \\ 
UDF12-4219-6278 & --- & --- & --- & 1464 & 649 & --- & --- & HUDF.z.2677 & UDFz-42196278 & HUDF-1442 \\ 
UDF12-3677-7535 & --- & --- & UDFz-36777536 & 1911 & 2894 & --- & --- & HUDF.z.7462 & UDFz-36777536 & HUDF-1473 \\ 
UDF12-4105-7155 & --- & --- & UDFz-41057156 & 2066 & 2013 & A017 & --- & --- & --- & --- \\ 
UDF12-3958-6564 & --- & --- & UDFz-39586565 & 1915 & 1445 & A033 & --- & --- & UDFz-39576564 & HUDF-1995 \\ 
UDF12-3744-6512 & --- & --- & UDFz-37446513 & 1880 & 1289 & A040 & --- & HUDF.z.4121 & UDFz-37446512 & HUDF-1632 \\ 
UDF12-3638-7162 & --- & --- & UDFz-36387163 & 1958 & 2032 & A016 & zD6 & HUDF.z.5659 & UDFz-36377163 & HUDF-1818 \\ 
% \smallskip \\
 \multicolumn{11}{c}{\bf Bright $Y_{105}$-dropouts}  \smallskip \\
UDF12-3879-7071 & UDF-983-964 & HUDF-480 & UDFz-38807073 & 835 & 1768 & A025 & zD2 & HUDF.z.5141 & UDFy-38807071 & HUDF-860 \\ 
UDF12-4470-6442 & --- & --- & UDFz-44716442 & 1107 & 1106 & A044 & zD7 & --- & UDFy-44706443 & HUDF-1173 \\ 
UDF12-3952-7173 & --- & --- & ---\tablenotemark{$\dagger1$} & 1422 & 2055 & B041 & ---\tablenotemark{$\dagger2$} & --- & UDFy-39537174 & --- \\ 
UDF12-4314-6284 & --- & --- & UDFz-43146285 & 1678 & 669 & A060 & zD5 & HUDF.z.2714 & UDFy-43136284 & HUDF-1419 \\ 
UDF12-3722-8061 & --- & --- & UDFz-37228061 & 1574 & 3053 & A003 & zD9 & --- & UDFy-37218061 & HUDF-1660 \\ 
UDF12-3813-5540 & --- & --- & UDFy-38135539 & 1721 & 125 & B115 & YD3 & HUDF.YD3 & UDFy-38125539 & HUDF-2003 \\ 
 \multicolumn{11}{c}{\bf Bright $z>8.5$ candidates}  \smallskip \\
UDF12-3954-6284 & --- & --- & --- & --- & --- & --- & --- & --- & UDFj-39546284 & --- \\
  \enddata 
\tablecomments{
(1) \cite{bouwens2008} ;
(2) \cite{oesch2009};
(3) \cite{oesch2010} and \cite{bouwens2010}; 
(4) \cite{mclure2009};
(5) \cite{finkelstein2009f};
(6) \cite{yan2010};
(7) \cite{bunker2010};
(8) \cite{wilkins2011} and \cite{lorenzoni2010};
(9) \cite{bouwens2011} and \cite{bouwens2011b};
(10) \cite{mclure2011}. 
UDF12-4258-6567 has been spectroscopically confirmed by \cite{fontana2010}.
}
\tablenotetext{$\dagger1$}{%
Close to UDFz-39557176.
}
\tablenotetext{$\dagger2$}{%
Close to zD4.
}
\end{deluxetable*} 
%ttttttttttttttttttttttttttttttttttttttttttttttttttttttttttttttttttttttttt%
\tabletypesize{\footnotesize}

\tabletypesize{\scriptsize}
%ttttttttttttttttttttttttttttttttttttttttttttttttttttttttttttttttttttttttt%
\begin{deluxetable*}{ccccccccc} 
\tablecolumns{9} 
\tablewidth{0pt} 
\tablecaption{Photometric Redshifts for Bright $z_{850}$- and $Y_{105}$-Dropout Galaxies Found in the HUDF\label{tab:sum_photoz_drop_hudf}}
\tablehead{
    \colhead{ID}     
    &  \colhead{ID(i)}    
    &  \colhead{$z_{\rm photo}$}    
    & \colhead{$\delta z_{\rm photo}$}  
    &  \colhead{ID(ii)}    
    &  \colhead{$z_{\rm photo}$}    
    & \colhead{$\delta z_{\rm photo}$}  
    &  \colhead{ID(iii)}    
    &  \colhead{$z_{\rm photo}$}    
}
\startdata 
 \multicolumn{9}{c}{\bf Bright $z_{850}$-dropouts}  \smallskip \\
UDF12-4258-6567 & 688 & 6.70 & (6.50-6.90) &    1441 & 6.83 & (6.70-6.98) &   40250 & 7.05 \\
UDF12-3746-6327 & 837 & 6.35 & (6.15-6.55) &    769 & 6.36 & (6.09-6.47) &     ---\tablenotemark{$\dagger1$} & ---\tablenotemark{$\dagger1$} \\
UDF12-4256-7314 & 1144 & 6.80 & (6.50-7.10) &  2432 & 6.86 & (6.70-7.01) & 40332 & 7.20 \\
UDF12-4219-6278 & 1464 & 6.30 & (5.95-6.75) &  649 & 6.40 & (6.25-6.58) &     20075 & 6.70 \\
UDF12-3677-7535 & 1911 & 6.40 & (6.20-6.60) &  2894 & 6.34 & (6.21-6.49) &   40396 & 6.50 \\
UDF12-4105-7155 & 2066 & 7.20 & (6.50-7.80) &  2013 & 6.70 & (6.48-7.03) & 40296 & 7.05 \\
UDF12-3958-6564 & 1915 & 6.40 & (6.15-6.65) &  1445 & 6.40 & (6.17-6.61) &   40252 & 6.80 \\
UDF12-3744-6512 & 1880 & 6.50 & (6.25-6.80) &  1289 & 6.38 & (6.22-6.55) &   40226 & 6.70 \\
UDF12-3638-7162 & 1958 & 6.50 & (6.25-6.80) &  2032 & 6.40 & (6.23-6.58) &   40309 & 6.55 \\
% \smallskip \\
 \multicolumn{9}{c}{\bf Bright $Y_{105}$-dropouts}  \smallskip \\
UDF12-3879-7071 & 835 & 7.20 & (6.90-7.50) &     1768 & 7.22 & (7.08-7.40) &   40278 & 7.65 \\
UDF12-4470-6442 & 1107 & 7.60 & (7.30-7.90) &   1106 & 7.32 & (7.12-7.53) &   40197 & 7.70 \\
UDF12-3952-7173 & 1422 & 7.60 & (7.00-8.05) &   2055 & 7.81 & (7.48-8.24) & 70284 & 7.85 \\
UDF12-4314-6284 & 1678 & 7.05 & (6.60-7.40) &   669 & 7.25 & (7.00-7.57) &      40143 & 7.30 \\
UDF12-3722-8061 & 1574 & 7.20 & (6.55-7.60) &   3053 & 7.40 & (7.12-7.70) &    40410 & 7.70 \\
UDF12-3813-5540 & 1721 & 8.45 & (7.75-8.85) &   125 & 8.61 & (8.19-8.89) &    70020 & 8.30
  \enddata 
\tablecomments{
(i) \cite{mclure2009};
(ii) \cite{finkelstein2009f};
(iii) \cite{mclure2012b}.
$\delta z_{\rm photo}$ denotes $1\sigma$ confidence interval.
}
\tablenotetext{$\dagger1$}{%
This object is not included in the catalog of \cite{mclure2012b}, 
because its photometric redshift is less than $6.5$.
}
\end{deluxetable*} 
%ttttttttttttttttttttttttttttttttttttttttttttttttttttttttttttttttttttttttt%
\tabletypesize{\footnotesize}

%sectioin2 {sec:observations} 
%section3 {sec:samples}
%section4 {sec:morphologies_of_galaxies}
%section5 {sec:results_discussion}
%section6 {sec:summary}

The outline of this paper is as follows. 
After describing the imaging data used in this study in Section \ref{sec:observations}, 
we summarize our dropout galaxy samples in Section \ref{sec:samples}. 
Our size analysis is described in Section \ref{sec:morphologies_of_galaxies}. 
In Section \ref{sec:results_discussion}, 
we investigate the size-luminosity relation and the size evolution 
and discuss the implications. 
A summary is given in Section \ref{sec:summary}. 
Throughout this paper, we use magnitudes in the AB system \citep{oke1983}
and assume a flat universe 
with ($\Omega_{\rm m}$, $\Omega_{\rm \Lambda}$, $h$) $= (0.3, \, 0.7, \, 0.7)$.
In this cosmological model, 
an angular dimension of $1.0$ arcsec corresponds to 
a physical dimension of 
$5.365$ kpc arcsec$^{-1}$ at $z=6.7$, 
$4.818$ kpc arcsec$^{-1}$ at $z=8.0$, 
$4.465$ kpc arcsec$^{-1}$ at $z=9.0$, 
and $3.683$ kpc arcsec$^{-1}$ at $z=11.9$. 
We express galaxy UV luminosities 
in units of the characteristic luminosity of $z \sim 3$ galaxies, 
$L^\ast_{z=3}$,
which corresponds to $M_{1600} = -21.0$ \citep{steidel1999}. 
The four WFC3/IR filters we use, F105W, F125W, F140W and F160W 
are denoted by $Y_{105}$, $J_{125}$, $J_{140}$ and $H_{160}$, respectively.
We also use four ACS filters, F435W, F606W, F775W and F850LP, 
which are denoted by $B_{435}$, $V_{606}$, $i_{775}$ and $z_{850}$, respectively.

%%%%%%%%%%%%%%%%%%%%%%%%%%%%%%%%%%%%%%%%%%%%%%%%%%%%%%%%%%%%%%%%%
%%%%%%%%%%%%%%%%%%%%%%%%%%%%%%%%%%%%%%%%%%%%%%%%%%%%%%%%%%%%%%%%%
\section{Observations} \label{sec:observations}
%%%%%%%%%%%%%%%%%%%%%%%%%%%%%%%%%%%%%%%%%%%%%%%%%%%%%%%%%%%%%%%%%
%%%%%%%%%%%%%%%%%%%%%%%%%%%%%%%%%%%%%%%%%%%%%%%%%%%%%%%%%%%%%%%%%

The primary data set used in this morphology analysis for $z \sim 7-12$ galaxies is 
the ultra-deep WFC3/IR observations taken for  
the UDF12 campaign 
combined with images taken for 
the UDF09 campaign (GO 11563; PI: G. Illingworth).
In the UDF09 campaign, 
WFC3/IR data was obtained over three fields: 
the HUDF main, and two parallel fields. 
The UDF12 campaign has obtained $128$ orbits of WFC3/IR data over the HUDF main field. 
We have combined all the exposures 
including the data from other HST programs 
(GO12060, 12061, 12062; PI: S. Faber, H. Ferguson; GO12099; PI: A. Riess).
In total, 
the observations over the HUDF main field include $253$ orbits 
(F105W: 100 orbits, F125W: 39 orbits, F140W: 30 orbits, F160W: 84 orbits).  
A more detailed description of the UDF12 data set is provided by \cite{koekemoer2012},
and the final reduced data are being made publicly available 
as High-Level Science Products\footnote{http://archive.stsci.edu/prepds/hudf12/} 
that are delivered to the Space Telescope Science Institute archive, 
and further details and current updates about the survey are provided 
at the project website\footnote{http://udf12.arizona.edu/}.

To minimize the effects of morphological $K$-correction 
and 
take the advantage of the UDF12 campaign,  
we measure sizes of galaxies in the images of the WFC3/IR band 
that is the closest to the rest-frame $1600-1700${\AA}.  
A stack of the PSF-matched $J_{125}$- and $J_{140}$-band images is used for $z_{850}$-dropouts, 
a stack of the PSF-matched $J_{140}$- and $H_{160}$-band images  for $Y_{105}$-dropouts, 
and 
the $H_{160}$-band image for candidates at $z>8.5$. 
Their $5 \sigma$ limiting magnitudes are 
$29.8$ ($J_{125}+J_{140}$), 
$29.7$ ($J_{140}+H_{160}$), 
and 
$29.5$ ($H_{160}$) 
within filter-matched apertures, 
which are $0.45-0.50$ arcsec in diameter \citep{ellis2012b}. 
We use images with a pixel scale of 0.03 arcsec pixel$^{-1}$.

%%%%%%%%%%%%%%%%%%%%%%%%%%%%%%%%%%%%%%%%%%%%%%%%%%%%%%%%%%%%%%%%%
%%%%%%%%%%%%%%%%%%%%%%%%%%%%%%%%%%%%%%%%%%%%%%%%%%%%%%%%%%%%%%%%%
\section{Samples} \label{sec:samples}
%%%%%%%%%%%%%%%%%%%%%%%%%%%%%%%%%%%%%%%%%%%%%%%%%%%%%%%%%%%%%%%%%
%%%%%%%%%%%%%%%%%%%%%%%%%%%%%%%%%%%%%%%%%%%%%%%%%%%%%%%%%%%%%%%%%

We investigate the sizes of $z \sim 7-12$ galaxies 
based on the the $z\sim7-8$ samples selected by \cite{schenker2012} 
and the $z>8.5$ samples whose photometric redshifts 
from SED fitting analysis are available in \cite{mclure2012b} \citep[see also][]{ellis2012b}. 
Here we briefly summarize how these galaxies are selected.

To select star-forming galaxies at $z \sim 7-8$, 
\cite{schenker2012} applied the dropout technique, 
which probes a blue UV spectrum and 
a spectral break blueward of Ly$\alpha$ due to IGM absorption. 
For $z \sim 7$ $z_{850}$-dropout galaxies, 
they first required a $3.5 \sigma$ detection in $Y_{105}$ 
plus one of the other filters which probe longer wavelengths  
($J_{125}$, $J_{140}$, or $H_{160}$). 
Then they applied the two color criteria:  
$z_{850} - Y_{105} > 0.7$ 
and  
$Y_{105} - J_{125} < 0.4$. 
Also the following criteria were used;  
(i) the significance is less than $2.0 \sigma$ in $B_{435}$, $V_{606}$, and $i_{775}$;  
(ii) the significance is not more than $1.5 \sigma$ in more than one band 
among $B_{435}$, $V_{606}$, and $i_{775}$;  
(iii) $\chi^2_{\rm opt} < 5.0$ \footnote{$\chi^2_{\rm opt}$ is defined by  
$\chi^2_{\rm opt} = \Sigma_i {\rm SGN}(f_i) \left( f_i/\sigma_i \right)^2$ 
where $f_i$ is the flux in band $i$, 
$\sigma_i$ is the uncertainty of $f_i$, 
and 
SGN$(f_i)$ is $1$ if $f_i > 0$ and $-1$ if $f_i < 0$, 
considering the bands shorter than Ly$\alpha$ \citep{bouwens2011}. 
For $z_{850}$-dropouts, $B_{435}$, $V_{606}$, and $i_{775}$ are considered, 
and 
for $Y_{105}$-dropouts,  $B_{435}$, $V_{606}$, $i_{775}$, $z_{850}$ are considered.}.
For $z \sim 8$ $Y_{105}$-dropout galaxies, 
they required a $3.5 \sigma$ detection in $J_{125}$ 
and one of the other filters which probe longer wavelengths, $J_{140}$ and $H_{160}$. 
From the detected objects, they selected dropouts which satisfy the two color criteria, 
$Y_{105} - J_{125} > 0.5$
and
$J_{125} - H_{160} < 0.4$, 
and the following criteria for the optical data;  
(i) the significance is less than $2.0 \sigma$ in $B_{435}$, $V_{606}$, $i_{775}$, and $z_{850}$;  
(ii) the significance is not more than $1.5 \sigma$ in more than one band 
among $B_{435}$, $V_{606}$, $i_{775}$, and $z_{850}$;  
(iii) $\chi^2_{\rm opt} < 5.0$.
By using these selection criteria, 
\cite{schenker2012} identified $47$ $z_{850}$-dropouts and $27$ $Y_{105}$-dropouts in the HUDF main field. 
\cite{mclure2012b} independently searched for galaxies at $z \gtrsim 7$ 
using the photometric redshift technique.  
The objects in their catalog with photometric redshifts $z_{\rm photo} \sim 7-9$ 
are well matched with the objects in the $z_{850}$- and $Y_{105}$-dropout catalogs 
constructed by \cite{schenker2012}.

In addition to the dropout galaxies, 
we study $z>8.5$ star-forming galaxy candidates reported by \cite{ellis2012b}. 
They located all sources by examining the stack of 
the final $J_{125}$-, $J_{140}$-, and $H_{160}$-band images 
and applied the photometric redshift technique \citep[see also,][]{mclure2012b}, 
making use of the full data set obtained by the UDF12 program and the previous programs. 
They also applied the dropout technique for the master catalog, 
searching for objects undetected at 
$2 \sigma$ in both $Y_{105}$ ($>31.0$ mag)
and in a combined ACS image.  
By both of the two techniques, 
they have found seven convincing $z > 8.5$ candidates.

Morphology measurements for galaxies require a significant detection 
in not only the central region of sources, but also the outer structures. 
Recently, \cite{mosleh2012} reported that, 
in order to recover the input sizes of their realistic simulations, 
a signal-to-noise (S/N) ratio of at least $10$ is required.
To obtain robust estimates on galaxy morphologies, 
we set a more strict criterion for S/N; 
we analyze our dropouts individually down to S/N of $15$. 
The number of $z_{850}$-dropouts and $Y_{105}$-dropouts 
with detection greater than $15 \sigma$ in $J_{125}+J_{140}$ and $J_{140}+H_{160}$
(about $28.5$ mag in the filter-matched apertures) 
is $9$ and $6$, respectively.

In order to extend our analysis to fainter magnitudes, 
we divide the samples into three luminosity bins, 
$L = (0.3-1) L^\ast_{z=3}$, 
$L = (0.12 - 0.3) L^\ast_{z=3}$, 
$L = (0.048 - 0.12) L^\ast_{z=3}$, 
based on their total magnitudes in 
$J_{125}+J_{140}$ for $z_{850}$-dropouts 
and 
$J_{140}+H_{160}$ for $Y_{105}$-dropouts.
Since it is difficult to establish reliable total magnitudes for faint
sources (S/N $< 15$) using \verb|GALFIT|, 
we subdivide the galaxies into luminosity bins 
based on aperture magnitudes which contain $70${\%} 
of a point-source flux, after making the appropriate 
aperture correction to $100${\%} of anticipated 
point-source flux \citep{mclure2012b}.
We make median-stacked images separately 
for the second and third brightest luminosity bins. 
The number of $z_{850}$-dropouts ($Y_{105}$-dropouts)   
with $L = (0.12 - 0.3) L^\ast_{z=3}$ is 
$8$ ($7$), 
while  
the number with $L = (0.048 - 0.12) L^\ast_{z=3}$ is $17$ ($13$). 
Note that, among the $8$ $z_{850}$-dropouts (the $7$ $Y_{105}$-dropouts) 
in the second brightest luminosity bin, 
$7$ ($3$) are individually detected at 
more than $15\sigma$ in $J_{125}+J_{140}$ ($J_{140}+H_{160}$).
We do not use stacked images for the objects in the brightest luminosity bin, 
since the numbers of the objects are small. 
We also tried stacking objects fainter than $L=0.048L^\ast_{z=3}$, 
but they did not provide meaningful size constraints.

Within the $z > 8.5$ sample, 
UDF12-3954-6284 has a relatively high photometric redshift, $z_{\rm photo} = 11.9$, 
while the others have $z_{\rm photo}= 8.6-9.5$ \citep{ellis2012b,mclure2012b}. 
Thus, we divide them into two sub-samples: 
one with UDF12-3954-6284 and the other with the remaining six objects. 
The average photometric redshift of the latter subsample is about $9.0$. 
Since most of these objects are quite faint, 
we make a stack of the $H_{160}$ images of the six $z \simeq 9$ objects, 
giving a detection with S/N $\sim 9$.

Note that the nature of the $z \sim 12$ source is still uncertain, 
because of its accompanying diffuse morphology (Section \ref{subsec:galfit_measurements})
and its luminosity, particularly given the lack of other detections 
beyond $z \sim 10.5$ \citep{ellis2012b}. 
Nevertheless, 
since no alternative, plausible explanation for this object has yet been proposed, 
we analyze this object as a $z \sim 12$ candidate individually.

In our following analysis, 
we treat the $9$ $z_{850}$-dropouts and $6$ $Y_{105}$-dropouts 
with $>15\sigma$ detections individually, 
and also the $4$ stacked objects at $z \sim 7-8$. 
In addition, we analyze the stacked $z \simeq 9$ object and the $z=11.9$ object.

Note that our bright dropouts have been found in the literature as summarized in Table \ref{tab:sum_drop_hudf}. 
Table \ref{tab:sum_photoz_drop_hudf} summarizes their photometric redshifts reported so far.

%FFFFFFFFFFFFFFFFFFFFFFFFFFFFFFFFFFFFFFFFFFFFFFFFFFFFFFFFFFFFFFFF%
\begin{figure*}
\begin{center}
   \includegraphics[scale=0.33]{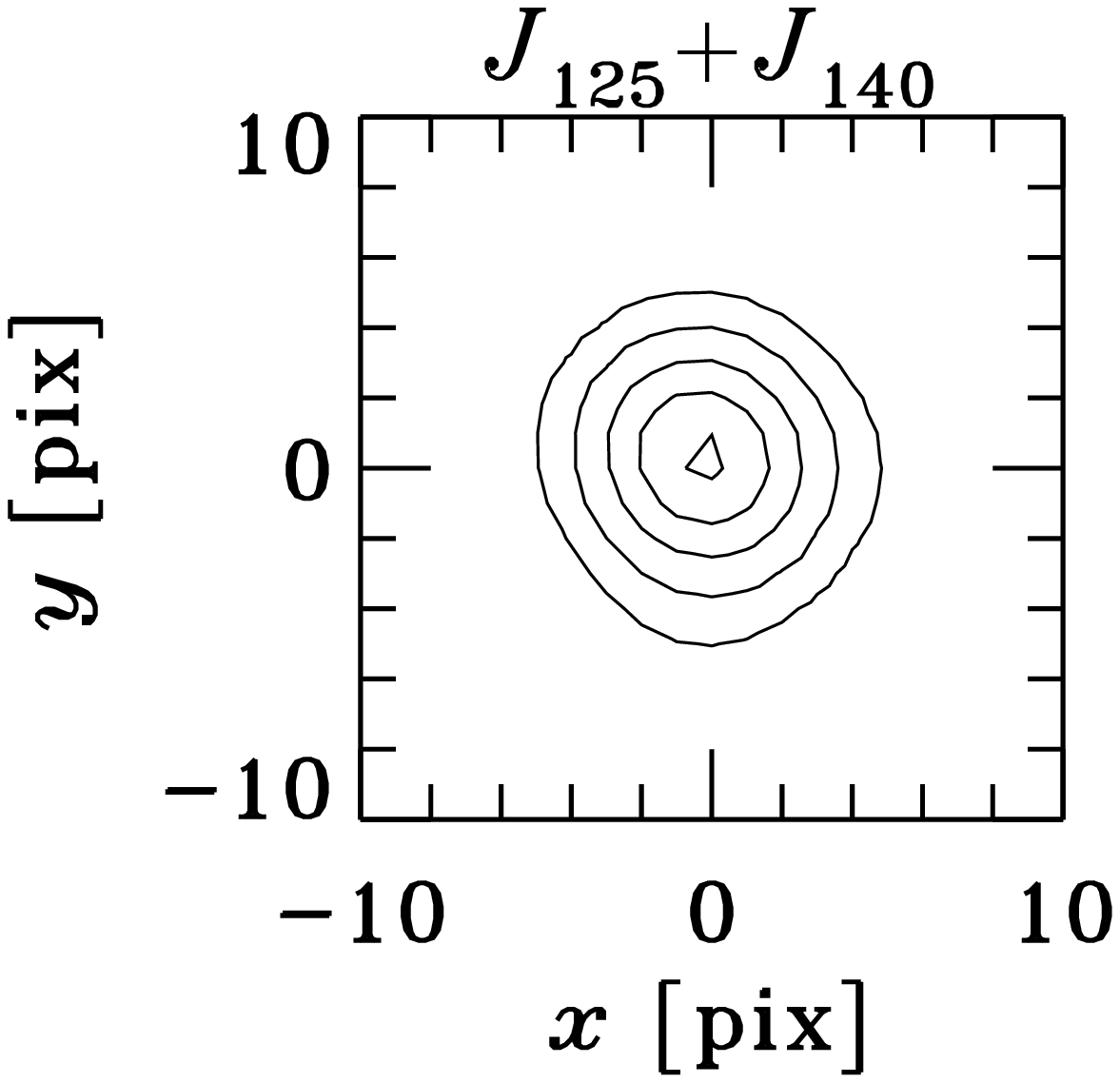}
   \includegraphics[scale=0.33]{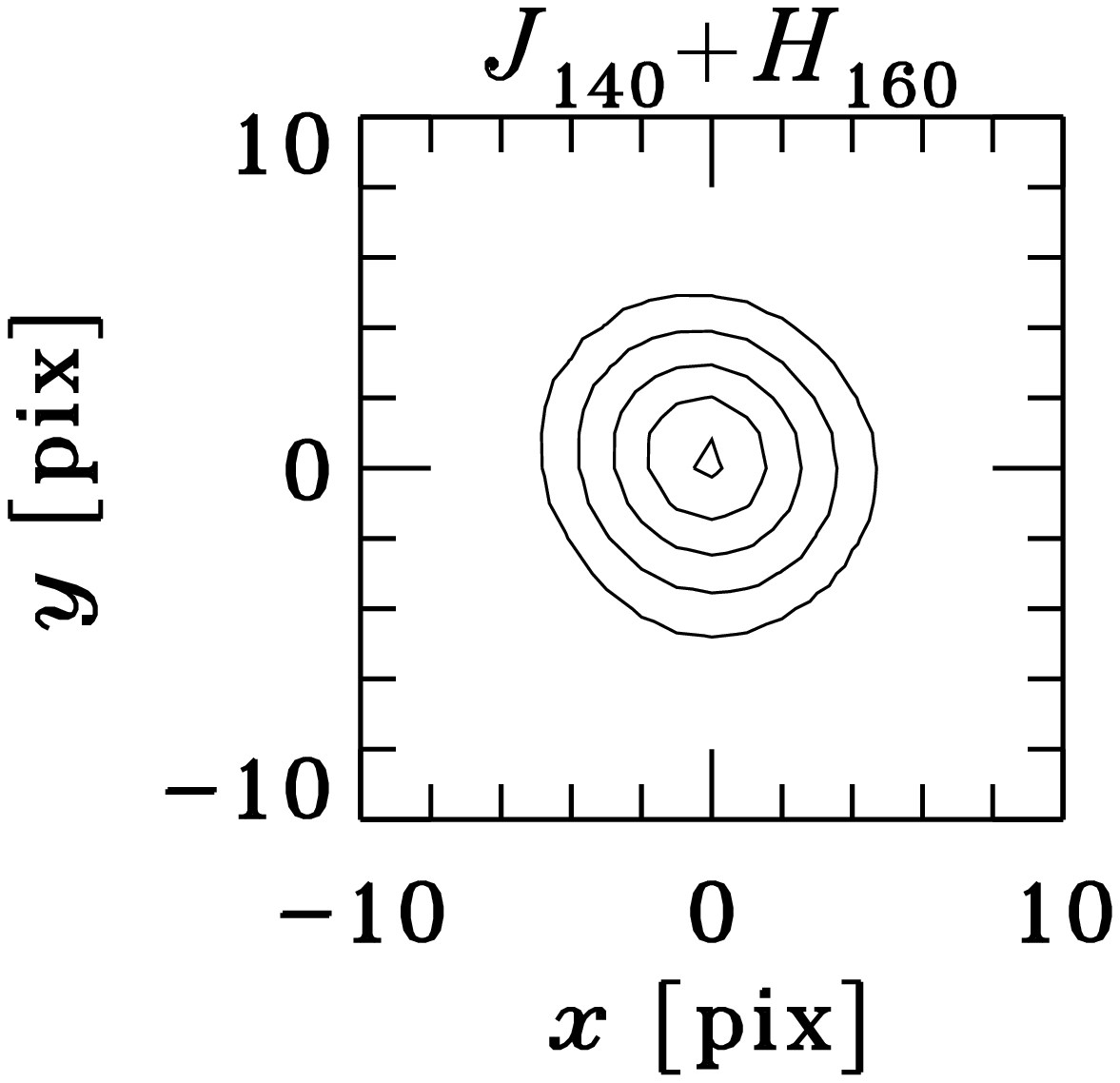}
   \includegraphics[scale=0.33]{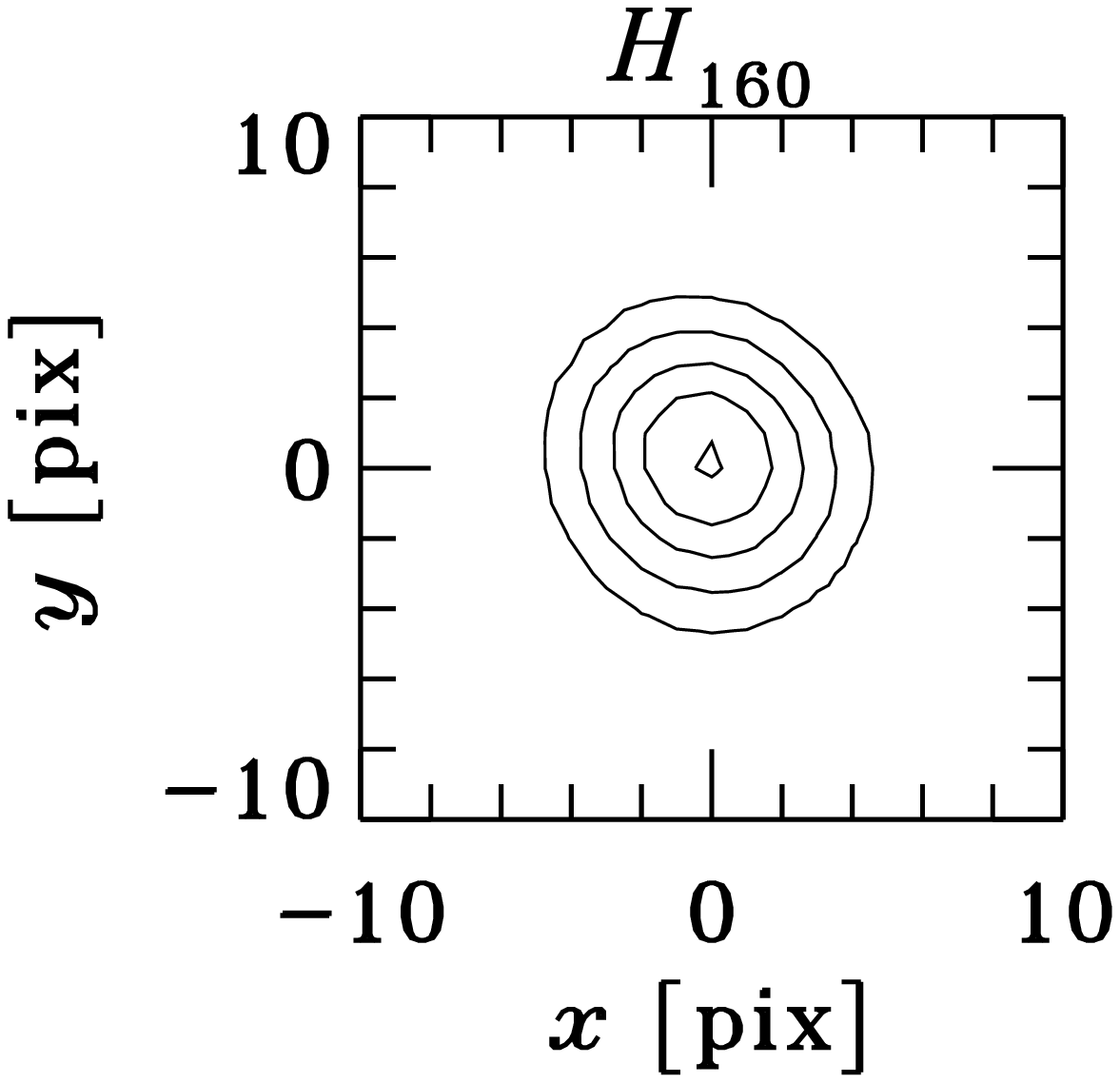}
 \caption[]
{
The contours of PSF images in $J_{125} + J_{140}$ (left), 
$J_{140} + H_{160}$ (middle), 
and $H_{160}$ (right).
The half-light radii of the PSFs are 
0.119 arcsec (3.97 pixels) in $J_{125} + J_{140}$, 
0.124 arcsec (4.14 pixels) in $J_{140} + H_{160}$. 
and 0.123 arcsec (4.12 pixels) in $H_{160}$. 
}
\label{fig:psf_images}
\end{center}
\end{figure*}
%FFFFFFFFFFFFFFFFFFFFFFFFFFFFFFFFFFFFFFFFFFFFFFFFFFFFFFFFFFFFFFFF%

%FFFFFFFFFFFFFFFFFFFFFFFFFFFFFFFFFFFFFFFFFFFFFFFFFFFFFFFFFFFFFFFF%
\begin{figure*}
   \includegraphics[scale=0.5]{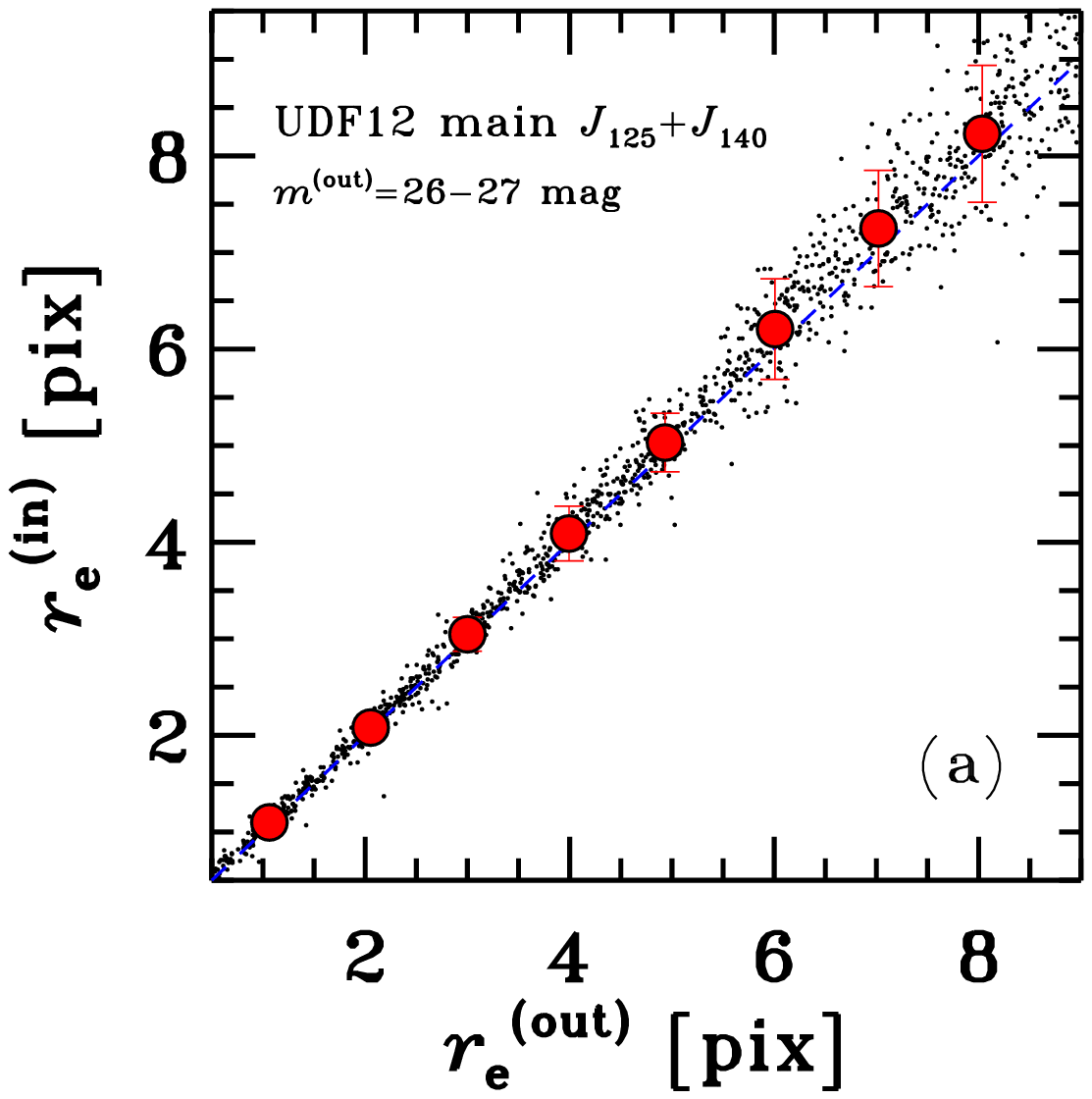}
   \includegraphics[scale=0.5]{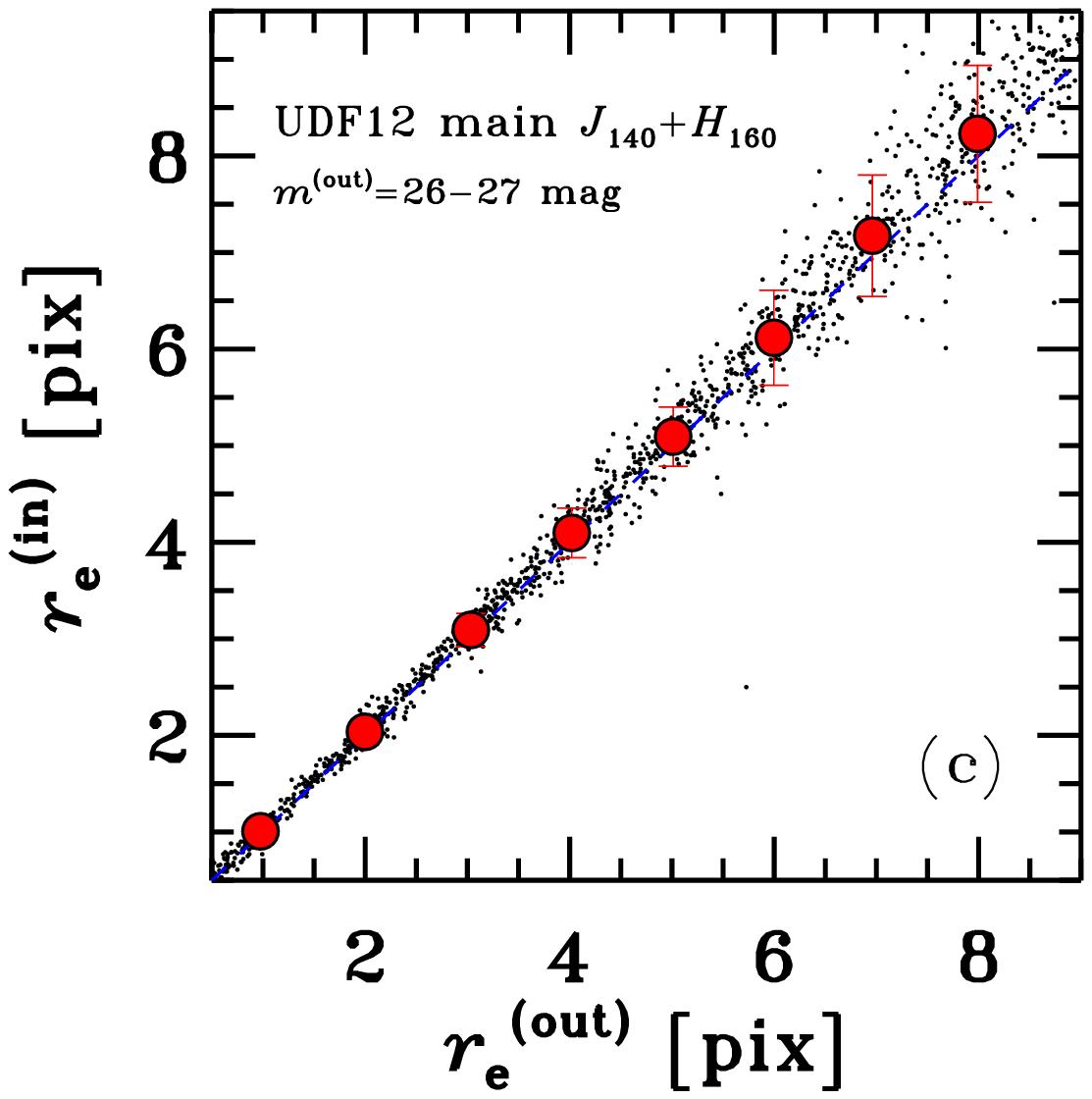}
   \includegraphics[scale=0.5]{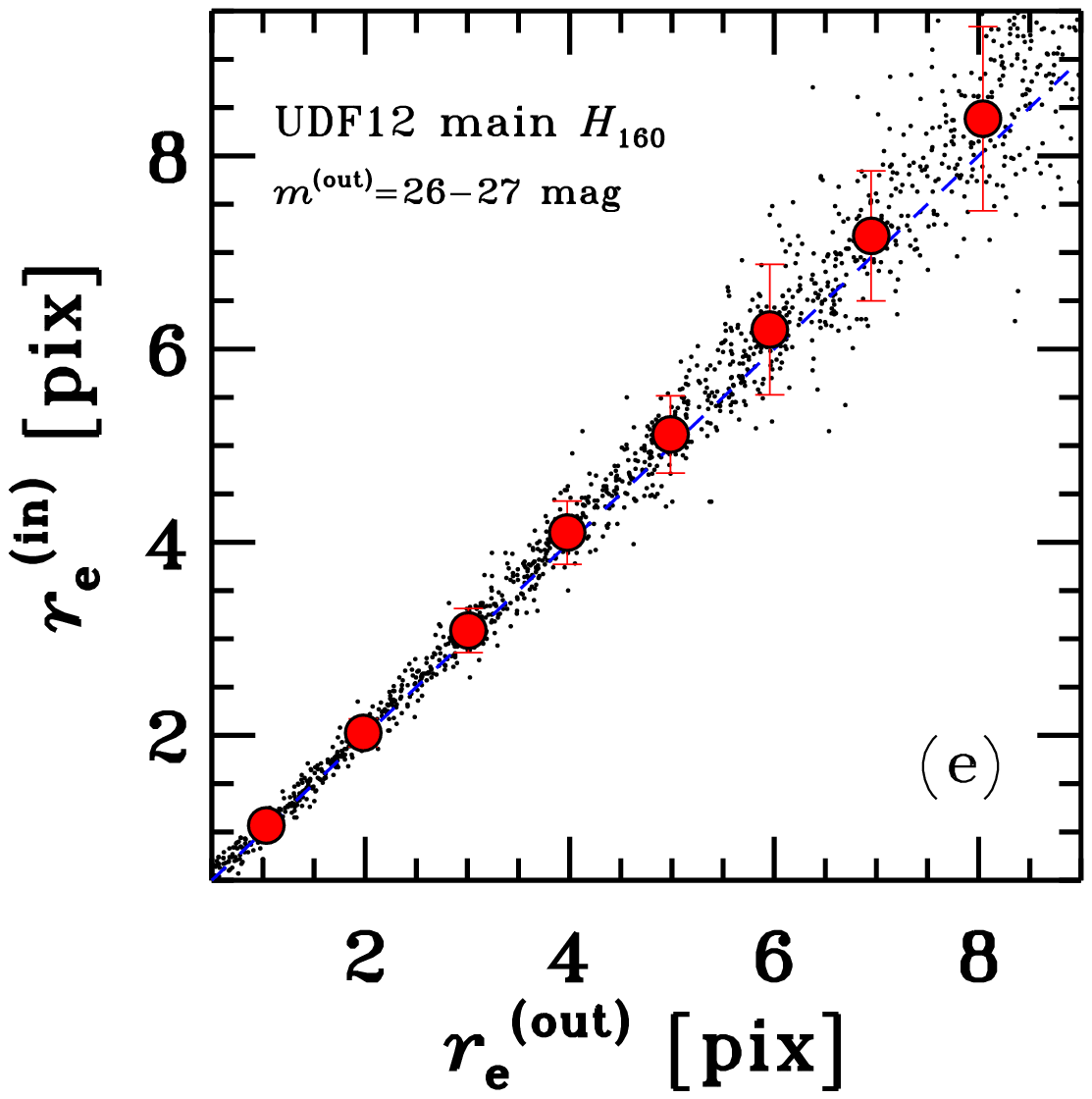} \\
   \includegraphics[scale=0.5]{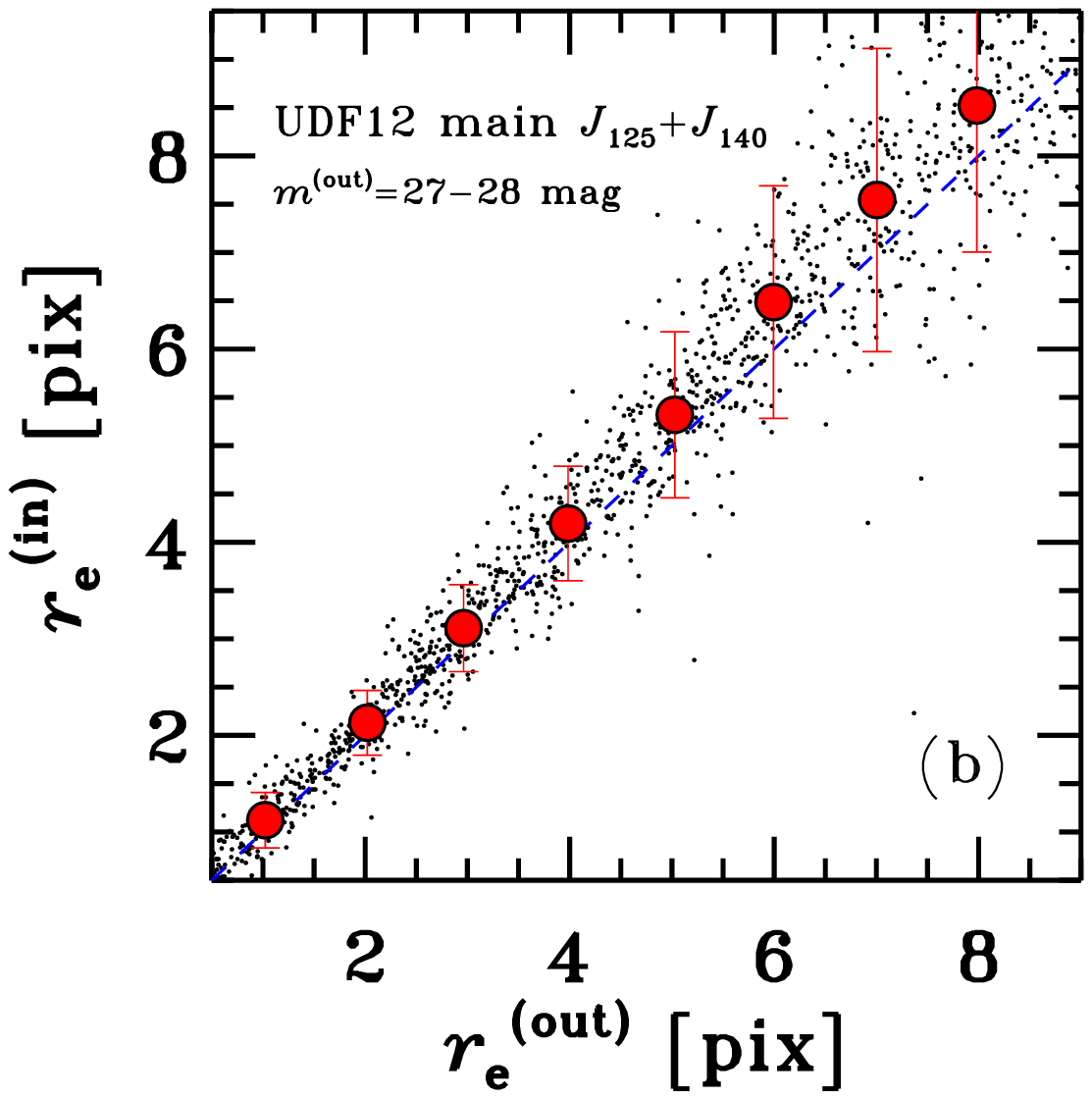}
   \includegraphics[scale=0.5]{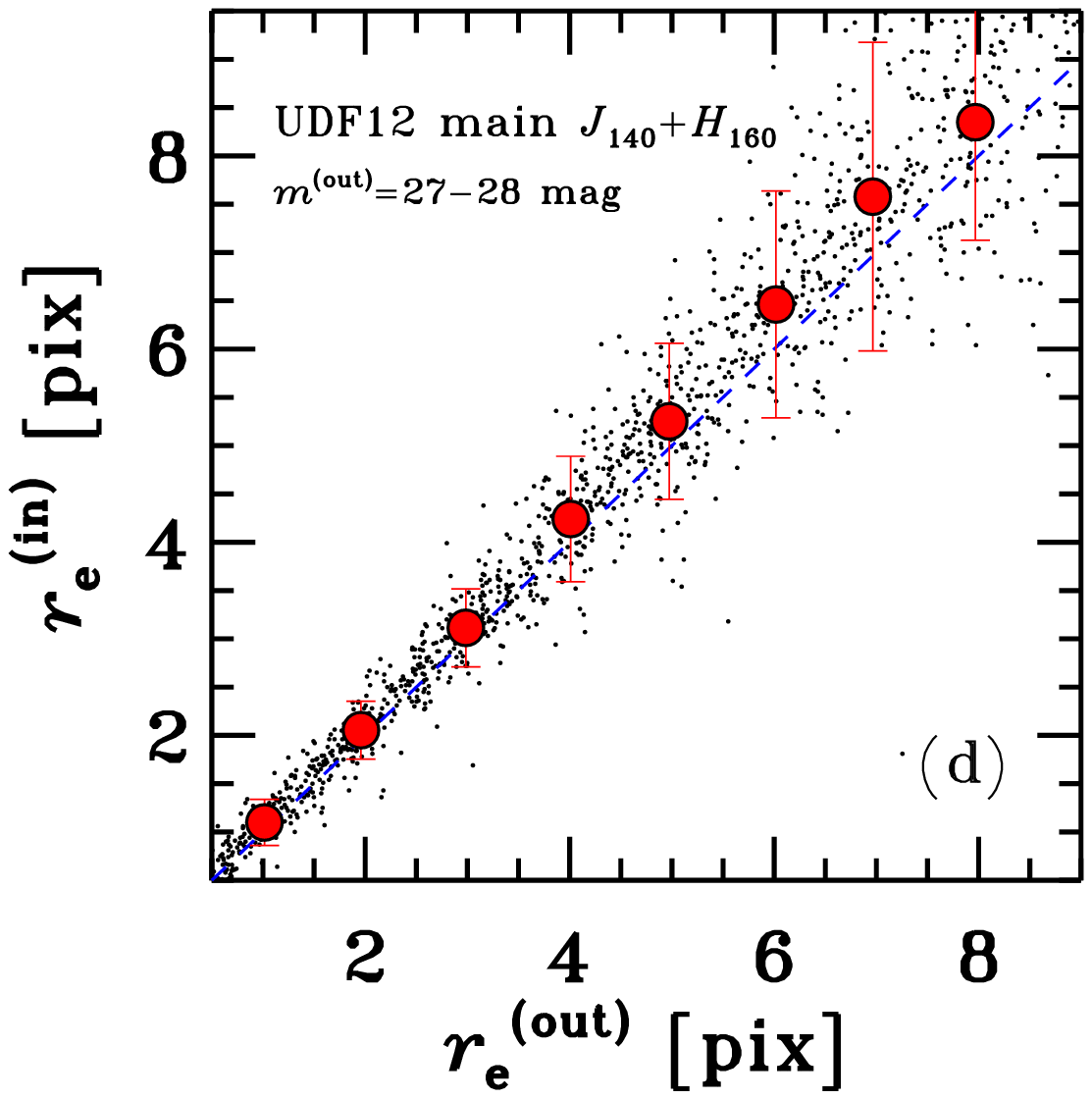}
   \includegraphics[scale=0.5]{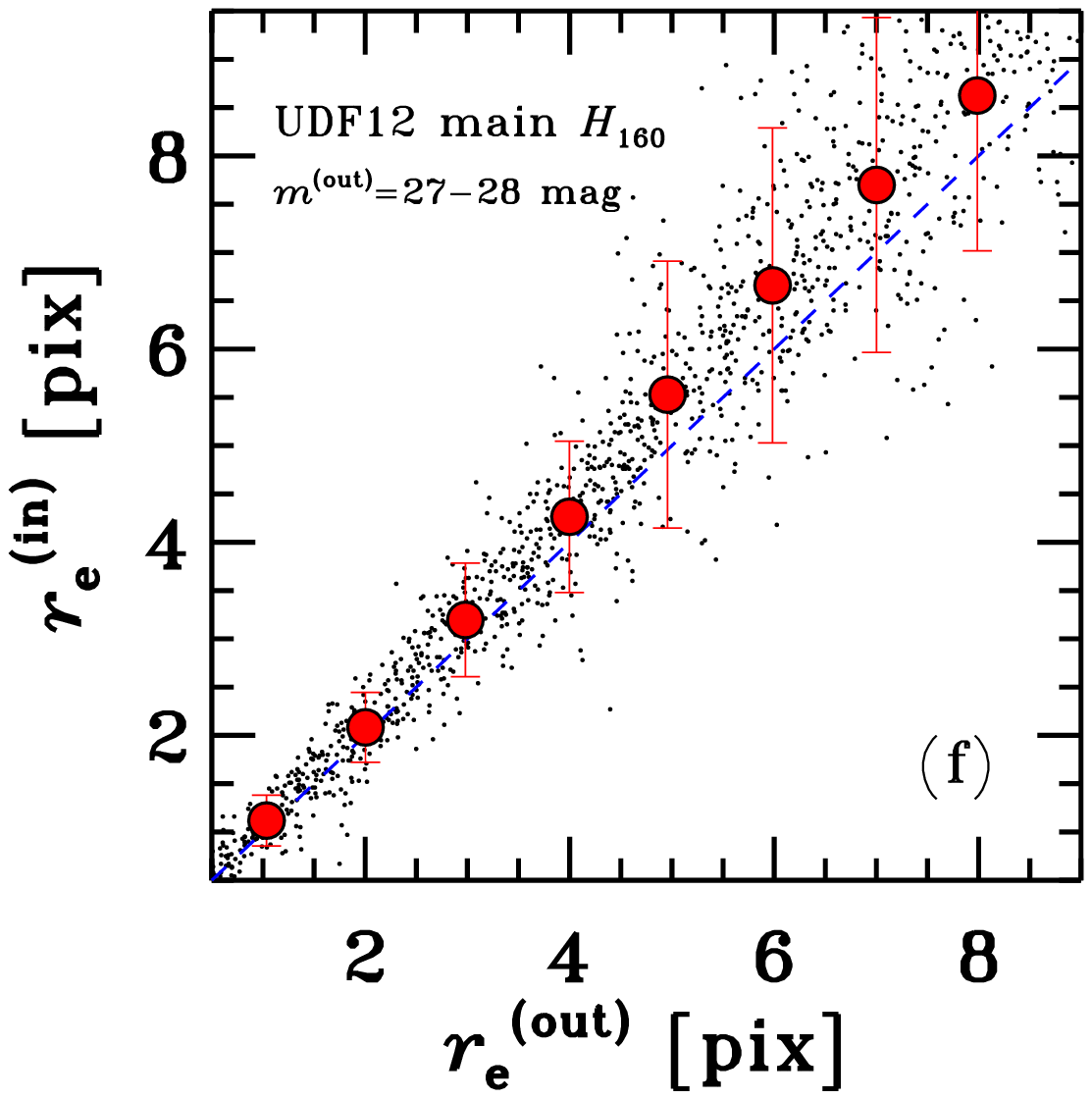}
 \caption[]
{
Panels (a)$-$(b), (c)$-$(d), and (e)$-$(f) show the results of our simulations 
for $z_{850}$-dropouts, $Y_{105}$-dropouts and $z>8.5$ candidates, respectively. 
These figures show input half-light radius $r_e^{\rm (in)}$
versus output radius $r_e^{\rm (out)}$ 
for a range of output magnitudes, 
$m^{\rm (out)} = 26-27$ (a, c, e) and $27-28$ mag (b, d, f). 
The red filled circles and the red error bars denote 
the average value and the relevant rms.
The blue dashed line shows the relation of $r_e^{\rm (in)}=r_e^{\rm (out)}$. 
}
\label{fig:simulations_re}
\end{figure*}
%FFFFFFFFFFFFFFFFFFFFFFFFFFFFFFFFFFFFFFFFFFFFFFFFFFFFFFFFFFFFFFFF%

%FFFFFFFFFFFFFFFFFFFFFFFFFFFFFFFFFFFFFFFFFFFFFFFFFFFFFFFFFFFFFFFF%
\begin{figure*}
   \includegraphics[scale=0.5]{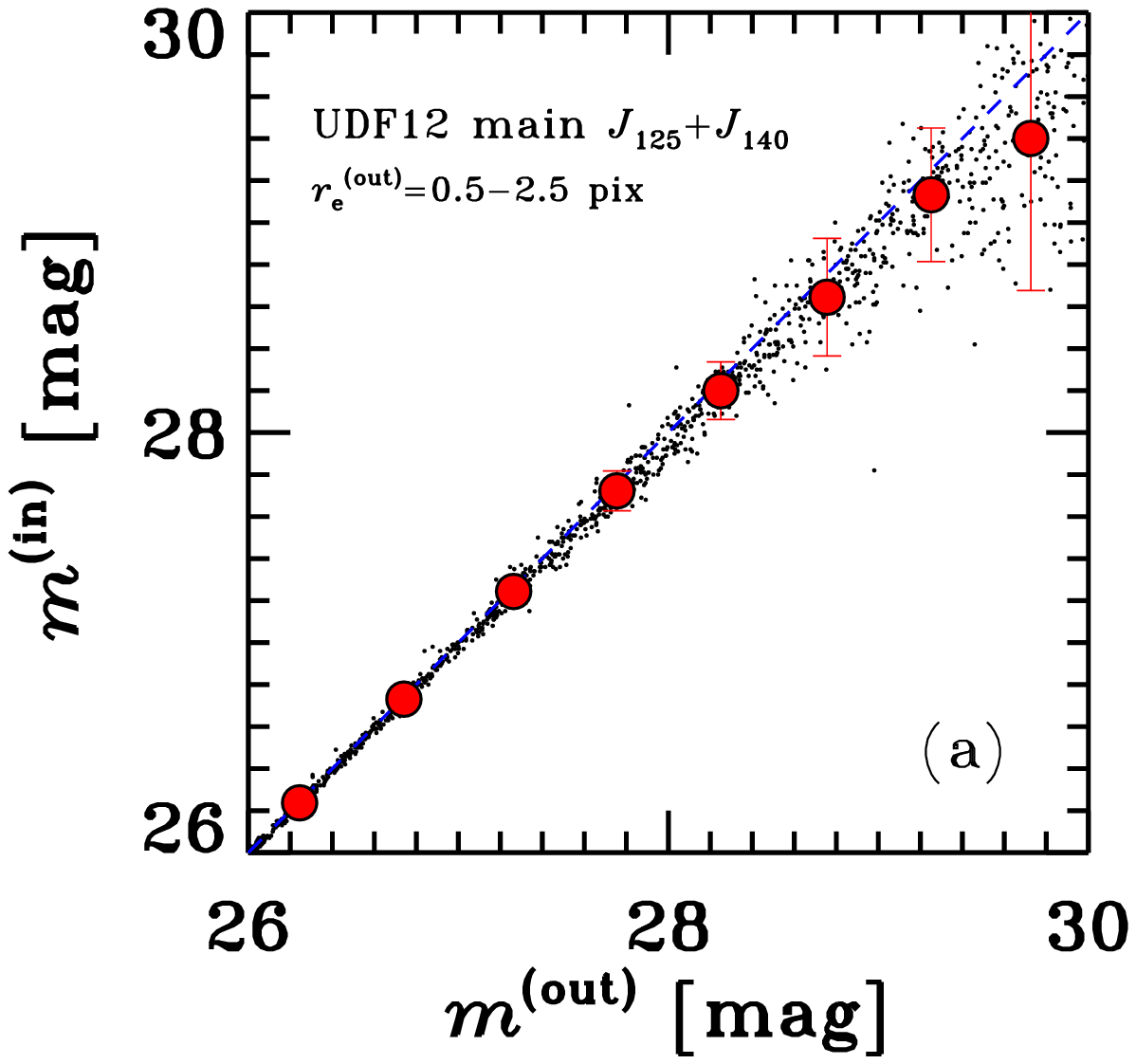} 
   \includegraphics[scale=0.5]{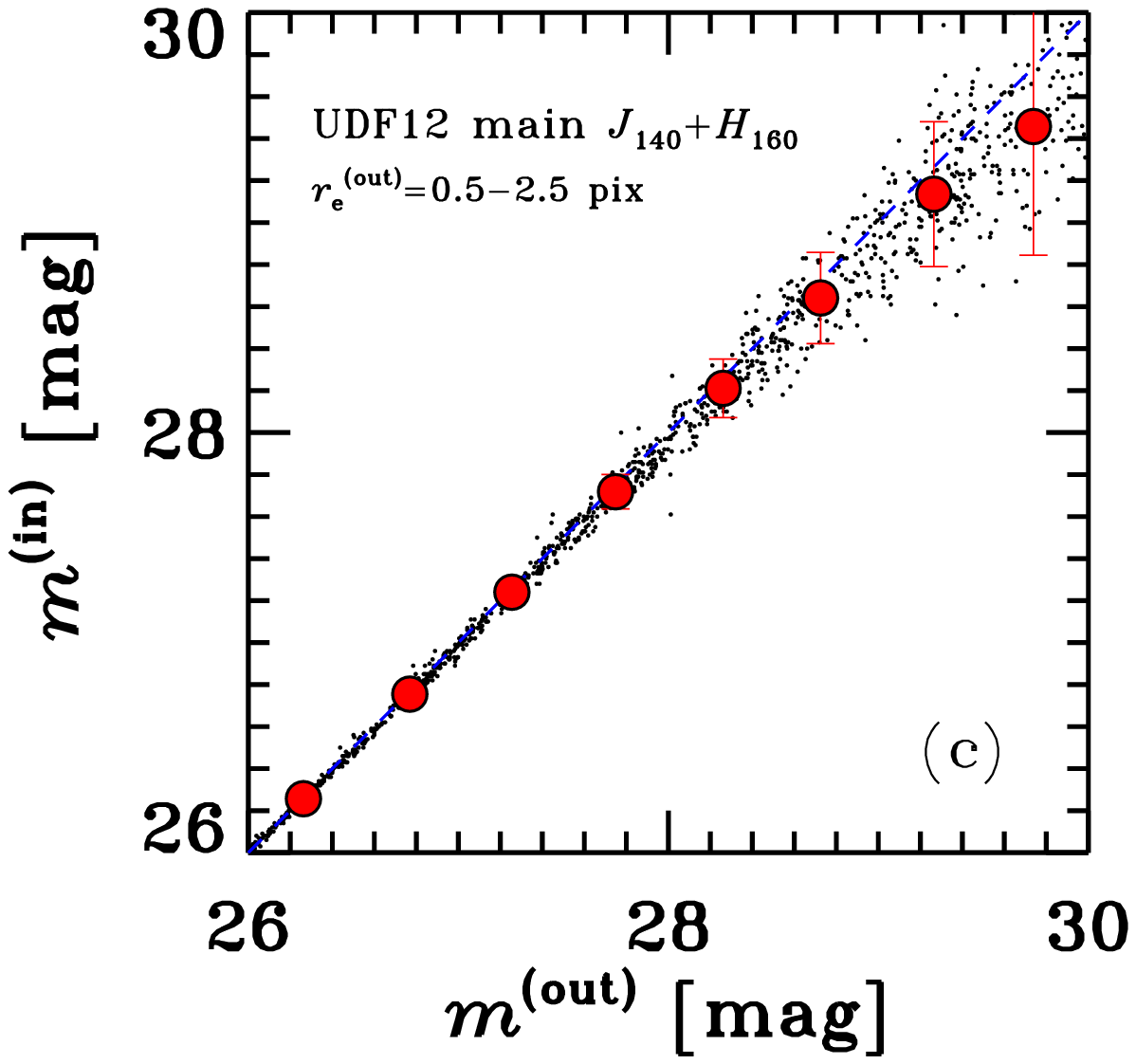}
   \includegraphics[scale=0.5]{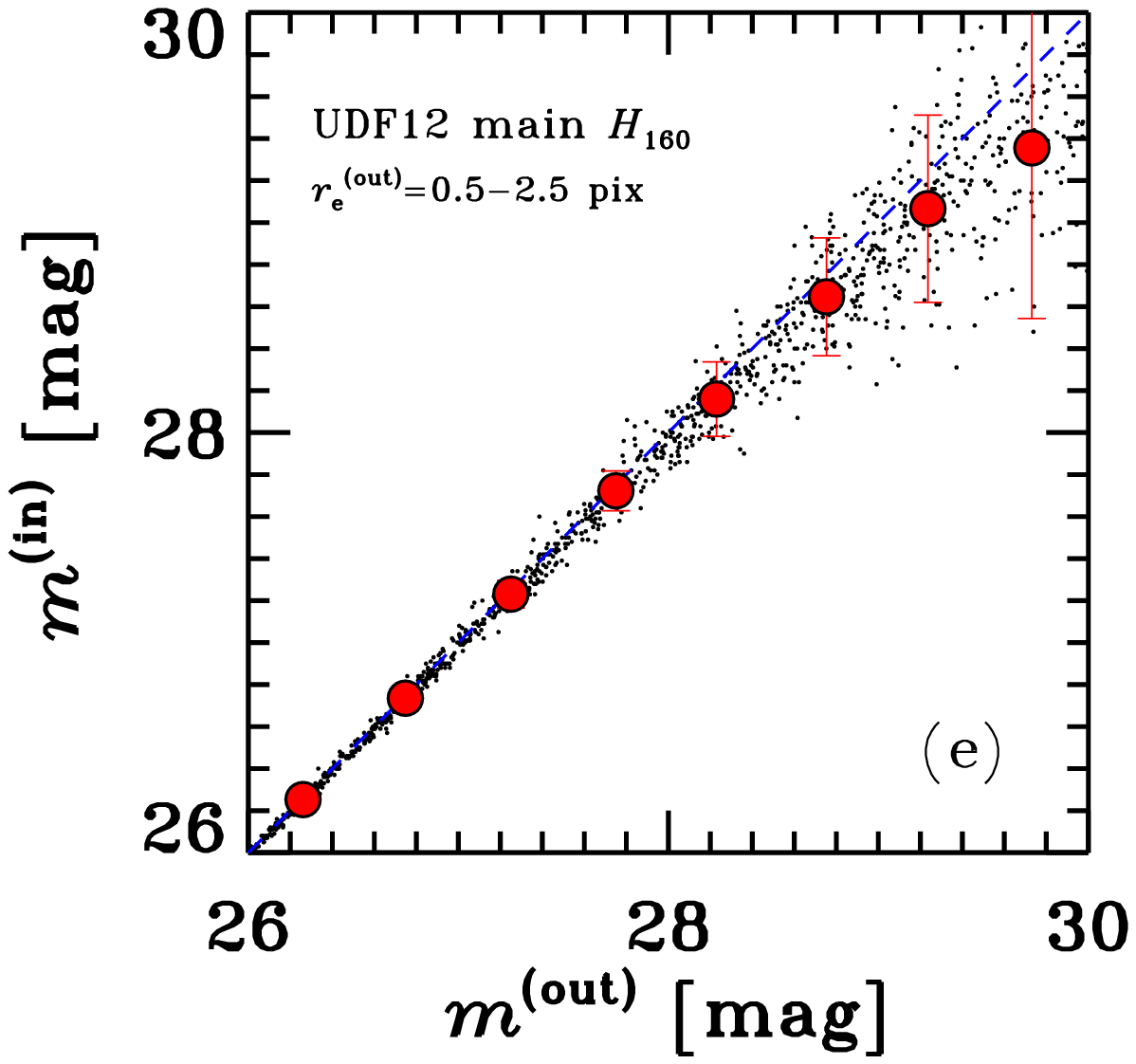} \\
   \includegraphics[scale=0.5]{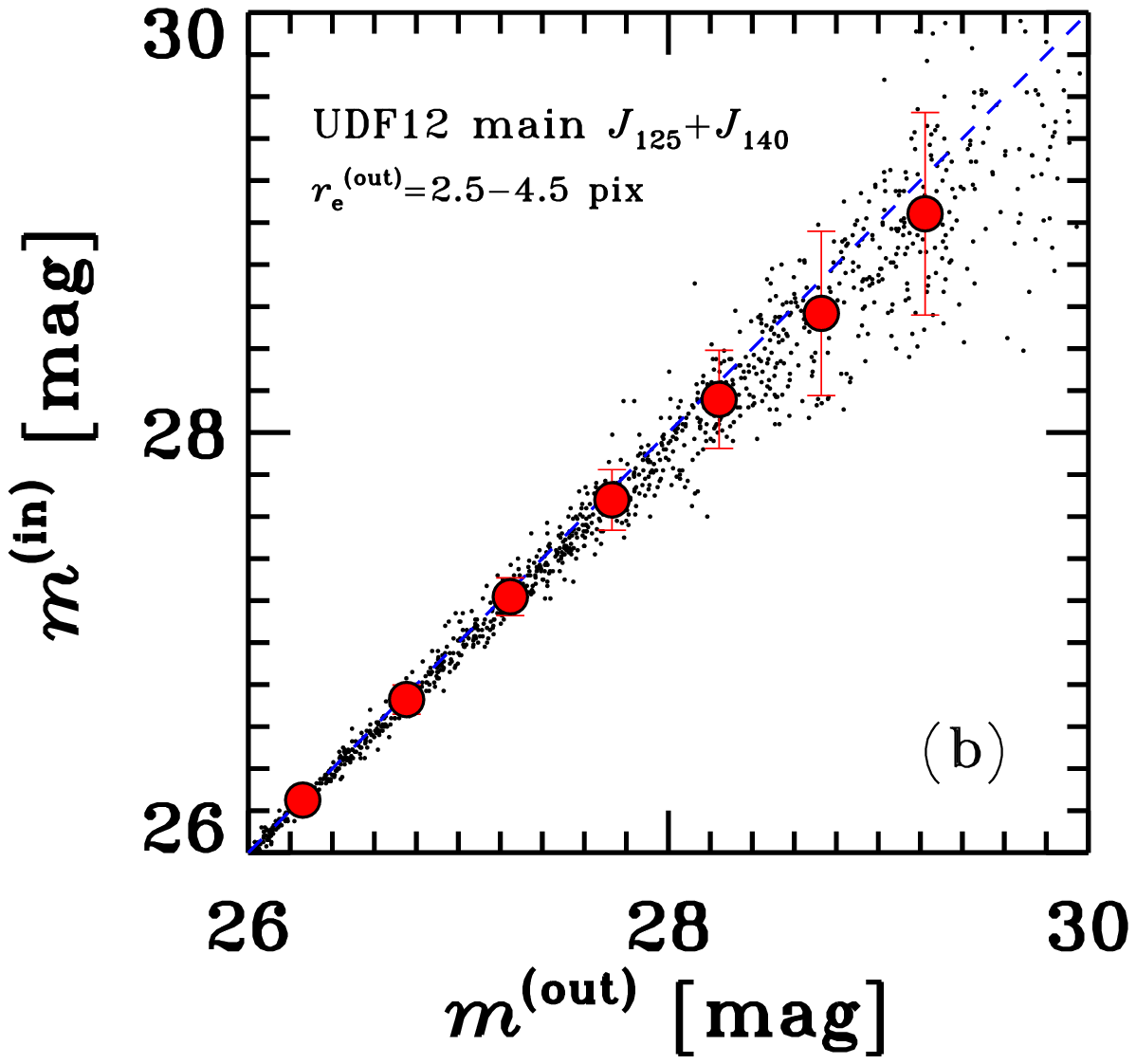} 
   \includegraphics[scale=0.5]{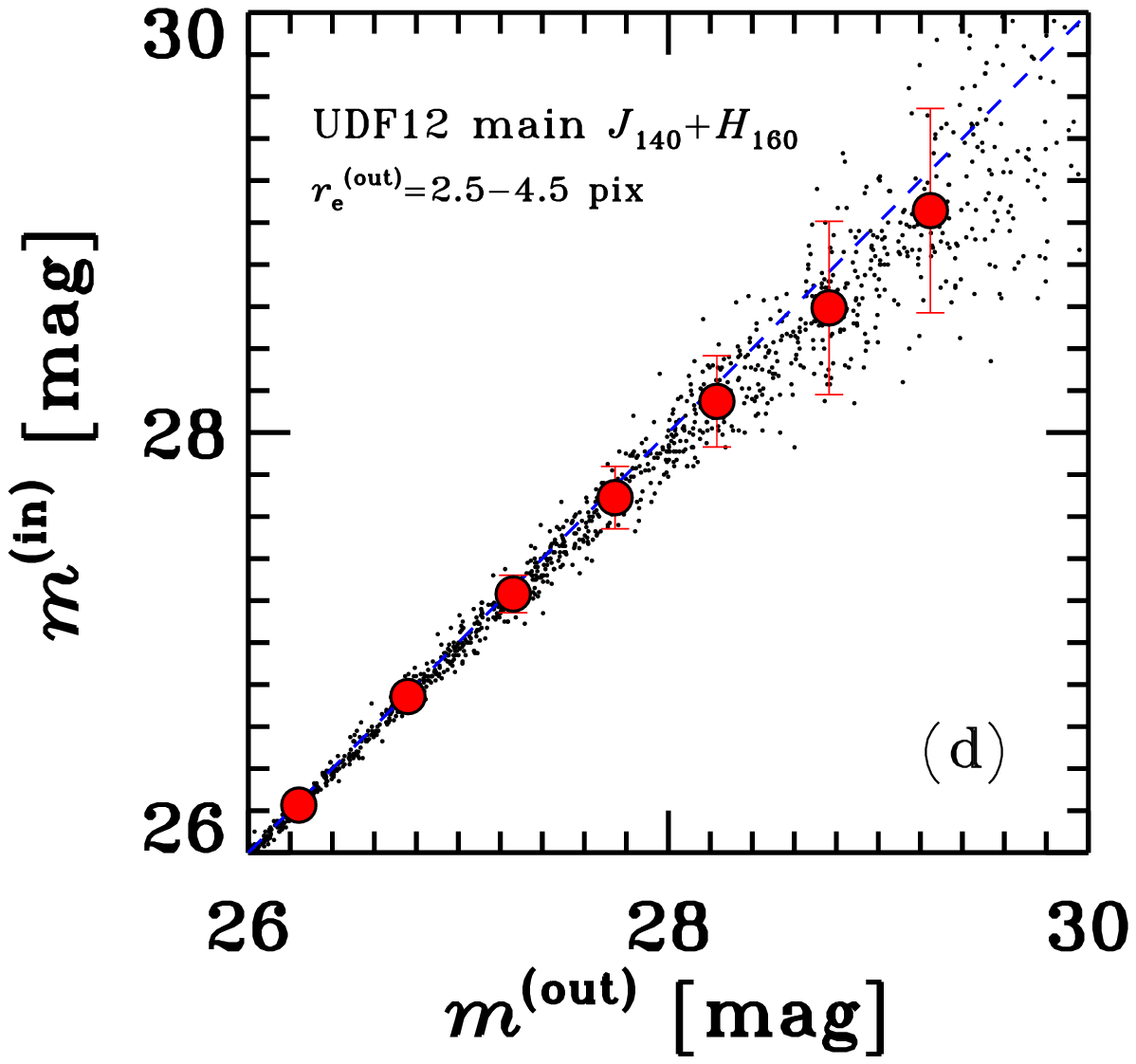}
   \includegraphics[scale=0.5]{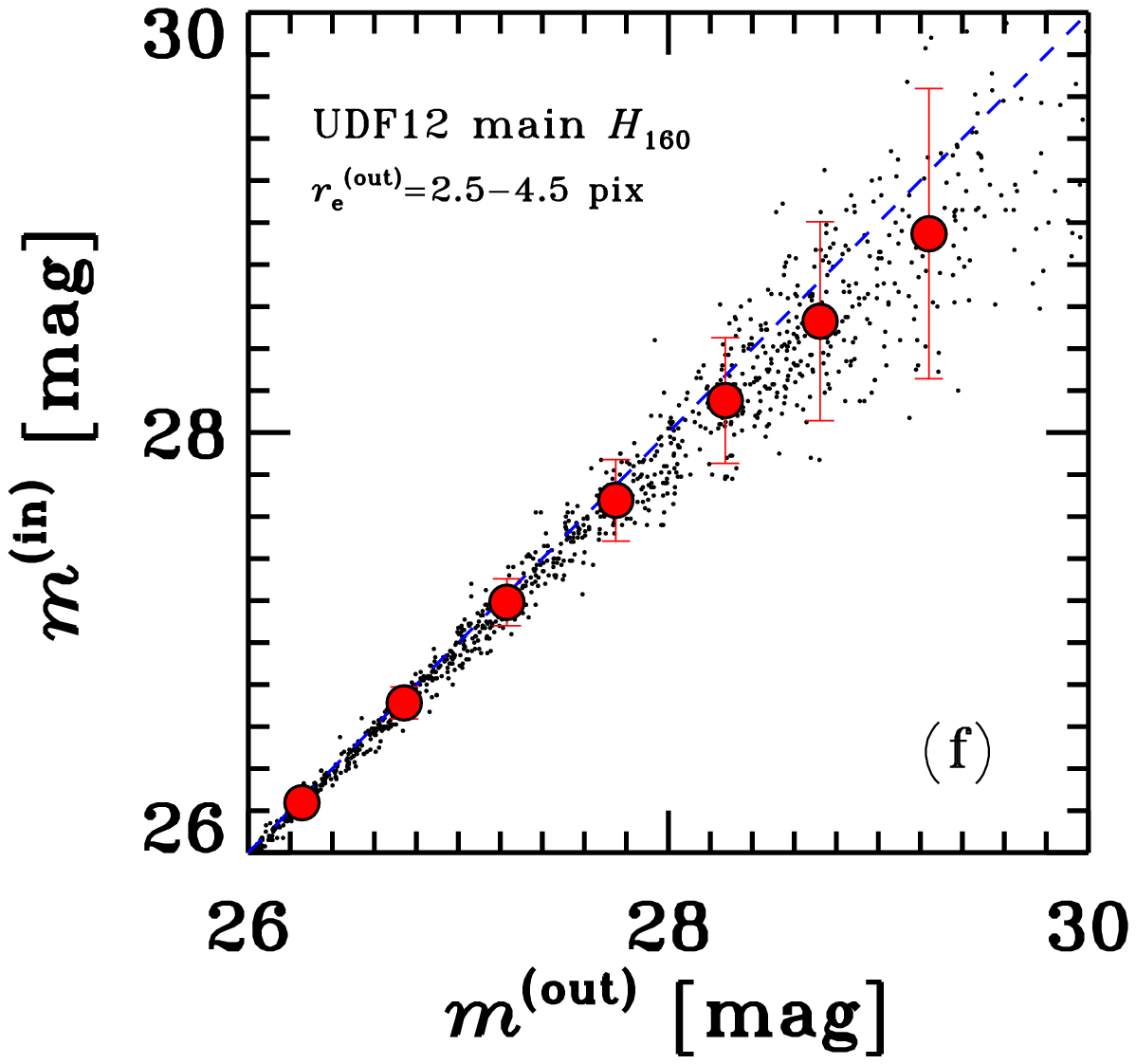}
 \caption[]
{
Panels (a)$-$(b),  (c)$-$(d), (e)$-$(f) show the results of our simulations 
for $z_{850}$-dropouts, $Y_{105}$-dropouts, and $z>8.5$ candidates, respectively.  
These figures show input magnitude $m^{\rm (in)}$
versus output magnitude $m^{\rm (out)}$ 
for a range of output half-light radii, 
$r_e^{\rm (out)} = 1-3$ (a, c, e) and $3-5$ pixels (b, d, f). 
The red filled circles and the red error bars denote 
the average value and the relevant rms.
The blue dashed line shows the relation of $m^{\rm (in)}=m^{\rm (out)}$. 
}
\label{fig:simulations_mag}
\end{figure*}
%FFFFFFFFFFFFFFFFFFFFFFFFFFFFFFFFFFFFFFFFFFFFFFFFFFFFFFFFFFFFFFFF%

%%%%%%%%%%%%%%%%%%%%%%%%%%%%%%%%%%%%%%%%%%%%%%%%%%%%%%%%%%%%%%%%%
%%%%%%%%%%%%%%%%%%%%%%%%%%%%%%%%%%%%%%%%%%%%%%%%%%%%%%%%%%%%%%%%%
\section{Sizes of Galaxies at $\lowercase{z} \sim 7-12$} \label{sec:morphologies_of_galaxies}
%%%%%%%%%%%%%%%%%%%%%%%%%%%%%%%%%%%%%%%%%%%%%%%%%%%%%%%%%%%%%%%%%
%%%%%%%%%%%%%%%%%%%%%%%%%%%%%%%%%%%%%%%%%%%%%%%%%%%%%%%%%%%%%%%%%

The S\'ersic power law \citep{sersic1968} is one of the most frequently used profiles 
to study galaxy morphology and has the following form: 
\begin{equation}
\Sigma (r)
	= \Sigma_e \exp \left( - b_n \left[ \left( \frac{r}{r_e} \right)^{1/n} -1 \right] \right), 
\end{equation}
where 
$\Sigma_e$ is the surface brightness at the half-light radius $r_e$, 
$n$ is the S\'ersic index, which is often referred to as the concentration parameter; 
larger $n$ values denote steeper inner profiles and highly extended outer wings. 
The parameter $r_e$ is the half-light radius, which holds half of the total flux inside. 
To make this definition true, the variable $b_n$ depends on $n$. 
We fit the two-dimensional surface brightness profile using 
the \verb|GALFIT| software version 3 \citep{peng2002,peng2010}, 
which convolves a galaxy model profile image with a PSF profile 
and optimizes the fits using Levenberg-Marquardt algorithm for $\chi^2$ minimization. 
The output parameters include 
the centroid coordinates of the objects, 
their total magnitude, 
half-light radius, 
S\'ersic index $n$, 
axis ratio, 
and position angle. 
The half-light radius provided by \verb|GALFIT| is 
the radius along the semi-major axis, $a$. 
For each galaxy, we calculate the circularized half-light radius, 
$r_e = a \sqrt{b/a}$, where $b/a$ is the axis ratio. 
The initial parameters used for profile fitting are provided by \verb|SExtractor| \citep{bertin1996}, 
and all of the parameters, except for the S\'ersic index, $n$, 
are allowed to vary during the fitting procedure. 
The S\'ersic index $n$ is fixed at $1.0$, which corresponds to the exponential profile\footnote{Although 
\cite{oesch2010b} set the S\'ersic index $n$ to be $1.5$ during their analysis, 
we confirm that measured sizes show very little difference if we use $n=1.5$.}. 
In this case, $b_n = 1.678$, 
which is obtained by solving the following equation: 
$\gamma (2n, b_n) = \Gamma(2n) /2$, 
where $\gamma$ is the incomplete gamma function, 
and $\Gamma$ is the gamma function.  
Noise images, required to weight individual pixels in the fit,  
are taken to be the root mean square (rms) maps generated from variance maps 
provided by the data reduction. 
We also use segmentation images which are produced by \verb|SExtractor|, 
to mask objects other than the object we are interested in during the profile fitting.

%%%%%%%%%%%%%%%%%%%%%%%%%%%%%%%%%%%%%%%%%%%%%%%%%%%%%%%%%%%%%%%%%
%%%%%%%%%%%%%%%%%%%%%%%%%%%%%%%%%%%%%%%%%%%%%%%%%%%%%%%%%%%%%%%%%
\subsection{Simulations of Systematic Effects} \label{subsec:systematic_effects}
%%%%%%%%%%%%%%%%%%%%%%%%%%%%%%%%%%%%%%%%%%%%%%%%%%%%%%%%%%%%%%%%%
%%%%%%%%%%%%%%%%%%%%%%%%%%%%%%%%%%%%%%%%%%%%%%%%%%%%%%%%%%%%%%%%%

Low surface brightness in the outskirts of a galaxy may not be correctly 
measured by \verb|GALFIT|, leading to systematically low measured 
half-light radii and/or total magnitudes.  
In order to quantify and correct for any such systematic effects, 
we use the following simulations.

First, we produce galaxy images 
whose S{\'e}rsic index $n$ is fixed at $1.0$, 
half-light radius $r_e$ is randomly chosen between $0.5$ and $10.5$ pixels, 
and total magnitude is randomly chosen between $26$ and $30$ mag. 
Note that axis ratios are fixed at $1$ during the simulations. 
This means that the systematic and statistical uncertainties will be 
larger than those obtained by our simulations, if output axis ratio is smaller than $1$\footnote{If 
axis ratios are parametrized, measurements of circularized radii 
are systematically underestimated by about $10$ {\%} and 
the statistical uncertainties in the measured circularized radii 
are about $40$ {\%} at S/N $= 15$ (Yuma et al. 2012 in preparation).}.  
Then we convolve them with a PSF image 
which is a composite of bright and unsaturated stellar objects 
in the HUDF \citep{pirzkal2005}. 
Figure \ref{fig:psf_images} shows 
the measured PSFs for the $J_{125}+J_{140}$, $J_{140}+H_{160}$, and $H_{160}$ images. 
The PSF-convolved galaxy images are inserted into empty regions 
of the original images 
before being analyzed in an identical manner to the true galaxy sample.

Figure \ref{fig:simulations_re} displays the results of size measurements 
of our simulated galaxies.
The panels show $r_e^{\rm (in)}$ vs. $r_e^{\rm (out)}$ for each image 
at two different magnitude ranges 
($26 <  m^{\rm (out)} < 27$ and $27 <  m^{\rm (out)} < 28$).  
We see that measurements for all images give low systematic offsets 
for objects with sizes smaller than $\sim 4$ pixels, 
although at larger sizes the profiles are progressively underestimated 
as the surface brightness of the objects decrease.  
The systematics are also seen to be larger for the fainter objects.
We also use these simulation results 
for estimating statistical errors in the measurements.

Figure \ref{fig:simulations_mag} shows the results for measured total magnitudes 
compared to input magnitude.  
This time the results are displayed in two size bins ($1 < r_e^{\rm (out)} < 3$ pixels, 
$3 < r_e^{\rm (out)} < 5$ pixels) for each image.  
The results for the smaller size bin show that the total measured magnitude is 
robust down to $\sim 28$ mag.  
For objects fainter than this the measured magnitude is systematically fainter 
than the intrinsic value, 
and the statistical errors increase.  
The trend is similar for both size bins but the results for larger objects show 
greater systematic offsets and statistical uncertainties.

%FFFFFFFFFFFFFFFFFFFFFFFFFFFFFFFFFFFFFFFFFFFFFFFFFFFFFFFFFFFFFFFF%
\begin{figure*}
\begin{center}
\includegraphics[scale=0.8]{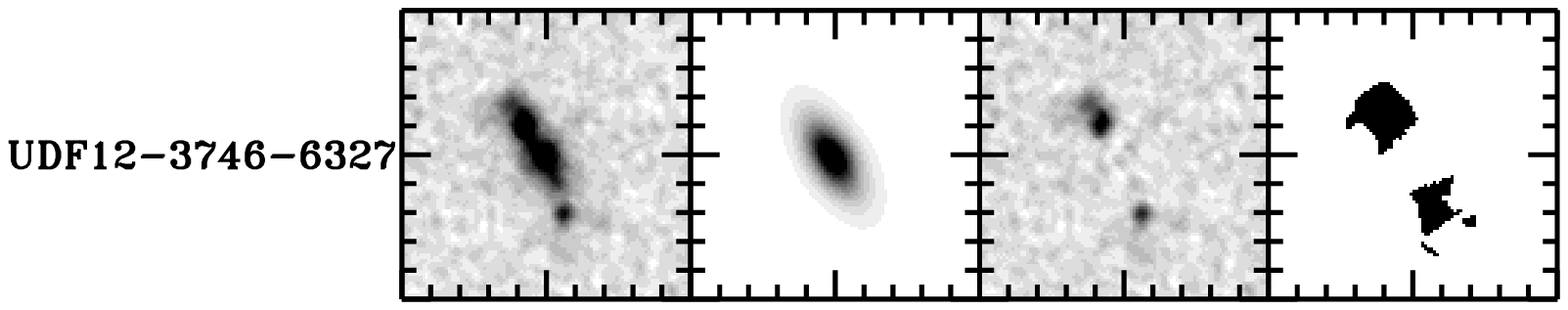} \\
\includegraphics[scale=0.8]{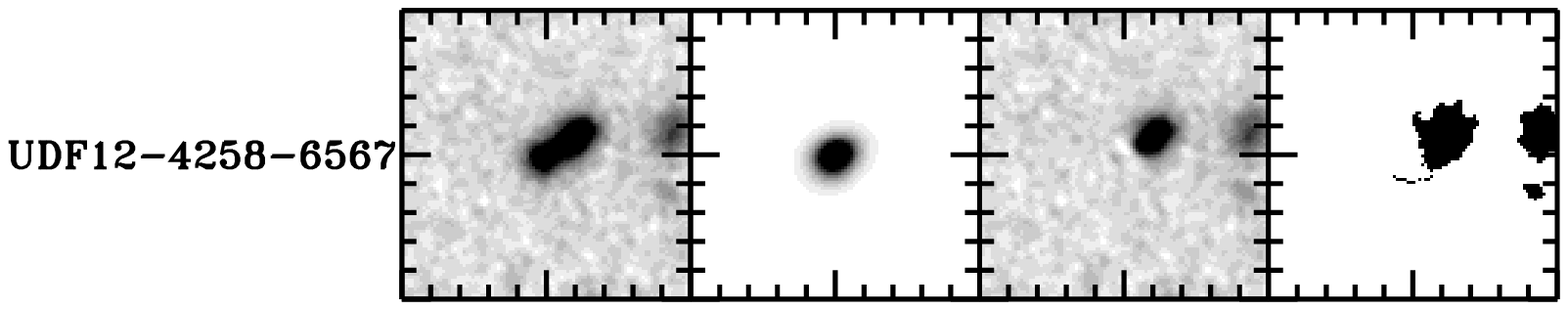} \\
\includegraphics[scale=0.8]{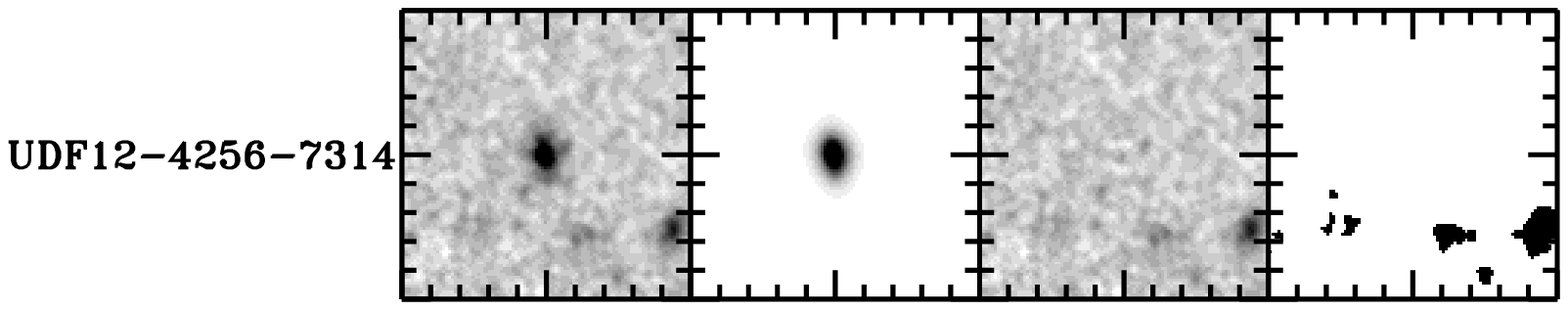} \\
\includegraphics[scale=0.8]{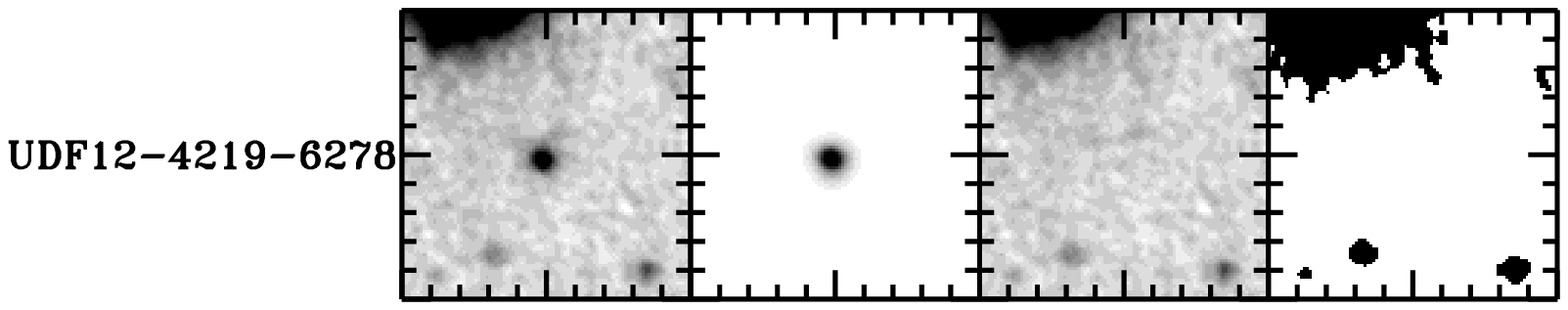} \\
\includegraphics[scale=0.8]{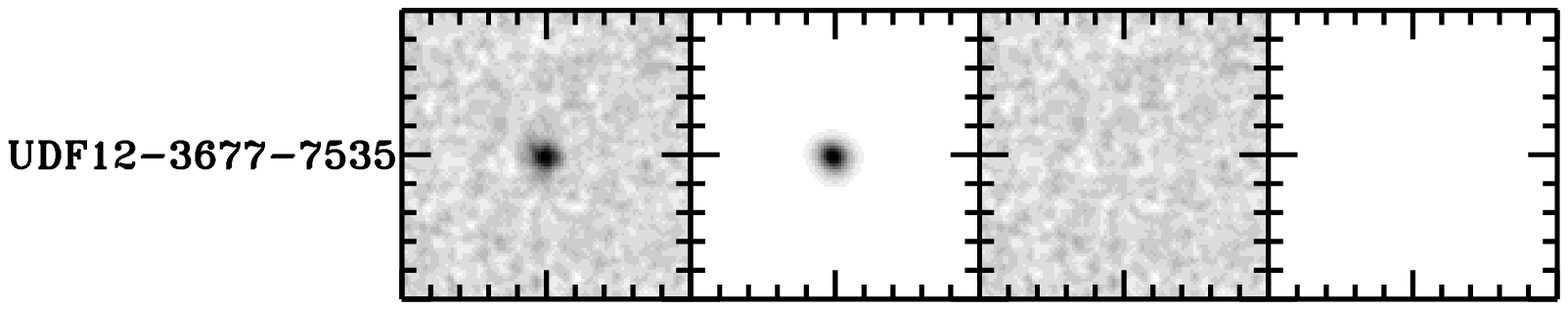} \\
\includegraphics[scale=0.8]{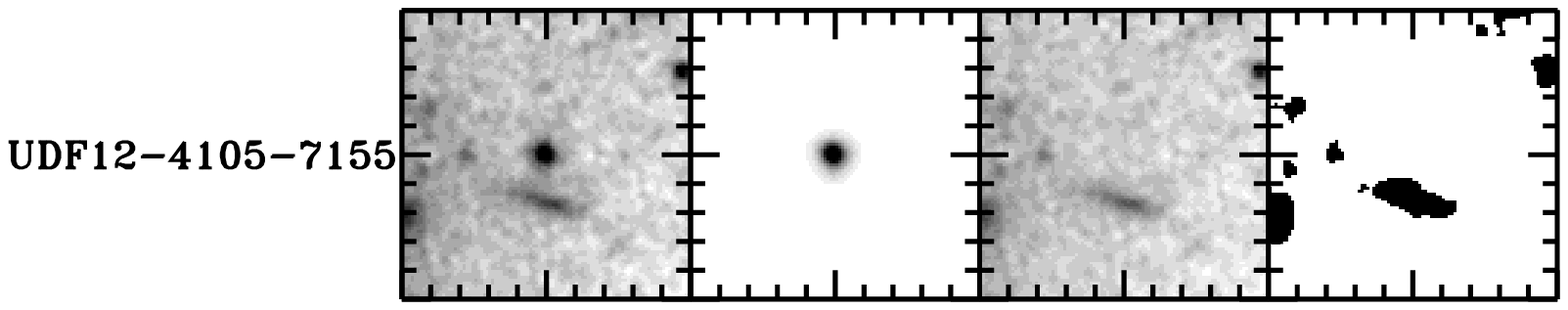} \\
\includegraphics[scale=0.8]{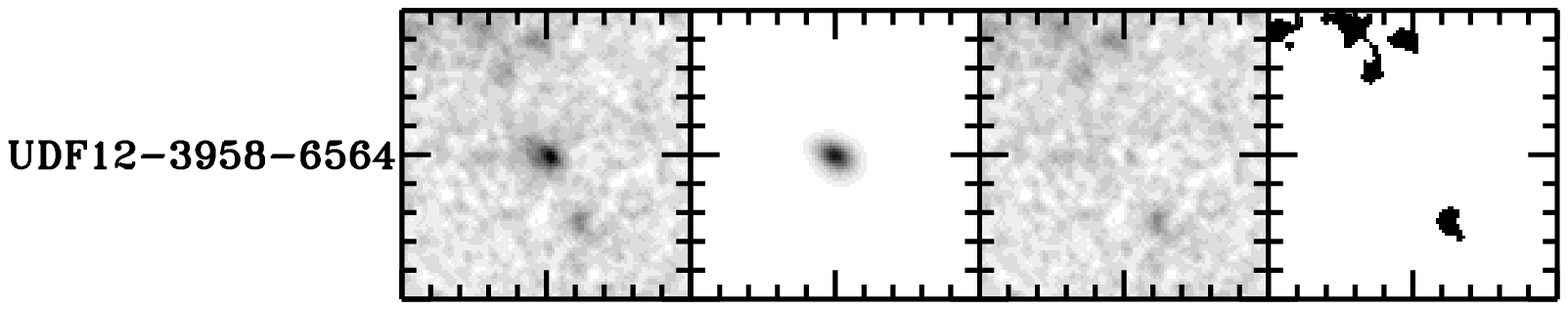} \\
\includegraphics[scale=0.8]{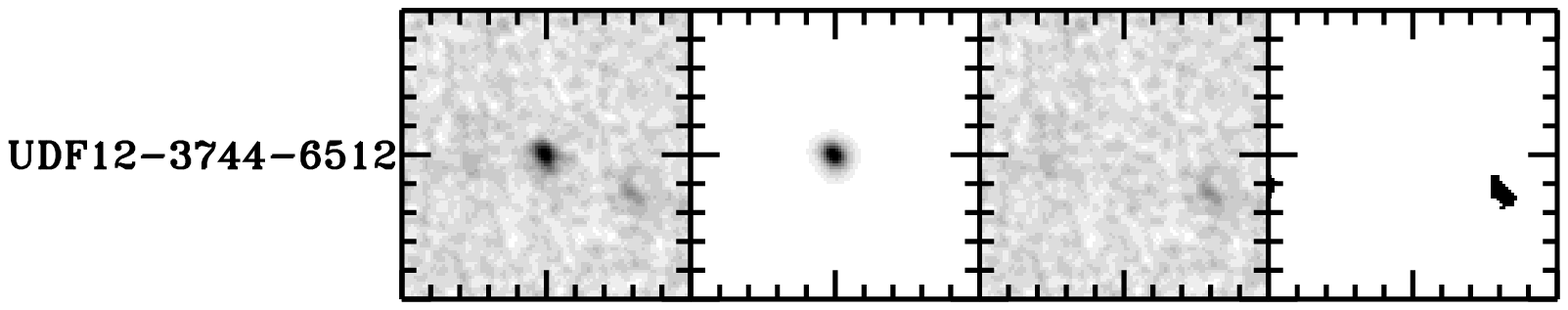} \\
\includegraphics[scale=0.8]{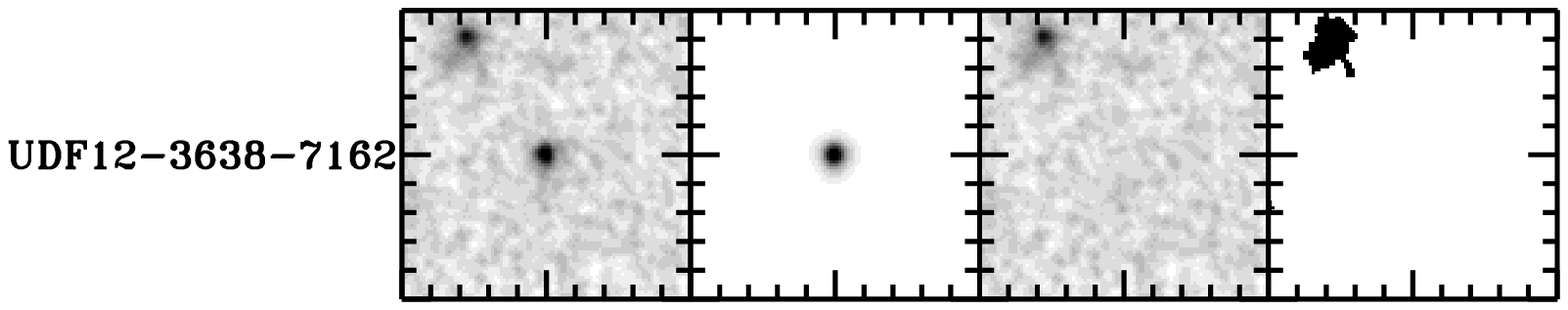} 
 \caption[]
{
S\'ersic profile fitting results for bright $z_{850}$-dropouts found in the HUDF main field. 
Shown, from left to right, are 
the $3'' \times 3''$ cut-outs of the original image, 
the best-fit model profile images, 
the residual images which are made by subtracting the best-fit images from the original ones, 
and 
the segmentation maps used for masking all the neighboring objects during the profile fitting. 
}
\label{fig:galfit_main_z1}
\end{center}
\end{figure*}
%FFFFFFFFFFFFFFFFFFFFFFFFFFFFFFFFFFFFFFFFFFFFFFFFFFFFFFFFFFFFFFFF%

%FFFFFFFFFFFFFFFFFFFFFFFFFFFFFFFFFFFFFFFFFFFFFFFFFFFFFFFFFFFFFFFF%
\begin{figure*}
\begin{center}
\includegraphics[scale=0.9]{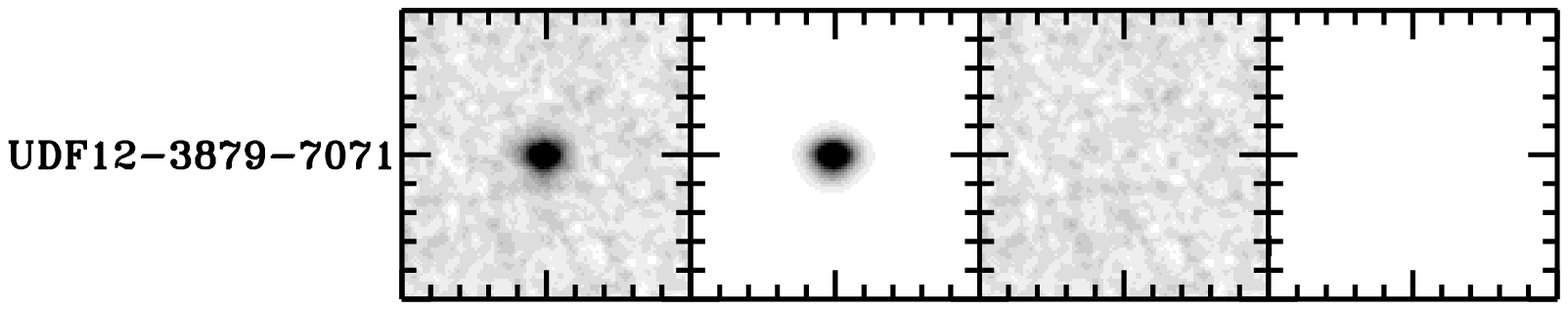} \\
\includegraphics[scale=0.9]{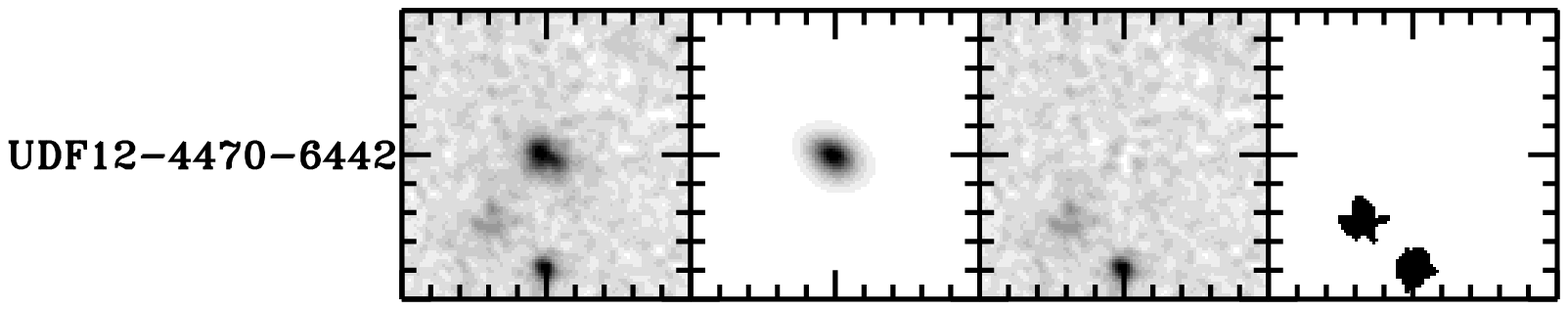} \\
\includegraphics[scale=0.9]{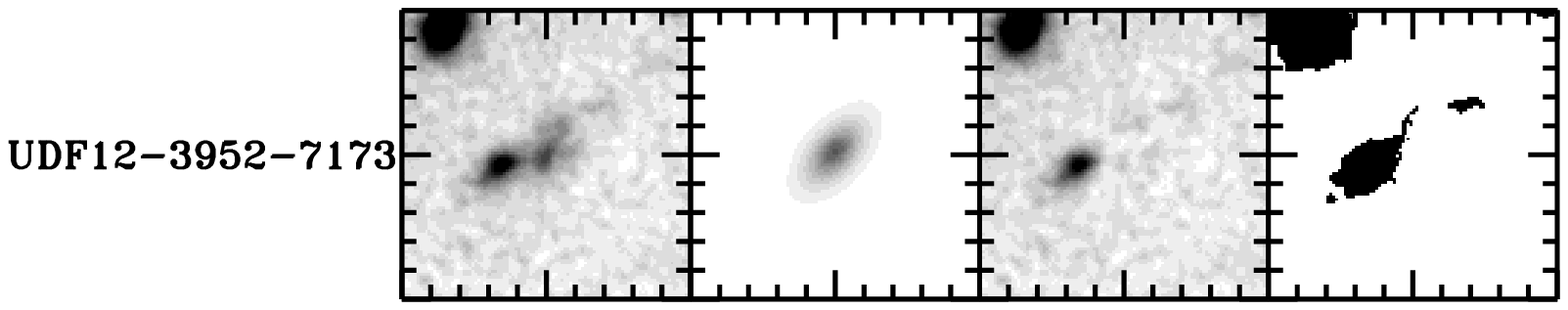} \\
\includegraphics[scale=0.9]{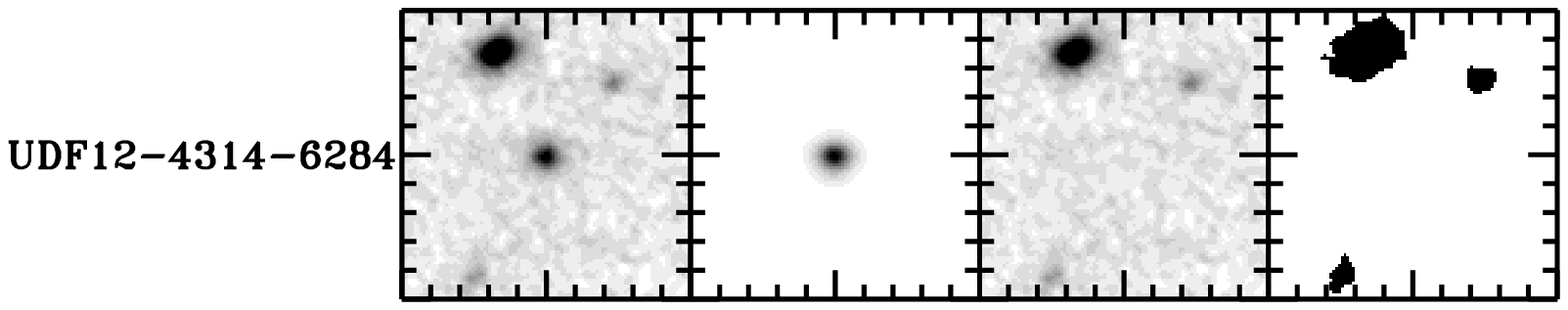} \\
\includegraphics[scale=0.9]{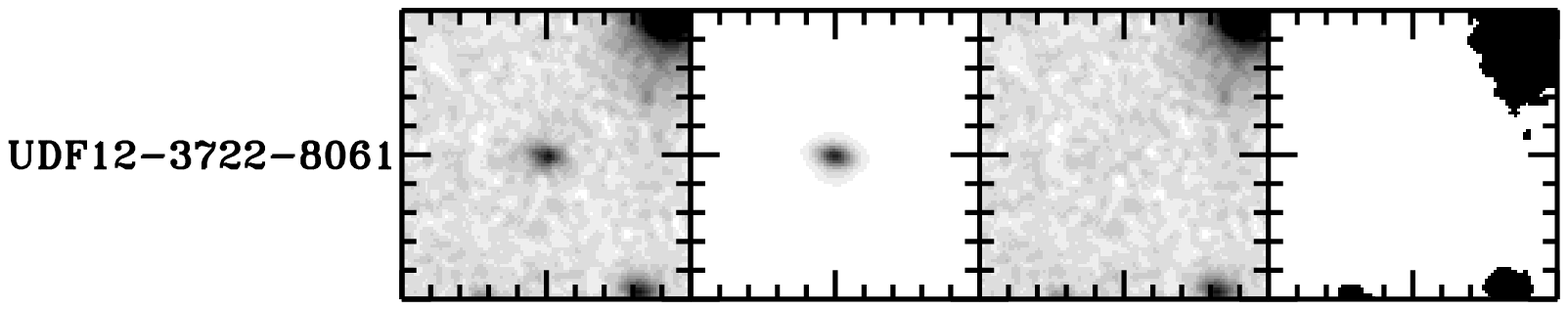} \\
\includegraphics[scale=0.9]{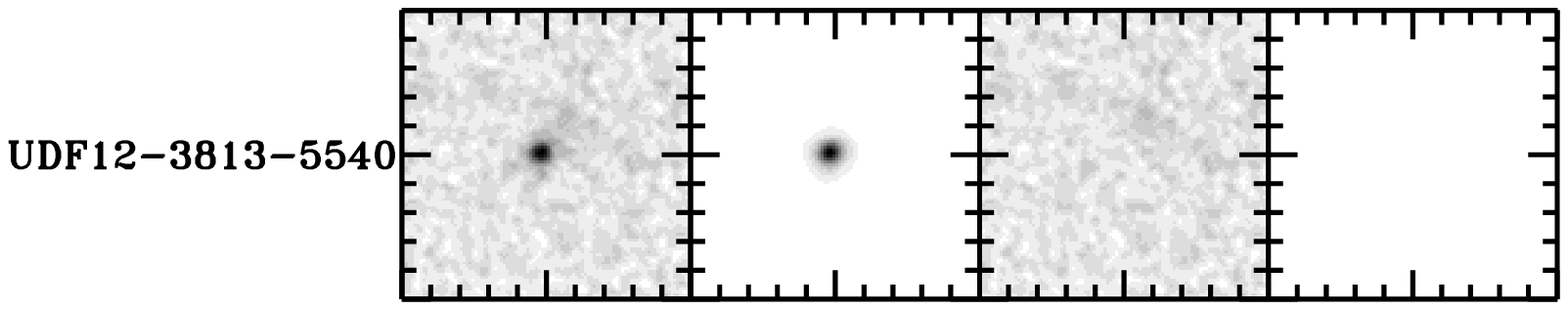} \\
 \caption[]
{
S\'ersic profile fitting results for bright $Y_{105}$-dropouts found in the HUDF main field. 
Shown, from left to right, are 
the $3'' \times 3''$ cut-outs of the original image, 
the best-fit model profile images, 
the residual images which are made by subtracting the best-fit images from the original ones, 
and 
the segmentation maps used for masking all the neighboring objects during the profile fitting. 
}
\label{fig:galfit_main_y1}
\end{center}
\end{figure*}
%FFFFFFFFFFFFFFFFFFFFFFFFFFFFFFFFFFFFFFFFFFFFFFFFFFFFFFFFFFFFFFFF%

%FFFFFFFFFFFFFFFFFFFFFFFFFFFFFFFFFFFFFFFFFFFFFFFFFFFFFFFFFFFFFFFF%
\begin{figure*}
\begin{center}
\includegraphics[scale=0.9]{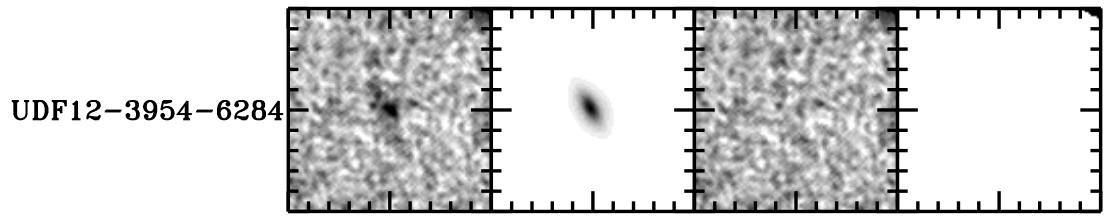} 
 \caption[]
{
S\'ersic profile fitting results for the $z \sim 12$ source, UDF12-3954-6284. 
Shown, from left to right, are 
the $3'' \times 3''$ cut-outs of the original image, 
the best-fit model profile images, 
the residual images which are made by subtracting the best-fit images from the original ones, 
and 
the segmentation maps used for masking all the neighboring objects during the profile fitting. 
}
\label{fig:galfit_main_hz1}
\end{center}
\end{figure*}
%FFFFFFFFFFFFFFFFFFFFFFFFFFFFFFFFFFFFFFFFFFFFFFFFFFFFFFFFFFFFFFFF%

%FFFFFFFFFFFFFFFFFFFFFFFFFFFFFFFFFFFFFFFFFFFFFFFFFFFFFFFFFFFFFFFF%
\begin{figure*}
\begin{center}
\includegraphics[scale=0.9]{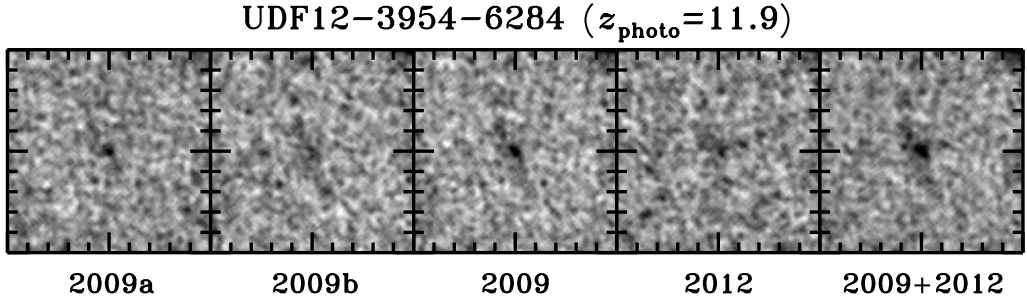} 
 \caption[]
{
$3'' \times 3''$ cut-outs of the $z \sim 12$ source UDF12-3954-6284 
from various subsets of the WFC3/IR $H_{160}$-band observations. 
From left to right, 
the first half of the 2009 dataset, 
the second half of the 2009 dataset, 
the full 53-orbit 2009 dataset, 
the 26-orbit 2012, 
and 
the full 84-orbit dataset (including 2009, 2012, and 
other exposures in this field). 
}
\label{fig:z12object}
\end{center}
\end{figure*}
%FFFFFFFFFFFFFFFFFFFFFFFFFFFFFFFFFFFFFFFFFFFFFFFFFFFFFFFFFFFFFFFF%

%FFFFFFFFFFFFFFFFFFFFFFFFFFFFFFFFFFFFFFFFFFFFFFFFFFFFFFFFFFFFFFFF%
\begin{figure*}
\begin{center}
\includegraphics[scale=0.9]{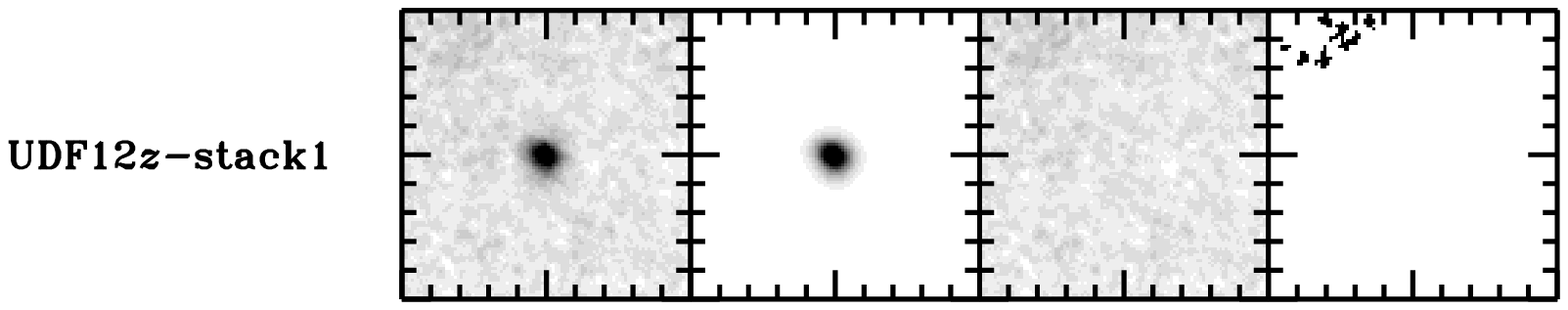} \\
\includegraphics[scale=0.9]{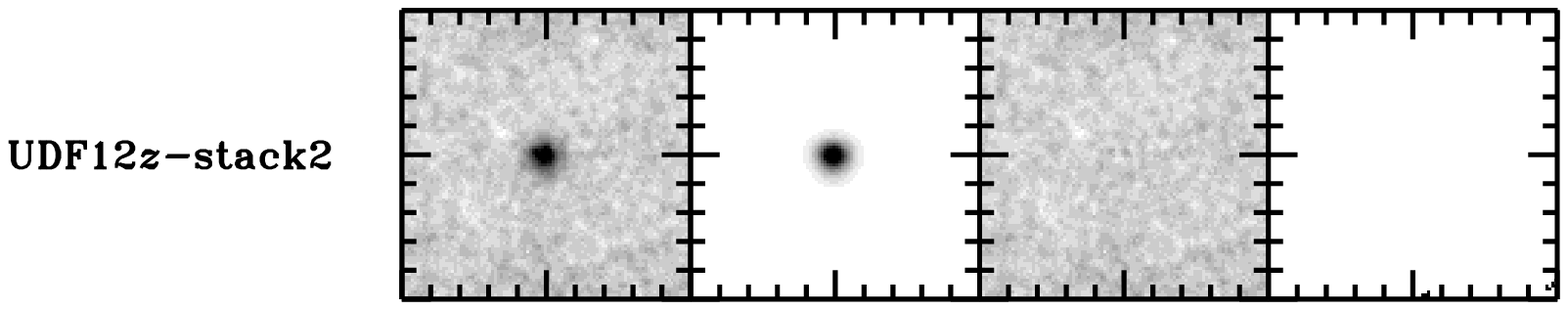}
 \caption[]
{
Same as Figure \ref{fig:galfit_main_z1}, 
except that the objects are the stacked $z_{850}$-dropouts, 
whose UV luminosities are 
$L = (0.12-0.3) L^\ast_{z=3}$ (top) and  
$L = (0.048-0.12) L^\ast_{z=3}$ (bottom). 
}
\label{fig:galfit_main_z_stack1}
\end{center}
\end{figure*}
%FFFFFFFFFFFFFFFFFFFFFFFFFFFFFFFFFFFFFFFFFFFFFFFFFFFFFFFFFFFFFFFF%

%FFFFFFFFFFFFFFFFFFFFFFFFFFFFFFFFFFFFFFFFFFFFFFFFFFFFFFFFFFFFFFFF%
\begin{figure*}
\begin{center}
\includegraphics[scale=0.9]{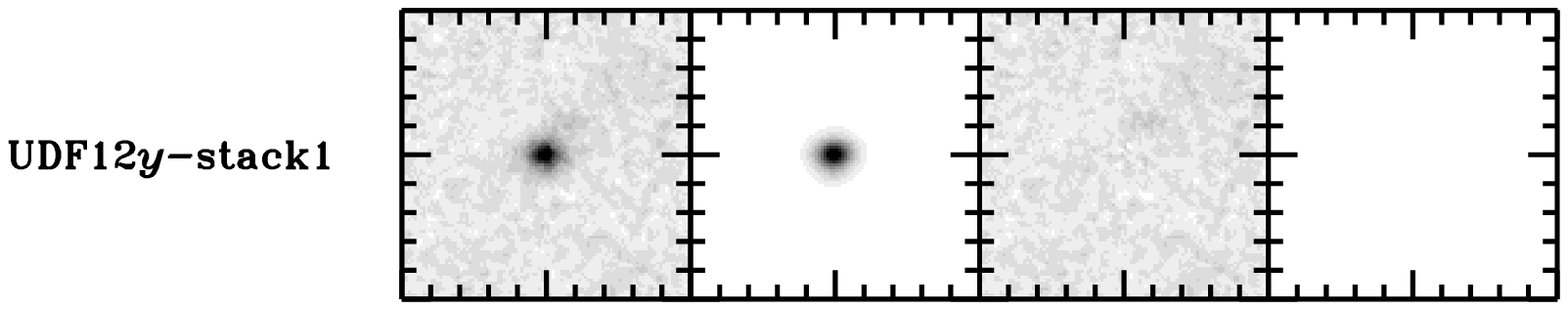} \\
\includegraphics[scale=0.9]{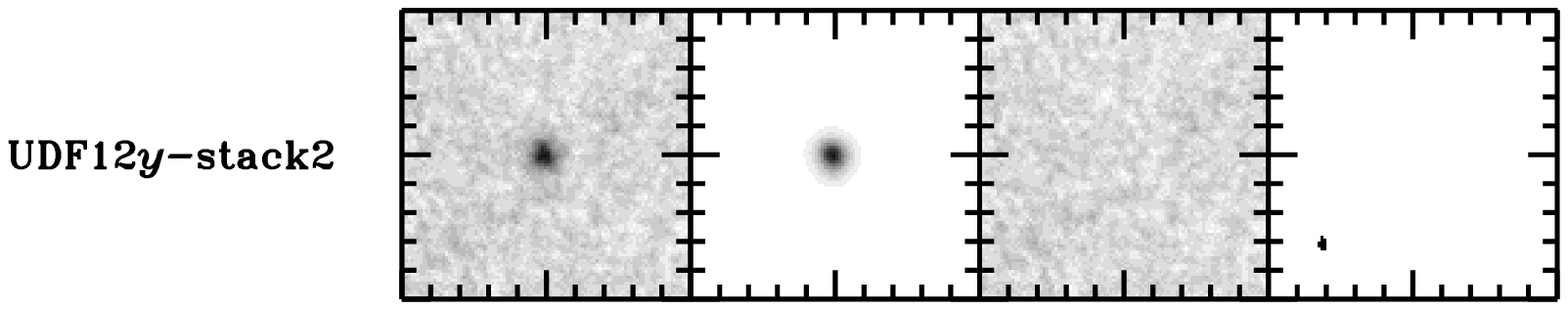} \\
 \caption[]
{
Same as Figure \ref{fig:galfit_main_z_stack1}, except that the objects are 
the stacked $Y_{105}$-dropouts. 
}
\label{fig:galfit_main_y_stack1}
\end{center}
\end{figure*}
%FFFFFFFFFFFFFFFFFFFFFFFFFFFFFFFFFFFFFFFFFFFFFFFFFFFFFFFFFFFFFFFF%

%FFFFFFFFFFFFFFFFFFFFFFFFFFFFFFFFFFFFFFFFFFFFFFFFFFFFFFFFFFFFFFFF%
\begin{figure*}
\begin{center}
\includegraphics[scale=0.9]{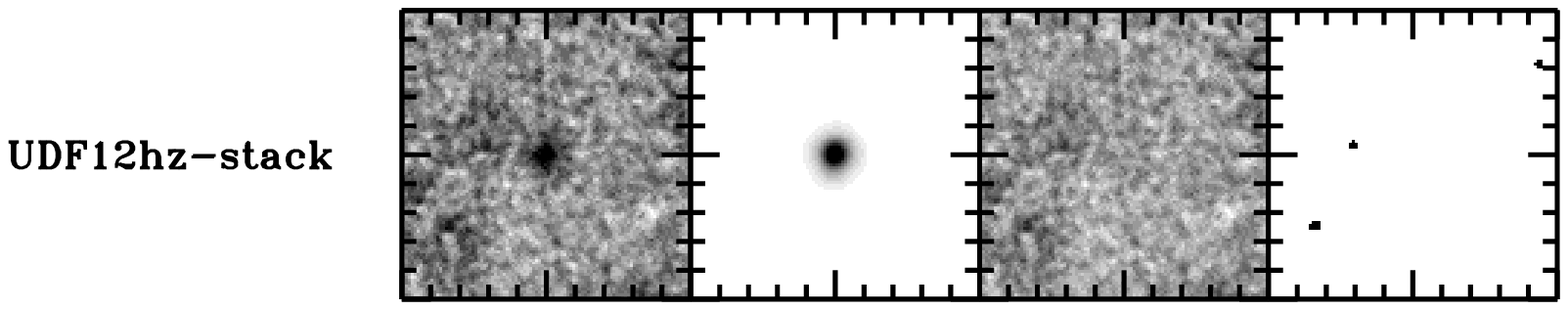} 
 \caption[]
{
Same as Figure \ref{fig:galfit_main_z_stack1}, 
except that the object is the stacked $z \sim 9$ object. 
}
\label{fig:galfit_main_hz_stack1}
\end{center}
\end{figure*}
%FFFFFFFFFFFFFFFFFFFFFFFFFFFFFFFFFFFFFFFFFFFFFFFFFFFFFFFFFFFFFFFF%

In summary, 
our simulations show that 
\verb|GALFIT| measurements of half-light radii and total magnitudes 
are systematically underestimated for faint objects.
We correct for systematic effects in the half-light radii and total magnitudes 
 using the measured offsets in Figures \ref{fig:simulations_re} and \ref{fig:simulations_mag}, 
 respectively. 
 Note that the errors on $r_e$ and total magnitude reported in this paper 
 are also taken from these simulations.

%%%%%%%%%%%%%%%%%%%%%%%%%%%%%%%%%%%%%%%%%%%%%%%%%%%%%%%%%%%%%%%%%
%%%%%%%%%%%%%%%%%%%%%%%%%%%%%%%%%%%%%%%%%%%%%%%%%%%%%%%%%%%%%%%%%
\subsection{GALFIT Measurements} \label{subsec:galfit_measurements}
%%%%%%%%%%%%%%%%%%%%%%%%%%%%%%%%%%%%%%%%%%%%%%%%%%%%%%%%%%%%%%%%%
%%%%%%%%%%%%%%%%%%%%%%%%%%%%%%%%%%%%%%%%%%%%%%%%%%%%%%%%%%%%%%%%%

We perform surface brightness profile fitting 
for our samples at $z \sim 7-12$, 
using \verb|GALFIT| and making use of our 
simulation results to correct for any systematic effects.
We analyze 
each of the objects with $>15\sigma$ detections individually 
($9$ $z_{850}$-dropouts, 
$6$ $Y_{105}$-dropouts), 
as well as  
the $z=11.9$ object, 
which is formally detected at $\sim 8 \sigma$. 
We extend the analysis to fainter magnitudes using stacked observations. 
The fainter $z_{850}$- and $Y_{105}$-dropouts 
are split into two luminosity bins 
before stacking 
($0.12 < L/L^\ast_{z=3} < 0.3$ and $0.048 < L/L^\ast_{z=3} < 0.12$), 
whereas we group all $z \sim 9$ candidates into a single stack.

Figure \ref{fig:galfit_main_z1} presents the results of S\'ersic profile fitting 
for the $9$ bright $z_{850}$-dropouts. 
Shown, from left to right, are the
$3'' \times 3''$ cut-outs of the original image, 
the best-fit model produced by \verb|GALFIT|, 
the residual images 
(original image $-$ best-fit profile) 
and 
the segmentation maps used for masking all the neighboring objects 
during the profile fitting.
Figure \ref{fig:galfit_main_y1} similarly shows 
the results for the $6$ bright $Y_{105}$-dropouts.  
All the objects are cleanly subtracted in the residual images. 
Note, however that three of the objects;  
two of the brightest $z_{850}$-dropouts, 
UDF12-4258-6567 and UDF12-3746-6327, 
as well as one of the $Y_{105}$-dropouts, 
UDF12-3952-7173 are significantly blended with neighboring objects 
in the original images.  
In addition, 
one of the $Y_{105}$-dropouts, 
UDF12-4470-6442 shows two cores. 
The uncertainties in the derived profile parameters for these objects 
will therefore be larger than for other isolated objects.

Figure \ref{fig:galfit_main_hz1} shows the profile fitting result for the $z \sim 12$ object, UDF12-3954-6284. 
Since the magnitude of the $z \sim 12$ object in $H_{160}$ measured with $0.50''$-diameter aperture 
is $29.2$ mag,  corresponding to S/N $\sim 8$, 
the profile fitting for this object is quite challenging. 
Actually, the best-fit model galaxy profile seems more elongated than 
that in the original image shown in Figure \ref{fig:galfit_main_hz1}, 
which would overestimate of its total magnitude. 
At least, the residual image in Figure \ref{fig:galfit_main_hz1} does not clearly 
show any noticeable residuals around the central position, 
although the uncertainties of the fitting parameters are relatively 
large as inferred from the moderate S/N ratio. 
If we measure the curve of growth for this object, 
using progressively larger circular apertures, 
we find that the magnitude saturates at $28.8$ mag 
within an aperture diameter $\sim 0.45''$.
We also find by this robust method that 
the half-light of the source is covered by about $0.35''$-diameter aperture, 
and after considering the PSF broadening effect, 
we obtain its half-light radius, $r_{\rm hl} = 0.45$ kpc, 
which is consistent with the \verb|GALFIT| measurement within $\sim 1 \sigma$, 
and is also nearly equal to 
the value reported by \cite{bouwens2012d}, $\sim 0.5$ kpc.

Additionally, we note that this object has an unusual morphology.
It is visually confirmed that the $z \sim 12$ object has a diffuse filamentary structure 
stretching from north-east to south-west, 
although the significance is very low. 
This has been already mentioned very recently by \cite{bouwens2012d}. 
Figure \ref{fig:z12object} shows the cutout $H_{160}$ images of this object 
from various subsets and the full data. 
The diffuse structure is seen in the full (2009$+$2012) data and in the 2009 data. 
The 2009b cutout also shows a low-S/N filament, 
and 
the 2009a cutout has a similar pattern along the same direction. 
If this diffuse filament is indeed associated with the source at $z \sim 12$, 
it corresponds to its bright UV continuum and/or Ly$\alpha$, 
which would suggest that this object is experiencing a major merger event, 
leading to their high star-formation activity. 
This star formation enhancement may explain 
the visibility of such a high-redshift galaxy.

%ttttttttttttttttttttttttttttttttttttttttttttttttttttttttttttttttttttttttt%
\begin{deluxetable*}{cccccccc} 
\tablecolumns{8} 
\tablewidth{0pt} 
\tablecaption{Surface Brightness Profile Fitting Results  for Bright $z_{850}$-dropouts \label{tab:sum_profile_fitting_zdrop}}
\tablehead{
    \colhead{Object ID}     
    &  \colhead{RA\tablenotemark{$\dagger1$}}  & \colhead{Decl.\tablenotemark{$\dagger1$}}    
    &  \colhead{$m_{\rm UV}^{\rm (ap)}$\tablenotemark{$\dagger2$}}  & \colhead{$n$\tablenotemark{$\dagger3$}}    
    & \colhead{$m_{\rm UV}$\tablenotemark{$\dagger4$}}  & \colhead{$M_{\rm UV}$\tablenotemark{$\dagger5$}}  
    & \colhead{$r_e$\tablenotemark{$\dagger6$}} \\
    \colhead{ } 
    & \colhead{[h:m:s]}  &  \colhead{[d:m:s]}  
    & \colhead{[mag]}  &  \colhead{ }  
    & \colhead{[mag]}  & \colhead{[mag]} 
    & \colhead{[kpc]}
}
\startdata 
 \multicolumn{8}{c}{S/N$>15$, $L/L^\ast_{z=3} = 0.3-1$}  \smallskip \\
UDF12-3746-6327 & 3:32:37.46 & $-27$:46:32.7  & 27.38 & 1.0 & $26.25 \pm 0.06$ & $-20.54 \pm 0.06$ & $1.09 \pm 0.09$ \\ 
UDF12-4258-6567 & 3:32:42.58 & $-27$:46:56.7  & 27.41 & 1.0 & $26.74 \pm 0.05$ & $-20.13 \pm 0.05$ & $0.50 \pm 0.03$ \\ \hline
 \multicolumn{8}{c}{S/N$>15$, $L/L^\ast_{z=3} = 0.12-0.3$}  \smallskip \\
UDF12-4256-7314 & 3:32:42.56 & $-27$:47:31.4  & 27.73 & 1.0 & $27.23 \pm 0.05$ & $-19.66 \pm 0.05$ & $0.38 \pm 0.05$ \\ 
UDF12-4219-6278 & 3:32:42.19 & $-27$:46:27.8  & 28.09 & 1.0 & $27.74 \pm 0.10$ & $-19.03 \pm 0.10$ & $0.26 \pm 0.07$ \\ 
UDF12-3677-7535 & 3:32:36.77 & $-27$:47:53.5  & 28.22 & 1.0 & $27.82 \pm 0.11$ & $-18.97 \pm 0.11$ & $0.38 \pm 0.09$ \\ 
UDF12-4105-7155 & 3:32:41.05 & $-27$:47:15.5  & 28.30 & 1.0 & $27.98 \pm 0.13$ & $-19.00 \pm 0.13$ & $0.20 \pm 0.07$ \\ 
UDF12-3958-6564 & 3:32:39.58 & $-27$:46:56.4  & 28.33 & 1.0 & $27.87 \pm 0.11$ & $-18.93 \pm 0.11$ & $0.42 \pm 0.10$ \\ 
UDF12-3744-6512 & 3:32:37.44 & $-27$:46:51.2  & 28.38 & 1.0 & $28.02 \pm 0.13$ & $-18.80 \pm 0.13$ & $0.15 \pm 0.09$ \\ 
UDF12-3638-7162 & 3:32:36.38 & $-27$:47:16.2  & 28.41 & 1.0 & $28.05 \pm 0.14$ & $-18.77 \pm 0.14$ & $0.11 \pm 0.08$ \\ \hline
%\smallskip \\
 \multicolumn{8}{c}{\bf stack}  \smallskip \\
 \multicolumn{3}{l}{UDF12$z$-stack1 ($L/L^\ast_{z=3} = 0.12-0.3$)} & 28.27 & 1.0 & $27.87 \pm 0.11$ & $-19.00 \pm 0.11$ & $0.31 \pm 0.09$ \\ 
 \multicolumn{3}{l}{UDF12$z$-stack2 ($L/L^\ast_{z=3} = 0.048-0.12$)} & 29.20 & 1.0 & $28.73 \pm 0.26$ & $-18.14 \pm 0.26$ & $0.36 \pm 0.15$
  \enddata 
\tablenotetext{$\dagger1$}{%
Coordinates are in J2000.
}
\tablenotetext{$\dagger2$}{%
Measured in $0.46''$-diameter aperture with the stack of the $J_{125}$ and $J_{140}$ images.
}
\tablenotetext{$\dagger3$}{%
S\'ersic index. This is fixed, not measured. 
}
\tablenotetext{$\dagger4$}{%
Total magnitude measured by \texttt{GALFIT}. The systematic effect is considered.
}
\tablenotetext{$\dagger5$}{%
Total absolute magnitude calculated using 
$z_{\rm photo}$ 
if available, 
otherwise $z=6.7$, which corresponds to the peak of the selection function for $z_{850}$-dropouts. 
}
\tablenotetext{$\dagger6$}{%
Circularized half-light radius $r_e = a \sqrt{b/a}$, 
where $a$ is radius along the major axis, 
and $b/a$ is axis ratio.
}
\end{deluxetable*} 
%ttttttttttttttttttttttttttttttttttttttttttttttttttttttttttttttttttttttttt%

%ttttttttttttttttttttttttttttttttttttttttttttttttttttttttttttttttttttttttt%
\begin{deluxetable*}{cccccccc} 
\tablecolumns{8} 
\tablewidth{0pt} 
\tablecaption{Surface Brightness Profile Fitting Results  for Bright $Y_{105}$-dropouts\label{tab:sum_profile_fitting_ydrop}}
\tablehead{
    \colhead{Object ID}     
    &  \colhead{RA\tablenotemark{$\dagger1$}}  & \colhead{Decl.\tablenotemark{$\dagger1$}}    
    &  \colhead{$m_{\rm UV}^{\rm (ap)}$\tablenotemark{$\dagger2$}}  & \colhead{$n$\tablenotemark{$\dagger3$}}    
    & \colhead{$m_{\rm UV}$\tablenotemark{$\dagger4$}}  & \colhead{$M_{\rm UV}$\tablenotemark{$\dagger5$}}  
    & \colhead{$r_e$\tablenotemark{$\dagger6$}}    \\
    \colhead{ } 
    & \colhead{[h:m:s]}  &  \colhead{[d:m:s]}  
    & \colhead{[mag]}  &  \colhead{ }  
    & \colhead{[mag]}  & \colhead{[mag]} 
    & \colhead{[kpc]}
}
\startdata 
 \multicolumn{8}{c}{S/N$>15$, $L/L^\ast_{z=3} = 0.3-1$}  \smallskip \\
UDF12-3879-7071 & 3:32:38.80 & $-27$:47:07.1  & 27.21 & 1.0 & $26.80 \pm 0.03$ & $-20.18 \pm 0.03$ & $0.36 \pm 0.02$ \\ 
UDF12-4470-6442 & 3:32:44.70 & $-27$:46:44.2  & 27.69 & 1.0 & $27.13 \pm 0.08$ & $-19.93 \pm 0.08$ & $0.60 \pm 0.09$ \\ 
UDF12-3952-7173 & 3:32:39.52 & $-27$:47:17.3  & 28.10 & 1.0 & $26.95 \pm 0.10$ & $-20.12 \pm 0.10$ & $1.05 \pm 0.19$ \\ \hline
 \multicolumn{8}{c}{S/N$>15$, $L/L^\ast_{z=3} = 0.12-0.3$}  \smallskip \\
UDF12-4314-6284 & 3:32:43.14 & $-27$:46:28.4  & 28.10 & 1.0 & $27.73 \pm 0.08$ & $-19.22 \pm 0.08$ & $0.34 \pm 0.07$ \\ 
UDF12-3722-8061 & 3:32:37.22 & $-27$:48:06.1  & 28.28 & 1.0 & $27.89 \pm 0.10$ & $-19.09 \pm 0.10$ & $0.20 \pm 0.06$ \\ 
UDF12-3813-5540 & 3:32:38.13 & $-27$:45:54.0  & 28.33 & 1.0 & $27.99 \pm 0.11$ & $-19.23 \pm 0.11$ & $0.09 \pm 0.05$ \\ \hline
% \smallskip \\
 \multicolumn{8}{c}{\bf stack}  \smallskip \\
 \multicolumn{3}{l}{UDF12$y$-stack1 ($L/L^\ast_{z=3} = 0.12-0.3$)} & 28.40 & 1.0 & $28.01 \pm 0.11$ & $-19.13 \pm 0.11$ & $0.35 \pm 0.13$ \\ 
 \multicolumn{3}{l}{UDF12$y$-stack2 ($L/L^\ast_{z=3} = 0.048-0.12$)} & 29.47 & 1.0 & $28.90 \pm 0.29$ & $-18.24 \pm 0.29$ & $0.36 \pm 0.13$ 
 \enddata 
\tablenotetext{$\dagger1$}{%
Coordinates are in J2000.
}
\tablenotetext{$\dagger2$}{%
Measured in $0.50''$-diameter aperture with the stack of the $J_{140}$ and $H_{160}$ images.
}
\tablenotetext{$\dagger3$}{%
S\'ersic index. This is fixed, not measured. 
}
\tablenotetext{$\dagger4$}{%
Total magnitude measured by \texttt{GALFIT}. The systematic effect is considered.
}
\tablenotetext{$\dagger5$}{%
Total absolute magnitude calculated using 
$z_{\rm photo}$ 
if available, 
otherwise $z=8.0$, which corresponds to the peak of the selection function for $Y_{105}$-dropouts. 
}
\tablenotetext{$\dagger6$}{%
Circularized half-light radius $r_e = a \sqrt{b/a}$, 
where $a$ is radius along the major axis, 
and $b/a$ is axis ratio.
}
\end{deluxetable*} 
%ttttttttttttttttttttttttttttttttttttttttttttttttttttttttttttttttttttttttt%

%ttttttttttttttttttttttttttttttttttttttttttttttttttttttttttttttttttttttttt%
\begin{deluxetable*}{cccccccc} 
\tablecolumns{8} 
\tablewidth{0pt} 
\tablecaption{Surface Brightness Profile Fitting Results  for $z>8.5$ Candidates\label{tab:sum_profile_fitting_hzdrop}}
\tablehead{
    \colhead{Object ID}     
    &  \colhead{RA\tablenotemark{$\dagger1$}}  & \colhead{Decl.\tablenotemark{$\dagger1$}}    
    &  \colhead{$m_{\rm UV}^{\rm (ap)}$\tablenotemark{$\dagger2$}}  & \colhead{$n$\tablenotemark{$\dagger3$}}    
    & \colhead{$m_{\rm UV}$\tablenotemark{$\dagger4$}}  & \colhead{$M_{\rm UV}$\tablenotemark{$\dagger5$}}  
    & \colhead{$r_e$\tablenotemark{$\dagger6$}}    \\
    \colhead{ } 
    & \colhead{[h:m:s]}  &  \colhead{[d:m:s]}  
    & \colhead{[mag]}  &  \colhead{ }  
    & \colhead{[mag]}  & \colhead{[mag]} 
    & \colhead{[kpc]}
}
\startdata 
UDF12-3954-6284 & 3:32:39.54 & $-27$:46:28.4  & 29.24 & 1.0 & $28.47 \pm 0.25$ & $-20.41 \pm 0.25$ & $0.32 \pm 0.14$ \\  \hline
% \smallskip \\
 \multicolumn{8}{c}{\bf stack}  \smallskip \\
 \multicolumn{3}{l}{UDF12hz-stack} & 29.69 & 1.0 & $29.11 \pm 0.49$ & $-18.21 \pm 0.49$ & $0.35 \pm 0.16$
  \enddata 
\tablenotetext{$\dagger1$}{%
Coordinates are in J2000.
}
\tablenotetext{$\dagger2$}{%
Measured in $0.50''$-diameter aperture with the $H_{160}$ image.
}
\tablenotetext{$\dagger3$}{%
S\'ersic index. This is fixed, not measured. 
}
\tablenotetext{$\dagger4$}{%
Total magnitude measured by \texttt{GALFIT}. The systematic effect is considered.
For UDF12-3954-6284, this might be overestimated, 
since the best-fit model galaxy profile seems more elongated than 
that in the original image shown in Figure \ref{fig:galfit_main_hz1}. 
If we measure the curve of growth for this source, 
the magnitude saturates at $28.8$ mag (See Section \ref{subsec:galfit_measurements}). 
}
\tablenotetext{$\dagger5$}{%
Total absolute magnitude. 
We calculate it with $z_{\rm photo}=11.9$ for UDF12-3954-6284, 
considering IGM absorption shortward of its Ly$\alpha$ wavelength. 
We use the average photometric redshift, $z_{\rm photo}=9.0$ for the stacked object. 
}
\tablenotetext{$\dagger6$}{%
Circularized half-light radius $r_e = a \sqrt{b/a}$, 
where $a$ is radius along the major axis, 
and $b/a$ is axis ratio.
}
\end{deluxetable*} 
%ttttttttttttttttttttttttttttttttttttttttttttttttttttttttttttttttttttttttt%

Figure \ref{fig:galfit_main_z_stack1} and \ref{fig:galfit_main_y_stack1} show
profile fitting results for the stacked $z_{850}$-dropouts and $Y_{105}$-dropouts, respectively, 
whose UV luminosities are 
$L=(0.12-0.3)L^\ast_{z=3}$ (top)
and 
$L=(0.048-0.12)L^\ast_{z=3}$ (bottom). 
Also shown in Figure \ref{fig:galfit_main_hz_stack1} is the profile fitting result 
for the stacked $z \simeq 9$ candidates.  
Note that we also make averaged (not median-stacked) images and 
perform profile fitting using \texttt{GALFIT}, 
which yield similar fitting results, 
although for some of the average stacks, 
\verb|GALFIT| does not provide a reasonable fit 
due to severe confusion with neighboring objects.

In the brightest luminosity bin, 
$L=(0.3-1)L^\ast_{z=3}$, 
we do not perform a stacking analysis, 
since the numbers of the dropouts are small 
(2 for $z_{850}$-dropouts and $3$ for $Y_{105}$-dropouts) 
and the stacked images are significantly confused by neighboring objects.
Instead, we calculate their average sizes and magnitudes; 
$r_e = 0.79 \pm 0.29$ kpc 
and 
$M_{\rm UV} = -20.33 \pm 0.20$ mag 
($z \sim 7$) 
and 
$r_e = 0.67 \pm 0.28$ kpc 
and
$M_{\rm UV} = -20.08 \pm 0.10$ mag 
($z \sim 8$).

The best-fit parameters are summarized 
in Table \ref{tab:sum_profile_fitting_zdrop} for the $z_{850}$-dropouts, 
Table \ref{tab:sum_profile_fitting_ydrop} for the $Y_{105}$-dropouts, 
and
Table \ref{tab:sum_profile_fitting_hzdrop} for the $z>8.5$ candidates. 
The weighted mean of half-light radii for 
the $z_{850}$-dropouts and $Y_{105}$-dropouts 
with $L=(0.05-1)L^\ast_{z=3}$ are 
$0.35 \pm 0.07$ kpc 
and 
$0.38 \pm 0.09$ kpc, respectively.  
In the next section, 
we present the size-luminosity relation, 
and 
investigate the redshift evolution of galaxy sizes and SFR surface densities 
based on these results.

%FFFFFFFFFFFFFFFFFFFFFFFFFFFFFFFFFFFFFFFFFFFFFFFFFFFFFFFFFFFFFFFF%
\begin{figure}
   \includegraphics[scale=0.7]{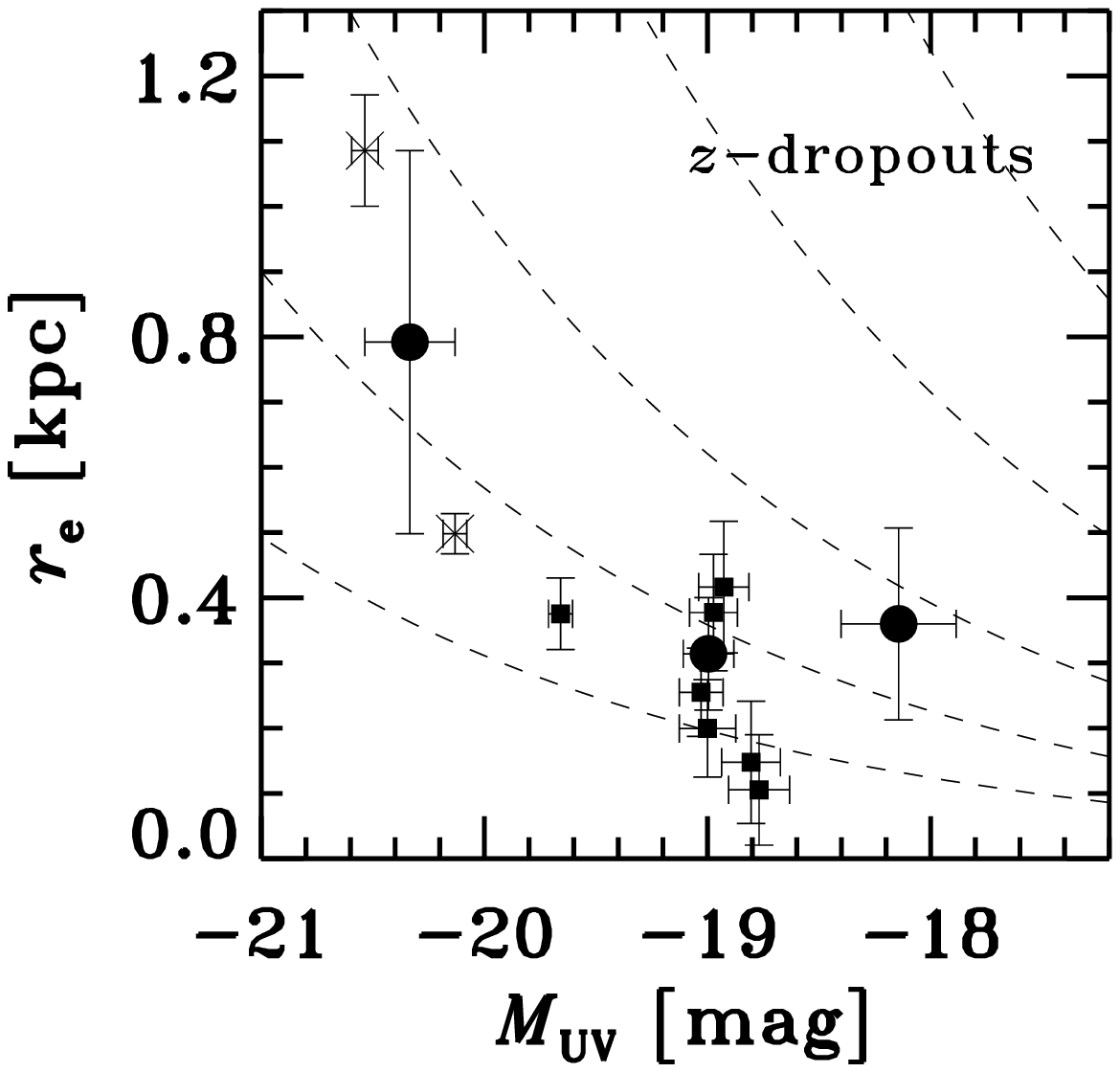} 
   \includegraphics[scale=0.7]{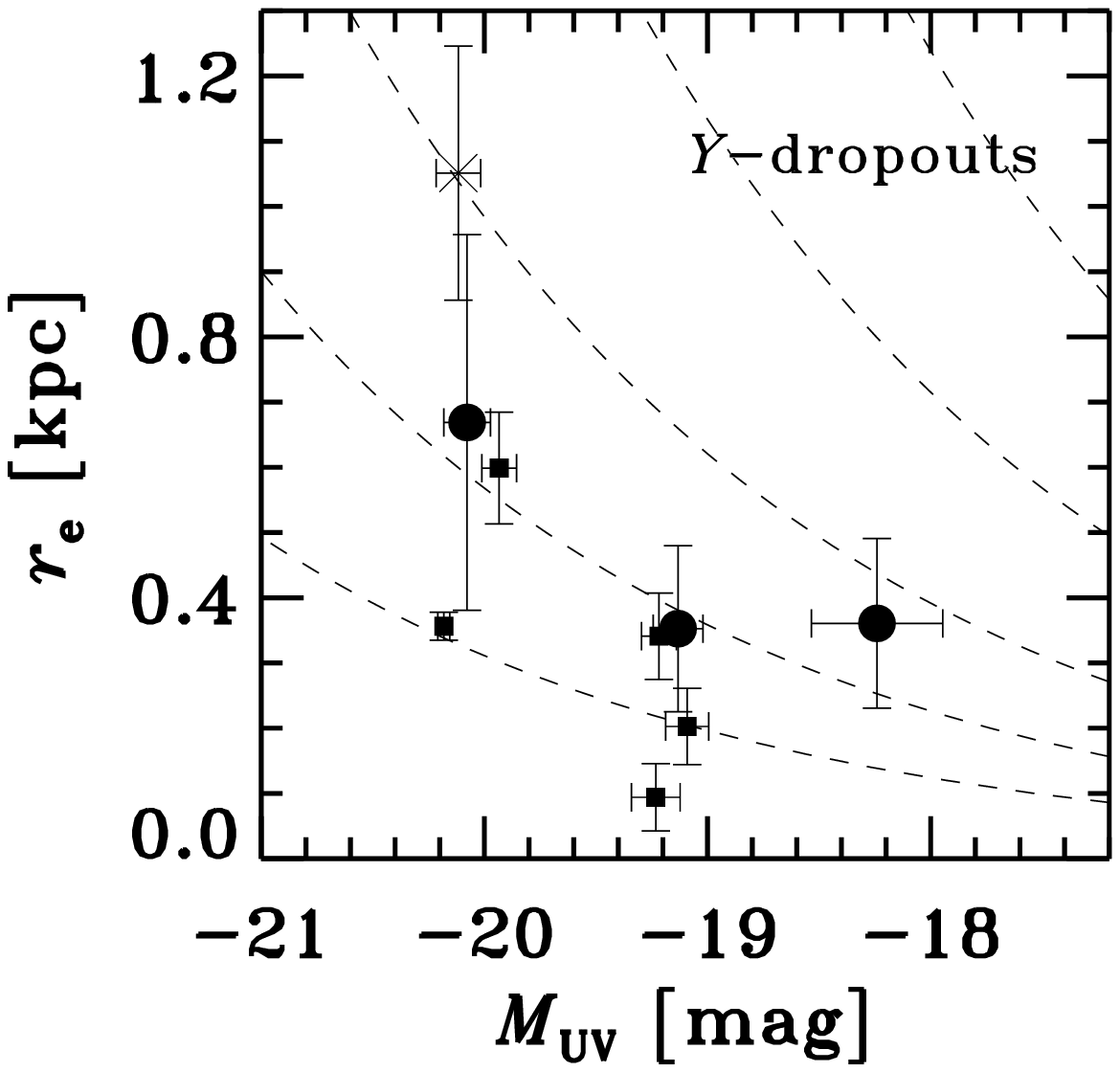}
 \caption[]
{
The size-luminosity relation for $z_{850}$-dropouts (top) and 
$Y_{105}$-dropouts {bottom}.
The filled circles correspond to the stacked objects 
with UV luminosities of 
$L=(0.12-0.3)L^\ast_{z=3}$, 
$L=(0.048-0.12)L^\ast_{z=3}$, 
and 
the averaged values of the objects with $L=(0.3-1)L^\ast_{z=3}$. 
The filled squares show the bright objects detected at $>15\sigma$ 
without any blending with neighboring sources, 
while 
the crosses show the objects detected in more than $15\sigma$ 
and blended with a neighboring source.  
The dashed curves in each figure correspond to 
a constant star-formation rate density 
$\Sigma_{\rm SFR}$ [$M_\odot$ yr$^{-1}$ kpc$^{-2}$]  
$= 0.1$, $0.3$, $1$, $3$, $10$ from top to bottom. 
}
\label{fig:Muv_re}
\end{figure}
%FFFFFFFFFFFFFFFFFFFFFFFFFFFFFFFFFFFFFFFFFFFFFFFFFFFFFFFFFFFFFFFF%

%%%%%%%%%%%%%%%%%%%%%%%%%%%%%%%%%%%%%%%%%%%%%%%%%%%%%%%%%%%%%%%%%
%%%%%%%%%%%%%%%%%%%%%%%%%%%%%%%%%%%%%%%%%%%%%%%%%%%%%%%%%%%%%%%%%
\section{Results and Discussion} \label{sec:results_discussion}
%%%%%%%%%%%%%%%%%%%%%%%%%%%%%%%%%%%%%%%%%%%%%%%%%%%%%%%%%%%%%%%%%
%%%%%%%%%%%%%%%%%%%%%%%%%%%%%%%%%%%%%%%%%%%%%%%%%%%%%%%%%%%%%%%%%

Our measurements of half-light radii in Tables
\ref{tab:sum_profile_fitting_zdrop}$-$\ref{tab:sum_profile_fitting_hzdrop}
show very small values, typically $\lesssim 0.5$ kpc (see also filled symbols 
in Figure \ref{fig:Muv_re}).
The average half-light radii of the dropouts 
are only $\simeq 0.3 - 0.4$ kpc at $z\sim 7-8$ 
(Section \ref{subsec:galfit_measurements})
and at $z>8.5$ 
(Table \ref{tab:sum_profile_fitting_hzdrop}).
The half-light radius of our $z\sim 12$ candidate 
is also remarkably small, $0.32\pm0.14$ kpc.
Even including the $1\sigma$ uncertainties,
these half-light radii are, coincidentally, just as large as 
those of giant molecular associations (GMAs)
with a mass of $\sim 10^7 M_\odot$
found in the local universe 
\citep[e.g.,][]{vogel1988,rand1990,tosaki2007}.

We investigate the relation between size and luminosity, i.e.
the size-luminosity relation, at each redshift.
Figure \ref{fig:Muv_re} presents the size-luminosity
relation for our $z_{850}$-dropout and $Y_{105}$-dropout galaxies
at $z\sim 7-8$. 
Our $z>8.5$ galaxy candidates are not shown, 
because we cannot constrain the relation with only two measurements 
(one from an individual object and one from the stack). 
In Figure \ref{fig:Muv_re}, fainter galaxies have
a smaller half-light radius than brighter galaxies.
This trend is the same as those of local galaxies
\citep{dejong2000} as well as
high-$z$ dropout galaxies at slightly lower
redshifts ($z\sim 6-7$), studied by \citet{grazian2012}.
Because the luminosity of a galaxy depends on
two physical quantities (surface brightness and size), 
one needs to clarify which quantity is dominant
in shaping the size-luminosity relation.
We define star-formation rate surface density, $\Sigma_{\rm SFR}$,
as the average star-formation rate in a circle whose radius is $r_e$,
\begin{equation}
\Sigma_{\rm SFR} \,[M_\odot \, {\rm yr}^{-1} \, {\rm kpc}^{-2}]
        \equiv \frac{{\rm SFR} / 2}{\pi r_e^2}.
\label{sfr_surface_density}
\end{equation}
In the case that dust extinction is negligible,
a rest-frame UV luminosity density
approximately correlates with a star-formation rate \citep{kennicutt1998},
\begin{equation}
{\rm SFR} \,[M_\odot \, {\rm yr}^{-1}]
        = 1.4 \times 10^{-28} L_\nu \, [{\rm erg}\,{\rm s}^{-1} {\rm Hz}^{-1}].
\label{sfr_luminosity}
\end{equation}
From equations (\ref{sfr_surface_density})$-$(\ref{sfr_luminosity}),
we obtain
\begin{equation}
M_{\rm UV}
        = - 2.5 \log \left( \frac{\Sigma_{\rm SFR} \,  r_e^2}{1.4 \times
10^{-28}  \cdot 2 \cdot \left( {\rm 10 pc} \, [{\rm cm}]\right)^2}
\right) -48.6.
\end{equation}
In Figure \ref{fig:Muv_re}, we show
constant star-formation rate surface densities
of $\Sigma_{\rm SFR} = 0.1$, $0.3$, $1$, $3$, and $10$
[$M_\odot$ yr$^{-1}$ kpc$^{-2}$] with dashed lines.
We find that most of individual galaxies and stacked
galaxies fall in the range of $\Sigma_{\rm SFR} \simeq 1-10$
within their uncertainties.
These results indicate that
both bright and faint $z\sim 7-8$ galaxies have
similar star-formation rate surface densities
of $\Sigma_{\rm SFR} \simeq 1-10$, and that
the size-luminosity relation at each redshift is mainly
determined by the size of galaxies
that have a similar star-formation rate surface density.

%FFFFFFFFFFFFFFFFFFFFFFFFFFFFFFFFFFFFFFFFFFFFFFFFFFFFFFFFFFFFFFFF%
\begin{figure}
   \includegraphics[scale=0.7]{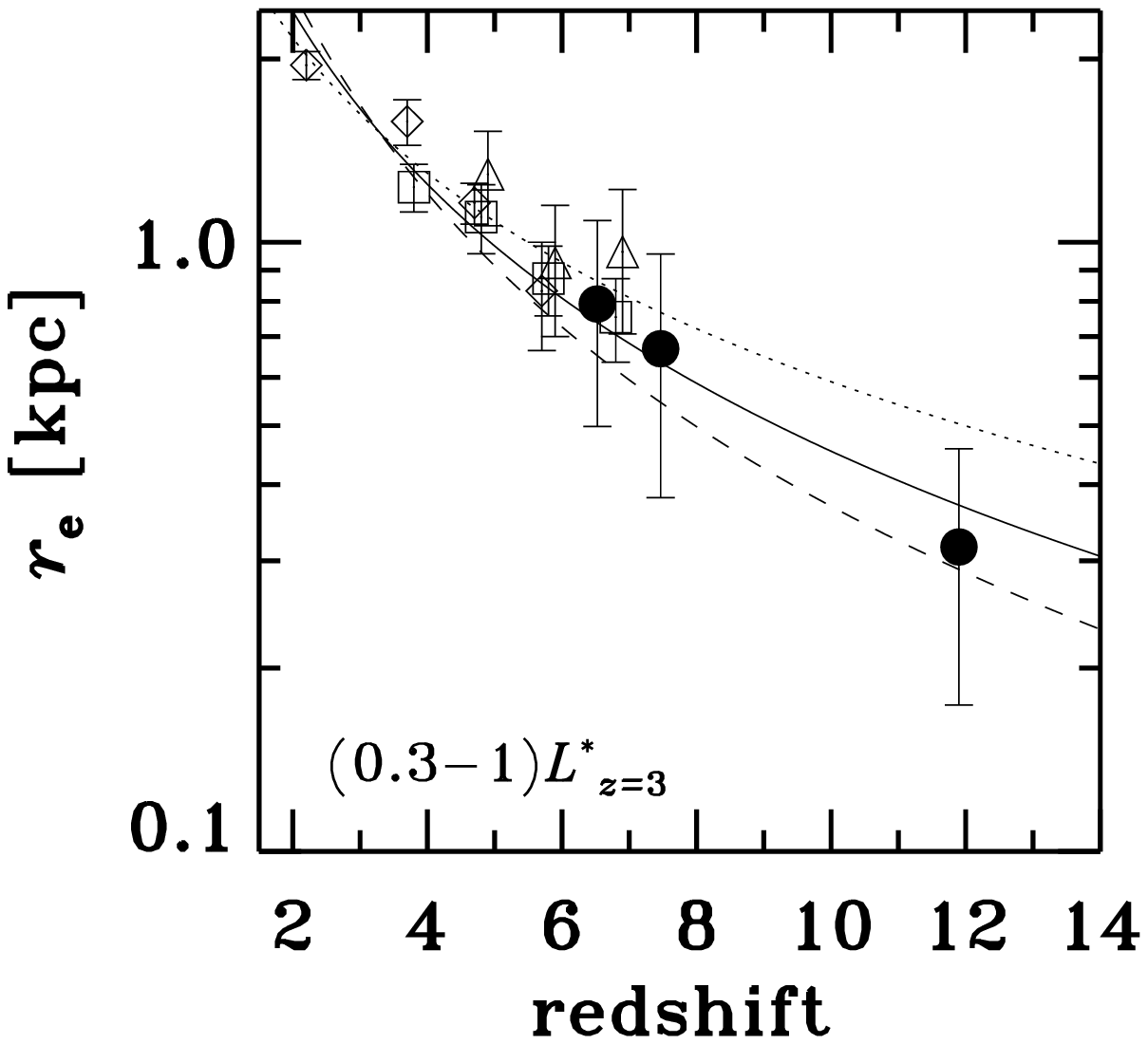}
   \includegraphics[scale=0.7]{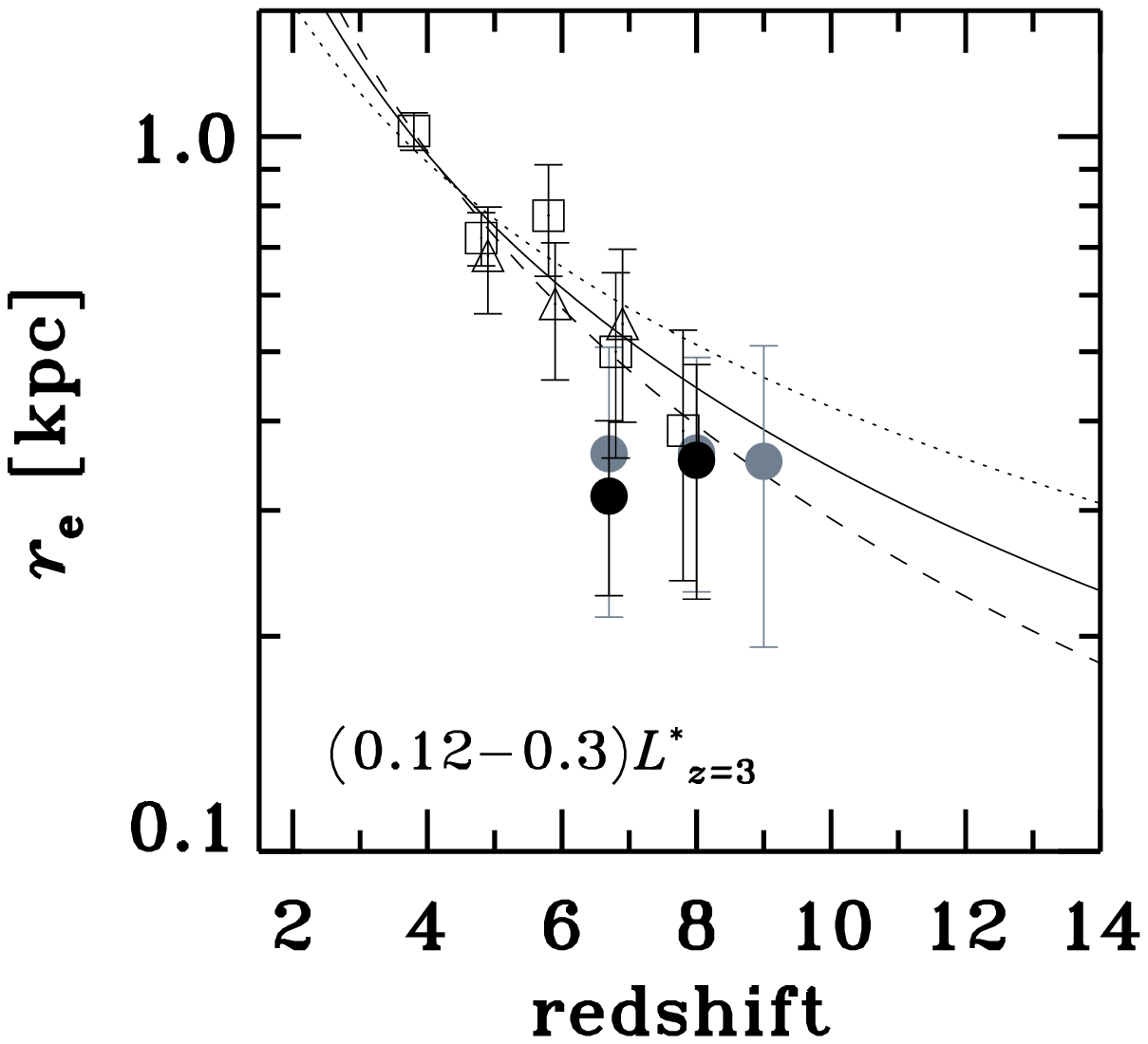}
 \caption[]
{
\textbf{Top}: 
Evolution of the half-light radius across the redshift range from $z \sim 2$ to $12$ 
in $(0.3-1) L^\ast_{z=3}$. 
The filled circles show 
the average sizes of our $z_{850}$-dropouts and $Y_{105}$-dropouts,  
and the size of the $z \sim 12$ object. 
The open symbols are taken from the literature; 
the open squares and triangles are dropout galaxies 
taken from \cite{oesch2010b}, 
the open diamonds are from \cite{bouwens2004}.  
After excluding the sample overlaps, 
we fit simple functions of $(1+z)^s$ with the data in both luminosity bins, 
and 
obtain $s=-1.28 \pm 0.13$. 
which is shown as the solid line. 
The dotted and dashed lines correspond to the case of $s=-1.0$ and $-1.5$, respectively.
\textbf{Bottom}:  
Evolution of the half-light radius across the redshift range from $z \sim 4$ to $8$ 
in $(0.12-0.3) L^\ast_{z=3}$. 
The open and filled black symbols denote the same as those in the top panel. 
The gray filled circles are dropout galaxies in the fainter luminosity bin, 
$(0.048-0.12) L^\ast_{z=3}$.
The solid, dotted, and dashed lines are the same as those in the top figure. 
}
\label{fig:redshift_re}
\end{figure}
%FFFFFFFFFFFFFFFFFFFFFFFFFFFFFFFFFFFFFFFFFFFFFFFFFFFFFFFFFFFFFFFF%

We then investigate the size evolution.
Since the half-light radius depends on luminosity, 
as displayed by the size-luminosity relation, we carefully
compare the half-light radii of our dropout galaxies
within a fixed magnitude range. Figure \ref{fig:redshift_re}
presents the average half-light radius as a function
of redshift for our dropout galaxies at $z\sim 7-12$
with $(0.3-1) L^\ast_{z=3}$ and $(0.12-0.3) L^\ast_{z=3}$,
together with dropout galaxies at $z \sim 4-8$ taken
from the literature. Our measurements of
average half-light radii are consistent with
those from the previous studies at $z \sim 7-8$, 
where we see overlap with previous measurements. 
Figure \ref{fig:redshift_re} indicates that
the average half-light radius decreases
with  redshift from $z \sim 4$ to $8$,
which is consistent with the claims of
previous studies (e.g. \citealt{ferguson2004,bouwens2004,
hathi2008,oesch2010b}).

UDF12 provides us with the deepest ever near-infrared images of the HUDF, 
allowing our study to extend the dynamic range of redshift 
in the size evolution analysis 
from $z\sim 8$ to $z\sim 12$,
and identifies that the decreasing trend holds 
up to $z\sim 12$ as shown in Figure \ref{fig:redshift_re}, 
if the putative $z \sim 12$ source is real.
Although the statistical uncertainty of measurement is large,
the half-light radius of $z\sim 12$ is
$r_e=0.32 \pm 0.14$ kpc in the luminosity bin
of $(0.3-1) L^\ast_{z=3}$, which is significantly
smaller than that of $z\sim 6$ by a factor of $3$.
Note that we can only plot the half-light radius at $z\sim 12$
on the panel of $(0.3-1) L^\ast_{z=3}$
in Figure \ref{fig:redshift_re},
because there are no $z\sim 12$ galaxies with
$(0.12-0.3) L^\ast_{z=3}$ in the UDF12 data.
Similarly, our stack of $z>8.5$ galaxies have
a luminosity fainter than $(0.12-0.3) L^\ast_{z=3}$,
which is too faint to be compared with the baseline of
the average half-light radii at $z\sim 4-6$. 
However,
our results of $z>7$ galaxies at these faint magnitudes, which
are shown as gray filled circles 
in the bottom panel of Figure \ref{fig:redshift_re}, 
are consistent with the decreasing trend of the half-light radius, 
albeit with large errors.

This decreasing trend may be explained by
the evolution of host dark halo radius. According
to the analytic model in the hierarchical structure
formation framework of $\Lambda$CDM (see, e.g.,
\citealt{mo1998,mo2002,ferguson2004}),
the virial radius of a dark matter halo is given by
\begin{equation}
r_{\rm vir}
        = \left( \frac{G M_{\rm vir}}{100 H(z)^2} \right)^{1/3},
\end{equation}
where $H(z)$ is the Hubble parameter
and $M_{\rm vir}$ is the virial mass of the halo.
The virial radius is also expressed as a function of 
the circular velocity of dark halo
by
\begin{equation}
r_{\rm vir}
        = \frac{v_{\rm vir}}{10 H(z)}.
\end{equation}
Since $H(z)$ is approximated by $\sim (1+z)^{3/2}$
in a flat universe at high redshifts,
the redshift evolution of the virial radius is
$r_{\rm vir} \propto H(z)^{-2/3}\sim (1+z)^{-1}$ for constant halo mass
and
$r_{\rm vir} \propto H(z)^{-1}\sim (1+z)^{-1.5}$ for constant velocity.

Figure \ref{fig:redshift_re} shows
the radius-redshift relation of dark matter halos 
for the case of constant halo mass
and
constant velocity. 
Previous studies 
investigating the radius-redshift relation 
in the redshift range $4 < z <8$ 
reach two different conclusions. 
\cite{bouwens2004,bouwens2006b} claim that the relation is
roughly $(1+z)^{-1}$, suggestive 
that the sizes of disks scale with constant halo mass, 
while \cite{ferguson2004} and \cite{hathi2008}
argue that $(1+z)^{-1.5}$, i.e., the constant
velocity case, is preferable.
We fit the radius-redshift relation 
over a wider range of redshift (extending to $z \sim 12$) 
with a function of $(1+z)^{s}$,
where $s$ is a free parameter.
We take into account 
our size measurements: 
for the brighter bin, 
the average sizes of $z_{850}$- and $Y_{105}$-dropouts 
and the size of the $z \sim 12$ source,  
and 
for the fainter bin, 
the measured sizes of the stacks of 
$z_{850}$- and $Y_{105}$-dropouts 
with $L=(0.12-0.3)L^\ast_{z=3}$. 
In addition, 
we use 
the results reported in the literature: 
the average sizes at $z=2.5$ \citep{bouwens2004}, 
the average sizes at $z=4-6$ \citep{oesch2010b}\footnote{For the fitting, 
we do not 
utilize the \texttt{GALFIT} measurements by \cite{oesch2010b},   
which they did not use as their fiducial ones. 
Note that 
the fitting result is consistent within $1 \sigma$,  
if we include the \texttt{GALFIT} measurements instead of their \texttt{SExtractor} measurements. 
}. 
We fit the following two functions to the data, 
$\log r_e = s \log (1+z) + a_1$ for $L=(0.3-1)L^\ast_{z=3}$
and 
$\log r_e = s \log (1+z) + a_2$ for $L=(0.12-0.3)L^\ast_{z=3}$, 
where $s$, $a_1$, and $a_2$ are free parameters. 
Varying the three parameters, 
we search for the best-fitting set of $(s,a_1,a_2)$ that minimizes $\chi^2$. 
The best-fit parameters are 
$s = -1.28 \pm 0.13$, 
$a_1 = 0.99 \pm 0.08$, 
and 
$a_2 = 0.87^{+0.09}_{-0.10}$. 
We have checked that exclusion of the the putative $z \simeq 12$ object
produces no significant change in our results, 
but it is nevertheless interesting that its size conforms with the trend 
established at slightly lower redshifts.
We note that these results are clearly consistent with 
the redshift trend derived by \cite{oesch2010b} from the early UDF09 data, 
as they reported $s = -1.12 \pm 0.17$ for galaxies with luminosities
in the range $L = (0.3-1)L^\ast_{z=3}$, 
and
$s = -1.32 \pm 0.52$ for the fainter galaxies with $L = (0.12-0.3) L^\ast_{z=3}$. 
However, our derived value for $s$ is more accurate, 
both 
because of the improved data and galaxy samples provided by UDF12, 
and 
because we have chosen to fit a single value of $s$ 
to both the bright and more modest luminosity galaxies.

It should be noted again that 
the above simple constant mass or content velocity models 
assume that the stellar to halo size ratio does not change 
over this redshift range \citep{mo1998}. 
To properly interpret our result, 
more realistic models are needed which carefully treat 
the stellar to halo size ratio, 
as well as 
consider effects on galaxy sizes from galaxy mergers, torquing, and feedback, 
based on a hierarchical galaxy formation scenario 
over the full redshift range \citep[e.g.,][]{somerville2008}. 
These size measurements of high-redshift galaxies provide 
a launching point for a theoretical understanding 
of the structure of such galaxies, 
which has only recently been attempted 
but is of critical importance in understanding their properties \citep[e.g.,][]{munoz2012}.

%FFFFFFFFFFFFFFFFFFFFFFFFFFFFFFFFFFFFFFFFFFFFFFFFFFFFFFFFFFFFFFFF%
\begin{figure}
   \includegraphics[scale=0.7]{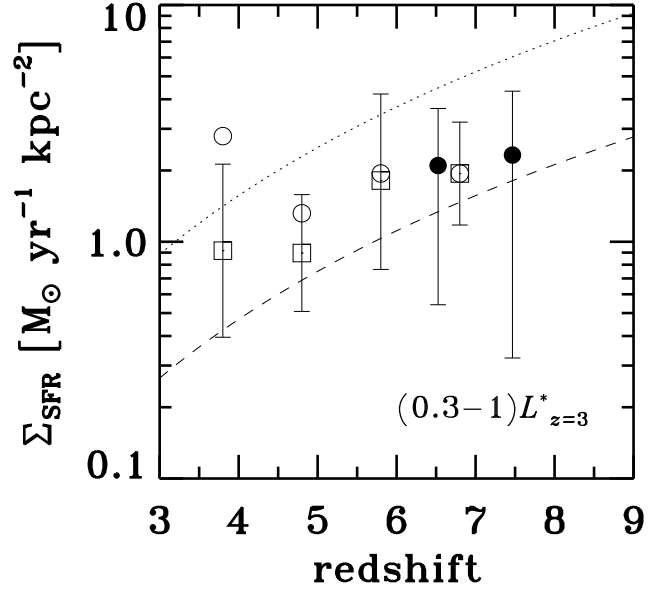}
 \caption[]
{
Evolution of the SFR surface density $\Sigma_{\rm SFR}$ 
as a function of redshift, for dropouts 
in the brightest luminosity bin, $L=(0.3-1) L^\ast_{z=3}$. 
The filled circles correspond to our $z_{850}$-dropouts and $Y_{105}$-dropouts. 
The open squares are taken from \cite{oesch2010b}, 
showing their dropout galaxies, 
while 
the open circles are the same objects but corrected for dust absorption. 
The dotted line and dashed line shows 
the case with a constant UV luminosity 
$L=L^\ast_{z=3}$ and $0.3L^\ast_{z=3}$, respectively, 
given that their half-light radii follow the simple relation of $(1+z)^s$, 
which is derived in Section \ref{sec:results_discussion} and shown in 
Figure \ref{fig:redshift_re}. 
}
\label{fig:redshift_SigmaSFR}
\end{figure}
%FFFFFFFFFFFFFFFFFFFFFFFFFFFFFFFFFFFFFFFFFFFFFFFFFFFFFFFFFFFFFFFF%

Figure \ref{fig:redshift_SigmaSFR} presents the average
star-formation surface density, $\Sigma_{\rm SFR}$,
as a function of redshift.
Note that the measurements
are shown up to $z\sim 8$ in Figure \ref{fig:redshift_SigmaSFR},
because the uncertainty
of the $z\sim 12$ measurements are too large to
place a meaningful constraint.
Our galaxies at $z>7$ have $\Sigma_{\rm SFR}\sim 2$.
$\Sigma_{\rm SFR}$ appears
to increase towards high redshifts. In fact, this increase of
$\Sigma_{\rm SFR}$ is
expected from the decreasing trend of galaxy size
at a given luminosity (Figure \ref{fig:redshift_re}).
Since $\Sigma_{\rm SFR}$ is proportional to
the UV luminosity density in the case of no dust extinction,
Figure \ref{fig:redshift_SigmaSFR} indicates that
UV luminosity surface brightness is higher for
$z\sim 7-8$ galaxies than for $z\sim 4-5$ galaxies
by a factor of $2-3$. Figure \ref{fig:redshift_SigmaSFR}
has data points of dust-corrected $\Sigma_{\rm SFR}$.
The dust-extinction corrected $\Sigma_{\rm SFR}$ is
significantly larger than uncorrected
$\Sigma_{\rm SFR}$ at $z\sim 4-5$, while almost
no difference is found at $z>6$.
\citet{oesch2010b} claim that the dust-extinction corrected
$\Sigma_{\rm SFR}$ is roughly constant over the
redshift range of $z\sim 3-7$. Extending their study,
we find this constant $\Sigma_{\rm SFR}$ up to $z\sim 8$.
Dotted and dashed lines in Figure \ref{fig:redshift_SigmaSFR} show 
the $\Sigma_{\rm SFR}$ values for $L^\ast_{z=3}$ and $0.3L^\ast_{z=3}$ 
expected from the best-fit size evolution of $(1+z)^s$.
These lines imply that one would find galaxies with extremely
high $\Sigma_{\rm SFR}$ at $z\gtrsim 10$ and beyond. 
A simple increase
of the average $\Sigma_{\rm SFR}$ is not expected, however. 
This is because the typical UV luminosity, $L^\ast$, is fainter at
higher redshifts and galaxies with $(0.3-1.0) L^\ast_{z=3}$
are quite rare at $z \gtrsim 10$. In this sense,
even higher-redshift galaxies cannot take
an extremely high average $\Sigma_{\rm SFR}$
beyond $\sim 3-10$ $M_\odot$ yr$^{-1}$ kpc$^{-2}$.
In the local universe, 
the SFR surface densities of normal disk galaxies are 
about $0.01 M_\odot$ yr$^{-1}$ kpc$^{-2}$, 
smaller than what we have found for $z \sim 4-8$ dropouts. 
The centers of normal disk galaxies, on the other hand,  
reach about $1 M_\odot$ yr$^{-1}$ kpc$^{-2}$ \citep[][see also \citealt{momose2010}]{kennicutt1998b}, 
which is comparable to $\Sigma_{\rm SFR}$ of $z \sim 4-8$ dropouts. 
Because local starbursts, especially nuclear starbursts, 
have high $\Sigma_{\rm SFR}$ up to $100-1000 M_\odot$ 
yr$^{-1}$ kpc$^{-2}$ \citep{kennicutt1998b}, 
the star-formation surface density of dropout galaxies 
at $z\sim 4-8$ is significantly smaller than
those of the extreme population found in the local universe, 
which indicate that star-formation in dropout galaxies
is not as rapid as that of local extreme starbursts.
Speculatively, 
because high-$z$ dropouts are metal and dust poor galaxies \citep[e.g.,][]{bouwens2012b}, 
gas cooling in a given amount of molecular clouds of 
high-$z$ dropouts would be less efficient than 
that of local starbursts.

%%%%%%%%%%%%%%%%%%%%%%%%%%%%%%%%%%%%%%%%%%%%%%%%%%%%%%%%%%%%%%%%%
%%%%%%%%%%%%%%%%%%%%%%%%%%%%%%%%%%%%%%%%%%%%%%%%%%%%%%%%%%%%%%%%%
\section{Summary} \label{sec:summary}
%%%%%%%%%%%%%%%%%%%%%%%%%%%%%%%%%%%%%%%%%%%%%%%%%%%%%%%%%%%%%%%%%
%%%%%%%%%%%%%%%%%%%%%%%%%%%%%%%%%%%%%%%%%%%%%%%%%%%%%%%%%%%%%%%%%

In this paper, 
we have presented sizes of dropout galaxy candidates at $z \sim 7-12$ 
identified by the 2012 Hubble Ultra Deep Field campaign. 
We have stacked the new F140W image 
with the existing F125W image and the deeper F160W image, 
to maximize the available depth at rest-frame wavelengths 
$\lambda_{\rm rest} \simeq 1600-1700${\AA} for 
$z_{850}$-dropout and $Y_{105}$-dropout samples respectively, 
allowing secure size measurement from $>15\sigma$ detections.  
The extremely deep F105W data ensures that $z>8$ candidates are robust, 
extending the redshift range of reliable objects.  
We have performed surface brightness profile fitting 
for our samples at $z \sim 7-12$. 
Our measurements have shown that 
the average half-light radii of galaxies are very small, 
$0.3-0.4$ kpc at $z \sim 7-12$. 
Such sizes are, perhaps coincidentally, comparable to 
the sizes of giant molecular associations in local star-forming galaxies.

Combining our new results at $z \simeq 7 - 12$ 
with existing average size measurements 
previously reported for dropout galaxies at $z \simeq 4 - 7$, 
we have investigated the size evolution of dropout galaxies.
We have confirmed the trend for size to decrease with increasing redshift 
(at a given luminosity) 
and have shown that this trend appears to extend out to $z \simeq 12$.
Motivated by the fact that, at least qualitatively, 
the sizes 
of the brighter ($0.3-1.0 L^\ast_{z=3}$)
and somewhat fainter ($0.12-0.3 L^\ast_{z=3}$) dropout galaxies 
show a similar 
trend with redshift, we have attempted to model the size evolution
of both samples together with a function of the form $(1+z)^{s}$ 
over the redshift range $z \simeq 4 - 12$. 
The result is a best-fitting parameter 
value of $s=-1.28\pm 0.13$, approximately
mid-way between the physically expected evolution 
for baryons embedded in dark halos of 
constant mass ($s=-1$) and constant velocity ($s=-1.5$).
This evolution is consistent with that derived by \cite{oesch2010b}, 
albeit our derived evolution with redshift is slightly steeper than 
that derived by \cite{oesch2010b} for the brighter galaxies. 
We have checked that our best-fitting value of $s$ is not significantly 
affected if the putative $z \simeq 12$ galaxy is excluded, but 
it is interesting that this object has a half-light radius which is 
perfectly consistent with our best-fitting relation.

We have also found that a clear size-luminosity relation, 
such as that found at lower redshift, is also evident in 
both our $z_{850}$- and $Y_{105}$-dropout samples. 
This relation can be interpreted in terms of a 
constant surface density of star formation over a range 
in luminosity of $0.05-1.0 L^\ast_{z=3}$. 
These size-redshift and size-luminosity relations 
suggest that galaxy sizes at $z>4$ are not simply decided 
by the evolution of the constant mass or velocity of the parent halo 
and/or 
follow in the evolution of the stellar to halo size ratio 
with a similar star-formation rate density.

Finally, our results also strengthen previous claims 
that the star-formation surface density in 
dropout galaxies is broadly unchanged from 
$z \simeq 4$ to $z \simeq 8$ at 
$\Sigma_{\rm SFR} \simeq 2\, {\rm M}_{\odot} \, {\rm yr}^{-1} \, {\rm kpc}^{-2}$. 
This value is $2-3$ orders of magnitude 
lower than that found in extreme starburst galaxies, 
but is very comparable to that seen today in the centers 
of normal disk galaxies. 
This provides further support for a steady smooth build-up 
of the stellar populations in galaxies in the young universe.

%%%%%%%%%%%%%%%%%%%%%%%%%%%%%%%%%%%%%%%%%%%%%%%%%%%%%%%%%%%%%%%%%
%%%%%%%%%%%%%%%%%%%%%%%%%%%%%%%%%%%%%%%%%%%%%%%%%%%%%%%%%%%%%%%%%
\section*{Acknowledgements}
%%%%%%%%%%%%%%%%%%%%%%%%%%%%%%%%%%%%%%%%%%%%%%%%%%%%%%%%%%%%%%%%%
%%%%%%%%%%%%%%%%%%%%%%%%%%%%%%%%%%%%%%%%%%%%%%%%%%%%%%%%%%%%%%%%%

We thank Rieko Momose and Suraphong Yuma for their helpful comments. 
This work was supported by 
Japan Society for the Promotion of Science (JSPS), 
Grants-in-Aid for Scientific Research (KAKENHI), 
Grant Numbers 24840010 and 23244025,
and World Premier International Research Center Initiative 
(WPI Initiative), MEXT, Japan.
EFCL, ABR and MC acknowledge the support of the UK Science 
and Technology Facilities Council.
US authors acknowledge financial support 
from the Space Telescope Science Institute 
under award HST-GO-12498.01-A.
JSD and RAAB acknowledge the support of the European Research Council
via the award of an Advanced Grant to JSD.
JSD also acknowledges the support of the Royal Society through a
Wolfson Research Merit Award.
SC acknowledges the support of the European Commission 
through the Marie Curie Initial Training Network ELIXIR.
This work is based on data 
from the \textit{Hubble Space Telescope} which is operated by NASA 
through the Space Telescope Science Institute 
via the association of Universities for Research in Astronomy, Inc. 
for NASA under contract NAS5-26555.

%%%%%%%%%%%%%%%%%%%%%%%%%%%%%%%%%%%%%%%%%%%%%%%%%%%%%%%%%%%%%%%%%
%%%%%%%%%%%%%%%%%%%%%%%%%%%%%%%%%%%%%%%%%%%%%%%%%%%%%%%%%%%%%%%%%

%\bibliographystyle{apj}
%\bibliography{apj-jour,../../../Papers/papers_gal/ms}

\begin{thebibliography}{50}
\expandafter\ifx\csname natexlab\endcsname\relax\def\natexlab#1{#1}\fi

\bibitem[{{Bertin} \& {Arnouts}(1996)}]{bertin1996}
{Bertin}, E., \& {Arnouts}, S. 1996, \aaps, 117, 393

\bibitem[{{Bouwens} {et~al.}(2004){Bouwens}, {Illingworth}, {Blakeslee},
  {Broadhurst}, \& {Franx}}]{bouwens2004}
{Bouwens}, R.~J., {Illingworth}, G.~D., {Blakeslee}, J.~P., {Broadhurst},
  T.~J., \& {Franx}, M. 2004, \apjl, 611, L1

\bibitem[{{Bouwens} {et~al.}(2006){Bouwens}, {Illingworth}, {Blakeslee}, \&
  {Franx}}]{bouwens2006b}
{Bouwens}, R.~J., {Illingworth}, G.~D., {Blakeslee}, J.~P., \& {Franx}, M.
  2006, \apj, 653, 53

\bibitem[{{Bouwens} {et~al.}(2008){Bouwens}, {Illingworth}, {Franx}, \&
  {Ford}}]{bouwens2008}
{Bouwens}, R.~J., {Illingworth}, G.~D., {Franx}, M., \& {Ford}, H. 2008, \apj,
  686, 230

\bibitem[{{Bouwens} {et~al.}(2010){Bouwens}, {Illingworth}, {Oesch},
  {Stiavelli}, {van Dokkum}, {Trenti}, {Magee}, {Labb{\'e}}, {Franx},
  {Carollo}, \& {Gonzalez}}]{bouwens2010}
{Bouwens}, R.~J., {Illingworth}, G.~D., {Oesch}, P.~A., {et~al.} 2010, \apjl,
  709, L133

\bibitem[{{Bouwens} {et~al.}(2011{\natexlab{a}}){Bouwens}, {Illingworth},
  {Labbe}, {Oesch}, {Trenti}, {Carollo}, {van Dokkum}, {Franx}, {Stiavelli},
  {Gonz{\'a}lez}, {Magee}, \& {Bradley}}]{bouwens2011b}
{Bouwens}, R.~J., {Illingworth}, G.~D., {Labbe}, I., {et~al.}
  2011{\natexlab{a}}, \nat, 469, 504

\bibitem[{{Bouwens} {et~al.}(2011{\natexlab{b}}){Bouwens}, {Illingworth},
  {Oesch}, {Labb{\'e}}, {Trenti}, {van Dokkum}, {Franx}, {Stiavelli},
  {Carollo}, {Magee}, \& {Gonzalez}}]{bouwens2011}
{Bouwens}, R.~J., {Illingworth}, G.~D., {Oesch}, P.~A., {et~al.}
  2011{\natexlab{b}}, \apj, 737, 90

\bibitem[{{Bouwens} {et~al.}(2012{\natexlab{a}}){Bouwens}, {Oesch},
  {Illingworth}, {Labbe}, {Magee}, {Smit}, {Franx}, {van Dokkum}, {Trenti},
  {Gonzalez}, \& {Carollo}}]{bouwens2012d}
{Bouwens}, R.~J., {Oesch}, P.~A., {Illingworth}, G.~D., {et~al.}
  2012{\natexlab{a}}, ArXiv e-prints (arXiv:1211.3105)

\bibitem[{{Bouwens} {et~al.}(2012{\natexlab{b}}){Bouwens}, {Illingworth},
  {Oesch}, {Franx}, {Labb{\'e}}, {Trenti}, {van Dokkum}, {Carollo},
  {Gonz{\'a}lez}, {Smit}, \& {Magee}}]{bouwens2012b}
{Bouwens}, R.~J., {Illingworth}, G.~D., {Oesch}, P.~A., {et~al.}
  2012{\natexlab{b}}, \apj, 754, 83

\bibitem[{{Bunker} {et~al.}(2010){Bunker}, {Wilkins}, {Ellis}, {Stark},
  {Lorenzoni}, {Chiu}, {Lacy}, {Jarvis}, \& {Hickey}}]{bunker2010}
{Bunker}, A.~J., {Wilkins}, S., {Ellis}, R.~S., {et~al.} 2010, \mnras, 409, 855

\bibitem[{{Castellano} {et~al.}(2010){Castellano}, {Fontana}, {Boutsia},
  {Grazian}, {Pentericci}, {Bouwens}, {Dickinson}, {Giavalisco}, {Santini},
  {Cristiani}, {Fiore}, {Gallozzi}, {Giallongo}, {Maiolino}, {Mannucci},
  {Menci}, {Moorwood}, {Nonino}, {Paris}, {Renzini}, {Rosati}, {Salimbeni},
  {Testa}, \& {Vanzella}}]{castellano2010}
{Castellano}, M., {Fontana}, A., {Boutsia}, K., {et~al.} 2010, \aap, 511, A20+

\bibitem[{{de Jong} \& {Lacey}(2000)}]{dejong2000}
{de Jong}, R.~S., \& {Lacey}, C. 2000, \apj, 545, 781

\bibitem[{{Dunlop} {et~al.}(2012){Dunlop}, {Rogers}, {McLure}, {Ellis},
  {Robertson}, {Koekemoer}, {Dayal}, {Curtis-Lake}, {Wild}, {Charlot},
  {Bowler}, {Schenker}, {Ouchi}, {Ono}, {Cirasuolo}, {Furlanetto}, {Stark},
  {Targett}, \& {Schneider}}]{dunlop2012}
{Dunlop}, J.~S., {Rogers}, A.~B., {McLure}, R.~J., {et~al.} 2012, MNRAS,
  submitted (arXiv:1212.0860)

\bibitem[{{Ellis} {et~al.}(2012){Ellis}, {McLure}, {Dunlop}, {Robertson},
  {Ono}, {Schenker}, {Koekemoer}, {Bowler}, {Ouchi}, {Rogers}, {Curtis-Lake},
  {Schneider}, {Charlot}, {Stark}, {Furlanetto}, \& {Cirasuolo}}]{ellis2012b}
{Ellis}, R.~S., {McLure}, R.~J., {Dunlop}, J.~S., {et~al.} 2012, ApJL, accepted
  (arXiv:1211.6804)

\bibitem[{{Ferguson} {et~al.}(2004){Ferguson}, {Dickinson}, {Giavalisco},
  {Kretchmer}, {Ravindranath}, {Idzi}, {Taylor}, {Conselice}, {Fall},
  {Gardner}, {Livio}, {Madau}, {Moustakas}, {Papovich}, {Somerville},
  {Spinrad}, \& {Stern}}]{ferguson2004}
{Ferguson}, H.~C., {Dickinson}, M., {Giavalisco}, M., {et~al.} 2004, \apjl,
  600, L107

\bibitem[{{Finkelstein} {et~al.}(2010){Finkelstein}, {Papovich}, {Giavalisco},
  {Reddy}, {Ferguson}, {Koekemoer}, \& {Dickinson}}]{finkelstein2009f}
{Finkelstein}, S.~L., {Papovich}, C., {Giavalisco}, M., {et~al.} 2010, \apj,
  719, 1250

\bibitem[{{Fontana} {et~al.}(2010){Fontana}, {Vanzella}, {Pentericci},
  {Castellano}, {Giavalisco}, {Grazian}, {Boutsia}, {Cristiani}, {Dickinson},
  {Giallongo}, {Maiolino}, {Moorwood}, \& {Santini}}]{fontana2010}
{Fontana}, A., {Vanzella}, E., {Pentericci}, L., {et~al.} 2010, \apjl, 725,
  L205

\bibitem[{{Grazian} {et~al.}(2012){Grazian}, {Castellano}, {Fontana},
  {Pentericci}, {Dunlop}, {McLure}, {Koekemoer}, {Dickinson}, {Faber},
  {Ferguson}, {Galametz}, {Giavalisco}, {Grogin}, {Hathi}, {Kocevski}, {Lai},
  {Newman}, \& {Vanzella}}]{grazian2012}
{Grazian}, A., {Castellano}, M., {Fontana}, A., {et~al.} 2012, \aap, 547, A51

\bibitem[{{Hathi} {et~al.}(2008){Hathi}, {Malhotra}, \& {Rhoads}}]{hathi2008}
{Hathi}, N.~P., {Malhotra}, S., \& {Rhoads}, J.~E. 2008, \apj, 673, 686

\bibitem[{{Kennicutt}(1998{\natexlab{a}})}]{kennicutt1998}
{Kennicutt}, Jr., R.~C. 1998{\natexlab{a}}, \araa, 36, 189

\bibitem[{{Kennicutt}(1998{\natexlab{b}})}]{kennicutt1998b}
---. 1998{\natexlab{b}}, \apj, 498, 541

\bibitem[{{Koekemoer} {et~al.}(2012){Koekemoer}, {Ellis}, {McLure}, {Dunlop},
  {Robertson}, {Ono}, {Schenker}, {Ouchi}, {Bowler}, {Rogers}, {Curtis-Lake},
  {Schneider}, {Charlot}, {Stark}, {Furlanetto}, {Cirasuolo}, {Wild}, \&
  {Targett}}]{koekemoer2012}
{Koekemoer}, A.~M., {Ellis}, R.~S., {McLure}, R.~J., {et~al.} 2012, ApJS,
  submitted (arXiv:1212.1448)

\bibitem[{{Lorenzoni} {et~al.}(2011){Lorenzoni}, {Bunker}, {Wilkins},
  {Stanway}, {Jarvis}, \& {Caruana}}]{lorenzoni2010}
{Lorenzoni}, S., {Bunker}, A.~J., {Wilkins}, S.~M., {et~al.} 2011, \mnras, 414,
  1455

\bibitem[{{McLure} {et~al.}(2010){McLure}, {Dunlop}, {Cirasuolo}, {Koekemoer},
  {Sabbi}, {Stark}, {Targett}, \& {Ellis}}]{mclure2009}
{McLure}, R.~J., {Dunlop}, J.~S., {Cirasuolo}, M., {et~al.} 2010, \mnras, 403,
  960

\bibitem[{{McLure} {et~al.}(2011){McLure}, {Dunlop}, {de Ravel}, {Cirasuolo},
  {Ellis}, {Schenker}, {Robertson}, {Koekemoer}, {Stark}, \&
  {Bowler}}]{mclure2011}
{McLure}, R.~J., {Dunlop}, J.~S., {de Ravel}, L., {et~al.} 2011, \mnras, 418,
  2074

\bibitem[{{McLure} {et~al.}(2012){McLure}, {McLure}, {McLure}, {McLure},
  {McLure}, {McLure}, {McLure}, {McLure}, {McLure}, \& {McLure}}]{mclure2012b}
{McLure}, R.~J., {et~al.} 2012, in preparation

\bibitem[{{Mo} {et~al.}(1998){Mo}, {Mao}, \& {White}}]{mo1998}
{Mo}, H.~J., {Mao}, S., \& {White}, S.~D.~M. 1998, \mnras, 295, 319

\bibitem[{{Mo} {et~al.}(1999){Mo}, {Mao}, \& {White}}]{mo1999}
---. 1999, \mnras, 304, 175

\bibitem[{{Mo} \& {White}(2002)}]{mo2002}
{Mo}, H.~J., \& {White}, S.~D.~M. 2002, \mnras, 336, 112

\bibitem[{{Momose} {et~al.}(2010){Momose}, {Okumura}, {Koda}, \&
  {Sawada}}]{momose2010}
{Momose}, R., {Okumura}, S.~K., {Koda}, J., \& {Sawada}, T. 2010, \apj, 721,
  383

\bibitem[{{Mosleh} {et~al.}(2012){Mosleh}, {Williams}, {Franx}, {Gonzalez},
  {Bouwens}, {Oesch}, {Labbe}, {Illingworth}, \& {Trenti}}]{mosleh2012}
{Mosleh}, M., {Williams}, R.~J., {Franx}, M., {et~al.} 2012, \apjl, 756, L12

\bibitem[{{Mu{\~n}oz} \& {Furlanetto}(2012)}]{munoz2012}
{Mu{\~n}oz}, J.~A., \& {Furlanetto}, S. 2012, \mnras, 426, 3477

\bibitem[{{Oesch} {et~al.}(2009){Oesch}, {Carollo}, {Stiavelli}, {Trenti},
  {Bergeron}, {Koekemoer}, {Lucas}, {Pavlovsky}, {Beckwith}, {Dahlen},
  {Ferguson}, {Gardner}, {Lilly}, {Mobasher}, \& {Panagia}}]{oesch2009}
{Oesch}, P.~A., {Carollo}, C.~M., {Stiavelli}, M., {et~al.} 2009, \apj, 690,
  1350

\bibitem[{{Oesch} {et~al.}(2010{\natexlab{a}}){Oesch}, {Bouwens}, {Carollo},
  {Illingworth}, {Trenti}, {Stiavelli}, {Magee}, {Labb{\'e}}, \&
  {Franx}}]{oesch2010b}
{Oesch}, P.~A., {Bouwens}, R.~J., {Carollo}, C.~M., {et~al.}
  2010{\natexlab{a}}, \apjl, 709, L21

\bibitem[{{Oesch} {et~al.}(2010{\natexlab{b}}){Oesch}, {Bouwens},
  {Illingworth}, {Carollo}, {Franx}, {Labb{\'e}}, {Magee}, {Stiavelli},
  {Trenti}, \& {van Dokkum}}]{oesch2010}
{Oesch}, P.~A., {Bouwens}, R.~J., {Illingworth}, G.~D., {et~al.}
  2010{\natexlab{b}}, \apjl, 709, L16

\bibitem[{{Oke} \& {Gunn}(1983)}]{oke1983}
{Oke}, J.~B., \& {Gunn}, J.~E. 1983, \apj, 266, 713

\bibitem[{{Ouchi} {et~al.}(2009){Ouchi}, {Mobasher}, {Shimasaku}, {Ferguson},
  {Fall}, {Ono}, {Kashikawa}, {Morokuma}, {Nakajima}, {Okamura}, {Dickinson},
  {Giavalisco}, \& {Ohta}}]{ouchi2009b}
{Ouchi}, M., {Mobasher}, B., {Shimasaku}, K., {et~al.} 2009, \apj, 706, 1136

\bibitem[{{Peng} {et~al.}(2002){Peng}, {Ho}, {Impey}, \& {Rix}}]{peng2002}
{Peng}, C.~Y., {Ho}, L.~C., {Impey}, C.~D., \& {Rix}, H.-W. 2002, \aj, 124, 266

\bibitem[{{Peng} {et~al.}(2010){Peng}, {Ho}, {Impey}, \& {Rix}}]{peng2010}
---. 2010, \aj, 139, 2097

\bibitem[{{Pirzkal} {et~al.}(2005){Pirzkal}, {Sahu}, {Burgasser}, {Moustakas},
  {Xu}, {Malhotra}, {Rhoads}, {Koekemoer}, {Nelan}, {Windhorst}, {Panagia},
  {Gronwall}, {Pasquali}, \& {Walsh}}]{pirzkal2005}
{Pirzkal}, N., {Sahu}, K.~C., {Burgasser}, A., {et~al.} 2005, \apj, 622, 319

\bibitem[{{Rand} \& {Kulkarni}(1990)}]{rand1990}
{Rand}, R.~J., \& {Kulkarni}, S.~R. 1990, \apjl, 349, L43

\bibitem[{{Robertson} {et~al.}(2012){Robertson}, {Robertson}, {Robertson}, {Robertson},
  {Robertson}, {Robertson}, {Robertson}, {Robertson}, {Robertson}, \& {Robertson}}]{robertson2012}
{Robertson}, B.~E., {et~al.} 2012, in preparation

\bibitem[{{Schenker} {et~al.}(2012){Schenker}, {Schenker}, {Schenker},
  {Schenker}, {Schenker}, {Schenker}, {Schenker}, {Schenker}, \&
  {Schenker}}]{schenker2012}
{Schenker}, M.~A., {et~al.} 2012, in preparation

\bibitem[{{Sersic}(1968)}]{sersic1968}
{Sersic}, J.~L. 1968, {Atlas de galaxias australes}

\bibitem[{{Somerville} {et~al.}(2008){Somerville}, {Barden}, {Rix}, {Bell},
  {Beckwith}, {Borch}, {Caldwell}, {H{\"a}u{\ss}ler}, {Heymans}, {Jahnke},
  {Jogee}, {McIntosh}, {Meisenheimer}, {Peng}, {S{\'a}nchez}, {Wisotzki}, \&
  {Wolf}}]{somerville2008}
{Somerville}, R.~S., {Barden}, M., {Rix}, H.-W., {et~al.} 2008, \apj, 672, 776

\bibitem[{{Steidel} {et~al.}(1999){Steidel}, {Adelberger}, {Giavalisco},
  {Dickinson}, \& {Pettini}}]{steidel1999}
{Steidel}, C.~C., {Adelberger}, K.~L., {Giavalisco}, M., {Dickinson}, M., \&
  {Pettini}, M. 1999, \apj, 519, 1

\bibitem[{{Tosaki} {et~al.}(2007){Tosaki}, {Shioya}, {Kuno}, {Hasegawa},
  {Nakanishi}, {Matsushita}, \& {Kohno}}]{tosaki2007}
{Tosaki}, T., {Shioya}, Y., {Kuno}, N., {et~al.} 2007, \pasj, 59, 33

\bibitem[{{Vogel} {et~al.}(1988){Vogel}, {Kulkarni}, \& {Scoville}}]{vogel1988}
{Vogel}, S.~N., {Kulkarni}, S.~R., \& {Scoville}, N.~Z. 1988, \nat, 334, 402

\bibitem[{{Wilkins} {et~al.}(2011){Wilkins}, {Bunker}, {Lorenzoni}, \&
  {Caruana}}]{wilkins2011}
{Wilkins}, S.~M., {Bunker}, A.~J., {Lorenzoni}, S., \& {Caruana}, J. 2011,
  \mnras, 411, 23

\bibitem[{{Yan} {et~al.}(2010){Yan}, {Windhorst}, {Hathi}, {Cohen}, {Ryan},
  {O'Connell}, \& {McCarthy}}]{yan2010}
{Yan}, H., {Windhorst}, R.~A., {Hathi}, N.~P., {et~al.} 2010, Research in
  Astronomy and Astrophysics, 10, 867

\end{thebibliography}

%%%%%%%%%%%%%%%%%%%%%%%%%%%%%%%%%%%%%%%%%%%%%%%%%%%%%%%%%%%%%%%%%
%%%%%%%%%%%%%%%%%%%%%%%%%%%%%%%%%%%%%%%%%%%%%%%%%%%%%%%%%%%%%%%%%
%%%%%%%%%%%%%%%%%%%%%%%%%%%%%%%%%%%%%%%%%%%%%%%%%%%%%%%%%%%%%%%%%

\end{document}